\documentclass[12pt]{article}

\usepackage{amsmath,amssymb,amsthm,bbm,dsfont,mathrsfs}
\usepackage{a4wide}
\usepackage{graphicx}

\DeclareMathAlphabet{\mathpzc}{OT1}{pzc}{m}{it}

\newtheorem{thm}{Theorem}[section]
\newtheorem{lem}{Lemma}[section]
\newtheorem{dfi}{Definition}[section]

\newcommand{\sK}{{\cal K}}

\newcommand{\sP}{{\cal P}}

\newcommand{\sV}{{\cal V}}

\newcommand{\va}{\scriptscriptstyle}
\newcommand{\vani}{\scriptstyle}
\newcommand{\R}{{\Bbb R}}
\newcommand{\C}{{\Bbb C}}
\newcommand{\Z}{{\Bbb Z}}
\newcommand{\N}{{\Bbb N}}
\newcommand{\phys}{\ \ =_{ \left.\right. _{\left.\right._{\!\!\!\!\!\!\!\!\!\!\!\!\!phys}}} \ \ }

\begin{document}

\title{Spin Foam Models for Quantum Gravity}

\author{Alejandro Perez\\
{} \\
 Center for Gravitational Physics and Geometry,\\ The
Pennsylvania
State University \\
University Park, PA 16802, USA\\
and\\
Erwin Schr\"odinger International\\ Institute for Mathematical
Physics\\
Boltzmanngasse 9,A-1090, Wien, Austria}
\date{\today}

\maketitle


\begin{abstract}

In this article we review the present status of the spin foam
formulation of non-perturbative (background independent) quantum
gravity. The article is divided in two parts. In the first part we
present a general introduction to the main ideas emphasizing their
motivations from various perspectives. Riemannian 3-dimensional
gravity is used as a simple example to illustrate conceptual issues and the main goals
of the approach. The main features of the various existing models for 4-dimensional
gravity are also presented here. We conclude with a discussion of
important questions to be addressed in four dimensions (gauge
invariance, discretization independence, etc.).

In the second part we concentrate on the definition of the
Barrett-Crane model. We present the main results obtained in this
framework from a critical perspective. Finally we review the combinatorial
formulation of spin foam models based on the dual group field
theory technology. We present the Barrett-Crane model in this
framework and review the finiteness results obtained for both its
Riemannian as well as its Lorentzian variants. 
\end{abstract}

\newpage

\newpage

\tableofcontents

\newpage

\section{Introduction}

Quantum gravity, the theory expected to reconcile the principles
of quantum mechanics and general relativity, remains a major
challenge in theoretical physics (for a review of the history of quantum gravity
see  \cite{c10}). The main lesson of general
relativity is that, unlike in any other interaction, space-time
geometry is fully dynamical. This special feature of gravity
precludes the possibility of representing fields on a fixed
background geometry and severely constrains the applicability of
standard techniques that are successful in the description of
other interactions. Although the necessity of a background
independent formulation of quantum gravity is widely recognized,
there is a current debate about the means by which it should be
implemented. In particular, it is not clear whether the
non renormalizability of perturbative quantum gravity should be
interpreted as an indication of the inconsistency of general
relativity at high energies, the inconsistency of the 
background-dependent framework applied to gravity, 
or a combination of both.

According to the (background dependent) perspective of
standard QFT \cite{wein}, non renormalizability signals the
inconsistency of the theory at high energies to be
corrected by a more fundamental theory in the UV regime. A
classical example of this is Fermi's four-fermion theory as an
effective description of the weak interaction. According to this
view different approaches to quantum gravity have been defined in
terms of modifications of general relativity based on
supersymmetry, higher dimensions, strings, etc. The finiteness
properties of the perturbative expansions (which are background
dependent from the onset) are improved in these theories; however,
the definition of a background independent quantization of such
modifications remains open.

The approach of non perturbative quantum gravity is based on a
different interpretation of the infinities in perturbative quantum
gravity: it is precisely the perturbative (background dependent)
techniques which are inconsistent with the fundamental nature of
gravity. This view is strongly suggested by the predictions of
the background independent canonical quantization of general
relativity defined by loop quantum gravity. Loop quantum
gravity (LQG) is a non perturbative formulation of quantum gravity
based on the connection formulation of general relativity (for an
reviews on the subject see  \cite{ash10,th4,c0,c9}). 
A great deal of progress has been made
within the theory. At the mathematical level, the main achievement
is the rigorous definition of the Hilbert space of quantum
geometry, the regularization of geometric operators and the
rigorous definition of the quantum Hamiltonian constraint (defining the quantum dynamics). 
States of quantum geometry are given by
polymer-like excitations supported on graphs (spin network
states). From the physical viewpoint its main prediction is the
discreteness of geometry at the Planck scale.
This provides a clear-cut understanding of the problem of UV divergences
in perturbative general relativity: at the Planck scale the
classical notion of space and time simply ceases to exist;
therefore, it is the assumption of a fixed smooth background
geometry (typically flat space-time) in perturbation theory that
becomes inconsistent at high energies. The theory successfully
incorporates interactions between quantum geometry and quantum
matter in a way that is completely free of divergences \cite{th5}.
The quantum nature of space appears as a physical regulator for
the other interactions.
 
Dynamics is governed by the quantum Hamiltonian constraint. Even
when this operator is rigorously defined \cite{th1} it is
technically difficult to characterize its solution space. This is
partly because the $3+1$-decomposition of space-time (necessary in
the canonical formulation) breaks the manifest $4$-diffeomorphism
invariance of the theory making awkward the analysis of
dynamics. The situation is somewhat analogous to that in standard
quantum field theory.  In the
Hamiltonian formulation of standard quantum field theory
manifest Lorentz invariance is lost due to a particular choice of
time slicing of Minkowski space-time. The formalism is certainly Lorentz
invariant, but one has to work harder to show it explicitly.
Manifest Lorentz invariance can be kept only in the Lagrangian
(path-integral) quantization making the (formal) path integral a powerful
device for analyzing relativistic dynamics.

Consequently, there has been growing interest in trying to define
dynamics in loop quantum gravity from a $4$-dimensional covariant
perspective. This has given rise to the so-called spin foam
approach to quantum gravity. Its main idea is the
construction of a rigorous definition of the path integral for
gravity based on the deep insights obtained in the canonical
framework of loop quantum gravity. In turn, the path integral
provides a device to explicitly solve the dynamics: path-integral
transition amplitudes can be shown to correspond to solutions of
the quantum Hamiltonian constraint.

The underlying discreteness discovered in loop quantum gravity is crucial: in spin
foam models the formal Misner-Hawking functional integral for
gravity is replaced by a sum over combinatorial objects given by
foam-like configurations (spin foams). A spin foam represents a possible history of
the gravitational field and can be interpreted as a set of transitions
through different quantum states of space. Boundary data in the path integral are given
by the polymer-like excitations (spin network states) 
representing $3$-geometry states in loop quantum gravity.
General covariance implies the absence of a meaningful notion of time and
transition amplitudes are to be interpreted as defining the
physical scalar product.

While the construction can be explicitly carried out in three
dimensions there are additional technical difficulties in four
dimensions. Various models have been proposed. A natural
question is whether the infinite sums over geometries defining
transition amplitudes would converge. In fact, there is no UV
problem due to the fundamental discreteness and potential
divergences are associated to the IR regime. There are recent
results in the context of the Barrett-Crane model showing that
amplitudes are well defined when the topology of the histories is
restricted in a certain way.

The aim of this article is to provide a comprehensive review of
the progress that has been achieved in the spin foam approach over
the last few years and provide as well a self contained
introduction for the interested reader that is not familiar with the
subject. The article is divided
into two fundamental parts. In the first part we present a general
introduction to the subject including a brief summary of LQG in
Section \ref{lqg}. We introduce the spin foam formulation from
different perspectives in Section \ref{sec:intsf}. In Section
\ref{sfm3d} we present a simple example of spin foam model:
Riemannian $3$-dimensional gravity. We use this example as the
basic tool to introduce the main ideas and to illustrate various
conceptual issues. We review the different proposed models for 
$4$-dimensional quantum gravity in Section \ref{sfm4d}. Finally,
in Section \ref{sci} we conclude the first part by analyzing the
various conceptual issues that arise in the approach. The first
part is a general introduction to the formalism; it is
self contained and could be read independently.

One of the simplest and most studied spin foam model for
$4$-dimensional gravity is the Barrett-Crane model \cite{BC2,BC1}.
The main purpose of the second part is to present a critical
survey of the different results that have been obtained in this
framework and its combinatorial generalizations \cite{fre2,a10,a9}
based on the dual group field theory (GFT) formulation. In Section
\ref{BCM} we present  a systematic derivation of the
Barrett-Crane model from the $Spin(4)$ Plebanski's formulation. This
derivation follows an alternative path from that of
 \cite{baez7,baez6}; here we emphasize the connection to a simplicial
action. 

Spin foams can be thought of as Feynman diagrams. In fact
a wide class of spin foam models can be derived from the
perturbative (Feynman) expansion of certain dual group field
theories (GFT) \cite{reis1,reis2}. A brief review of the main ideas
involved is presented in Section \ref{sec:gft-sf}. We conclude the
second part by studying the GFT formulation of the Barrett-Crane
model for both Riemannian and Lorentzian geometry. 
The definition of the actual models and the
sketch of the corresponding finiteness proofs \cite{a7,a2,a22} are
given in Section \ref{BCGFT}.

\part{Main ideas}

\section{Loop Quantum Gravity and Quantum Geometry}\label{lqg}

Loop quantum gravity is a rigorous realization of the quantization
program established in the 60's by Dirac, Wheeler, De-Witt, among
others (for recent reviews see \cite{ash10,th4,c9}). The technical
difficulties of Wheeler's `geometrodynamics' are circumvent by the
use of connection variables instead of metrics \cite{ash,
ash1,barbero}. At the kinematical level, the formulation is
similar to that of standard gauge theories. The fundamental
difference is however the absence of any non-dynamical background
field in the theory.

The configuration variable is an $SU(2)$-connection $A_a^i$
on a 3-manifold $\Sigma$ representing space. The canonical momenta
are given by the densitized triad $E_i^a$. The latter encode the
(fully dynamical) Riemannian geometry of $\Sigma$ and are the
analog of the `electric fields' of Yang-Mills theory.

In addition to diffeomorphisms there is the local $SU(2)$ gauge freedom that
rotates the triad and transforms the connection
in the usual way. According to Dirac, gauge freedoms result in
constraints among the phase space variables which conversely are
the generating functionals of infinitesimal gauge transformations.
In terms of connection variables the constraints are
\begin{equation}\label{constro}
{\cal G}_{i}={\cal D}_a E^a_i=0,\ \ \ \ {\cal C}_a=E_k^b
F^k_{ba}=0, \ \ \ \ {\cal S}=\epsilon^{ijk}E^a_iE^b_j F_{ab\, k}+
\cdots=0,
\end{equation}
where ${\cal D}_a$ is the covariant derivative and $F_{ba}$ is the
curvature of $A_a^i$.  ${\cal G}_{i}$ is the familiar Gauss
constraint---analogous to the Gauss law of electromagnetism---generating 
infinitesimal $SU(2)$ gauge transformations, ${\cal
C}_{a}$ is the vector constraint generating space-diffeomorphism, 
and ${\cal S}$ is the scalar constraint generating
`time' reparameterization (there is an additional term that 
we have omitted for simplicity).

Loop quantum gravity is defined using Dirac quantization. One
first represents (\ref{constro}) as operators in an auxiliary
Hilbert space $\cal H$  and then solves the constraint equations
\begin{equation}\label{constroq}
\hat {\cal G}_{i}\Psi=0,\ \ \ \ \hat {\cal C}_a \Psi=0, \ \ \ \
\hat {\cal S}\Psi=0.
\end{equation}
The Hilbert space of solutions is the so-called physical Hilbert
space ${\cal H}_{phys}$. In a generally covariant system quantum
dynamics is fully governed by constraint equations. In the case of
loop quantum gravity they represent {\em quantum Einstein's
equations}.

States in the auxiliary Hilbert space are represented by wave
functionals of the connection $\Psi(A)$ which are square
integrable with respect to a natural diffeomorphism invariant
measure, the Ashtekar-Lewandowski measure \cite{ash3} (we denote
it ${\cal L}^2[{\cal A}]$ where ${\cal A}$ is the space of
(generalized) connections). This space can be decomposed into a
direct sum of orthogonal subspaces ${\cal H}=\bigoplus_{\gamma}
{\cal H}_{\gamma}$ labeled by a graph $\gamma$ in $\Sigma$. The
fundamental excitations are given by the holonomy $h_{\ell}(A)\in
SU(2)$ along a path $\ell$ in $\Sigma$:
\begin{equation}
h_{\ell}(A)={\cal P}\ {\rm exp}\ \int_{\ell} A.
\end{equation}
Elements of ${\cal H}_{\gamma}$ are given by functions
\begin{equation}
\Psi_{f,\gamma}(A)=f(h_{\ell_{1}}(A),h_{\ell_{2}}(A),\dots,h_{\ell_{n}}(A)),
\end{equation}
where $h_{\ell}$ is the holonomy along the links $\ell\in \gamma$
and $f:SU(2)^n\rightarrow \C$ is (Haar measure) square integrable.
They are called {\em cylindrical functions} and represent a dense
set in $\cal H$ denoted $Cyl$.

\begin{figure}[h]
\centering {\includegraphics[width=10cm]{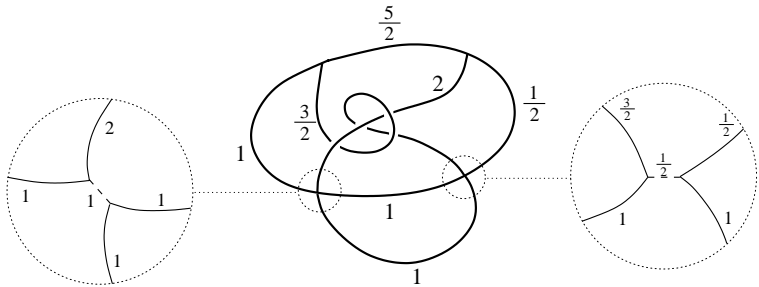}}
\caption{Spin-network state: At 3-valent nodes the intertwiner is
uniquely specified by the corresponding spins. At 4 or higher
valent nodes an intertwiner has to be specified. Choosing an
intertwiner corresponds to decompose the $n$-valent node in terms
of $3$-valent ones adding new virtual links (dashed lines) and
their corresponding spins. This is illustrated explicitly in the
figure for the two $4$-valent nodes.}\label{spinn}
\end{figure}

Gauge transformations generated by the Gauss constraint act
non-trivially at the endpoints of the holonomy, i.e., at nodes of
graphs. The Gauss constraint (in (\ref{constro})) is solved by
looking at $SU(2)$ gauge invariant functionals of the connection
(${\cal L}^2[{\cal A}]/{\cal G}$). The fundamental gauge invariant
quantity is given by the holonomy around closed loops. An
orthonormal basis of the kernel of the Gauss constraint is defined
by the so called spin network states 
$\Psi_{\va \gamma, \{j_{\ell}\},\{\iota_{n}\}}(A)$ \cite{reis8, c4, baez10}. Spin-networks
\footnote{Spin-networks were introduced by 
Penrose \cite{pen} in a attempt to define $3$-dimensional
Euclidean quantum geometry from the combinatorics of angular momentum in QM. 
Independently they have been used in lattice gauge theory \cite{ko,fur} as a natural basis
for gauge invariant functions on the lattice. 
For an account of their applications in various contexts see \cite{lee2}.}
are defined by a graph $\gamma$ in $\Sigma$, a collection of spins
$\{j_{\ell}\}$---unitary irreducible representations of $SU(2)$---associated 
with links $\ell\in \gamma$ and a collection of $SU(2)$
intertwiners $\{\iota_{n}\}$ associated to nodes $n \in \gamma$
(see Figure \ref{spinn}). The spin-network gauge invariant wave
functional $\Psi_{\va \gamma, \{j_{\ell}\},\{\iota_{n}\}}(A)$ is
constructed by first associating an $SU(2)$ matrix in the
$j_{\ell}$-representation to the holonomies $h_{\ell}(A)$
corresponding to the link $\ell$, and then contracting the
representation matrices at nodes with the corresponding
intertwiners $\iota_n$, namely
\begin{equation}
\Psi_{\va \gamma, \{j_{\ell}\},\{\iota_{n}\}}(A)=\prod_{n \in
\gamma} \ \iota_n\ \prod_{\ell \in \gamma}\  j_{\ell}[h_{\va
\ell}(A)],
\end{equation}
where $ j_{\ell}[h_{\va
\ell}(A)]$ denotes the corresponding $j_{\ell}$-representation matrix
evaluated at corresponding link holonomy
and the matrix index contraction is left implicit.

The solution of the vector constraint is more subtle \cite{ash3}.
One uses group averaging techniques and the existence of the
diffeomorphism invariant measure. The fact that zero lies in the
continuous spectrum of the diffeomorphism constraint implies
solutions to correspond to generalized states. These are not in
$\cal H$ but are elements of the topological dual
$Cyl^*$\footnote{According to the triple $Cyl\subset {\cal H}
\subset Cyl^*$.}. However, the intuitive idea is quite simple: 
solutions to the vector constraint are given by equivalence classes
of spin-network states up to diffeomorphism. Two spin-network states
are considered equivalent if their underlying graphs can be deformed into each 
other by the action of a diffeomorphism. 

This can be regarded as an indication that the smooth spin-network category
could be replaced by something which is more combinatorial in nature so that 
diffeomorphism invariance becomes a derived property of the classical limit. 
LQG has been modified along these lines by replacing the smooth 
manifold structure of the standard theory by
the weaker concept of piecewise linear manifold \cite{za2}. In this context,
graphs defining spin-network states can be completely characterized using 
the combinatorics of cellular decompositions of space. Only a discrete 
analog of the diffeomorphism symmetry survives which can be 
dealt with in a fully combinatorial manner.
We will take this point of view when we introduce the notion 
of spin foam in the following section.

\subsection*{Quantum geometry}

The generalized states described above solve all of the constraints
(\ref{constro}) but the scalar constraint. They are regarded as
quantum states of the Riemannian geometry on $\Sigma$. They define
the kinematical sector of the theory known as {\em quantum
geometry}.

Geometric operators acting on spin network states  can
be defined in terms of the fundamental triad 
operators $\hat E^a_i$. The simplest of such operators is the area of a surface
$S$ classically given by
\begin{equation}
A_{S}(E)=\int_S dx^2 \sqrt{{\rm Tr}[n_an_bE^aE^b]}
\end{equation}
where $n$ is a co normal. The geometric operator $\hat A_{S}(E)$
can be rigorously defined by its action on spin network states \cite{lee1,c3,ash2}. 
The area operator gives a clear geometrical
interpretation to spin-network states: the fundamental
1-dimensional excitations defining a spin-network state can be
thought of as quantized `flux lines' of area. More precisely, if
the surface $S\subset \Sigma$ is punctured by a spin-network link
carrying a spin $j$, this state is an eigenstate of $\hat A_S(E)$
with eigenvalue proportional to $\ell^2_P \sqrt{j(j+1)} $. In
the generic sector---where no node lies on the surface---the
spectrum takes the simple form
\begin{equation}\label{aarreeaa}
a_S(\{j\})=8 \pi \iota \ell^2_P \sum_i \sqrt{j_i(j_i+1)},
\end{equation}
where $i$ labels punctures and $\iota$ is the Imirzi parameter \cite{immi} \footnote{The
Imirzi parameter $\iota$ is a free parameter in the theory. This ambiguity is purely quantum
mechanical (it disappears in the classical limit). It has to be
fixed in terms of physical predictions. The computation of black hole
entropy in LQG fixes the value of $\iota$ (see \cite{ash9}).}. $a_S(\{j\})$ is the sum of single
puncture contributions. The general form of the spectrum including
the cases where nodes lie on $S$ has been computed in closed
form \cite{ash2}.

The spectrum of the volume operator is also discrete \cite{lee1,c3,loll1,ash22}. If we define
the volume operator $\hat V_\sigma(E)$ of a $3$-dimensional region
$\sigma\subset \Sigma$ then non vanishing eigenstates
are given by spin-networks containing $n$-valent nodes in
$\sigma$ for $n > 3$. Volume is concentrated in nodes.

\subsection*{Quantum dynamics}

In contrast to the Gauss and vector constraints, the scalar
constraint does not have a simple geometrical meaning. This makes
its quantization more involved. Regularization choices have to be
made and the result is not unique. After Thiemann's first rigorous
quantization \cite{th2} other well defined possibilities have been
found \cite{c00,pul1,pul4}. This ambiguity affects dynamics
governed by
\begin{equation}\label{wd}
\hat {\cal S}\Psi=0.
\end{equation}

The difficulty in dealing with the scalar constraint is not
surprising. The vector constraint---generating space
diffeomorphisms---and the scalar constraint---generating time
reparameterizations---arise from the underlying $4$-diffeomorphism
invariance of gravity. In the canonical formulation the $3+1$
splitting breaks the manifest $4$-dimensional symmetry. The price
paid is the complexity of the time re-parameterization constraint
$\cal S$. The situation is somewhat reminiscent of that in
standard quantum field theory where manifest Lorentz invariance is
lost in the Hamiltonian formulation \footnote{There is however an
additional complication here: the canonical constraint algebra
does not reproduce the 4-diffeomorphism Lie algebra. This
complicates the geometrical meaning of $S$.}.

From this perspective, there has been growing interest in
approaching the problem of dynamics by defining a covariant
formulation of quantum gravity. The idea is that (as in the QFT
case) one can keep manifest $4$-dimensional covariance in the path
integral formulation. The spin foam approach is an attempt to define
the path integral quantization of gravity using what we have learn
from LQG.

In standard quantum mechanics path integrals provide the solution
of dynamics as a device to compute the time evolution operator.
Similarly, in the generally covariant context it provides a tool
to find solutions to the constraint equations (this has been
emphasized formally in various places: in the case of gravity see
for example \cite{hart1}, for a detailed discussion of this in the
context of quantum mechanics see \cite{c6}). We will come back to
this issue later.

\begin{figure}[h]
\centerline{\hspace{0.5cm} \(\begin{array}{c}
\includegraphics[height=3cm]{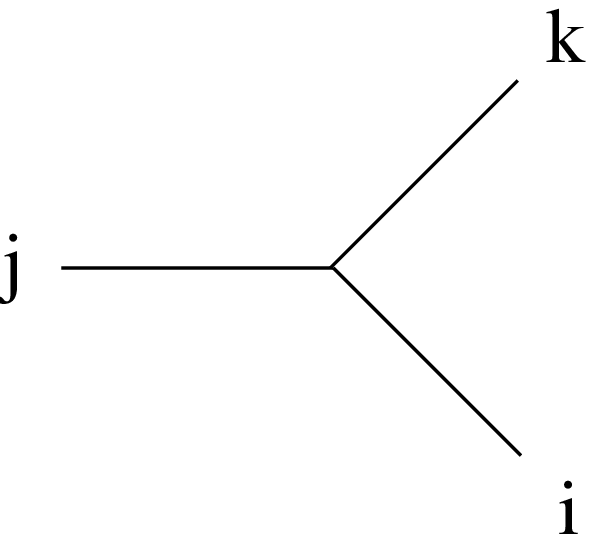}
\end{array}\ \   \rightarrow \ \
\begin{array}{c}
\includegraphics[height=3cm]{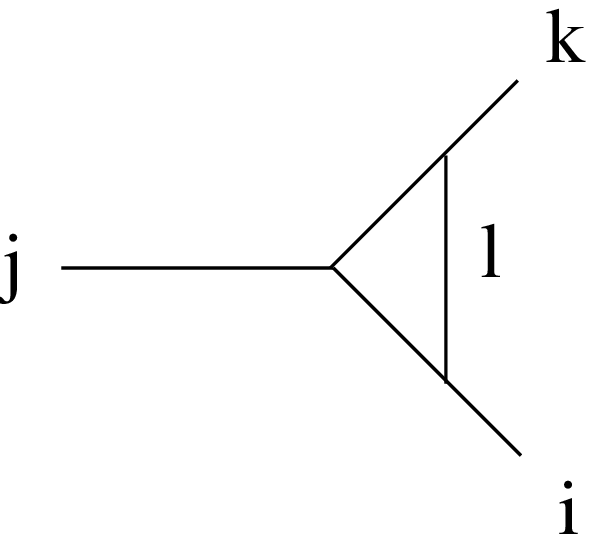}
\end{array} \) }
\caption{A typical transition generated by the action of the scalar constraint}
\label{branch}
\end{figure}

Let us finish by stating some properties of $\hat S$ that do not
depend on the ambiguities mentioned above. One is the discovery
that smooth loop states naturally solve the scalar constraint
operator \cite{jac,c8}. This set of states is clearly to small to
represent the physical Hilbert space (e.g., they span a zero
volume sector). However, this implies that $\hat S$ acts only on
spin network nodes. Its action modifies spin networks at nodes by
creating new links according to Figure \ref{branch} \footnote{This is
not the case in all the available definitions of the scalar
constraints as for example the one defined in \cite{pul1,pul4}.}.
This is crucial in the construction of the spin foam approach of
the next section.

\section{Spin Foams and the path integral for gravity}\label{sec:intsf}

The possibility of defining quantum gravity using Feynman's path
integral approach has been considered long ago by Misner and later
extensively studied by Hawking, Hartle and others \cite{hart2,haw1}. Given a
4-manifold $\cal M$ with boundaries $\Sigma_1$ and $\Sigma_2$, and
denoting by $G$ the space of metrics on ${\cal M}$, the transition
amplitude between $\left|\left[q_{ab}\right]\right>$ on $\Sigma_1$
and $\left|\left[q^{\prime}_{ab}\right]\right>$ on $\Sigma_2$ is
formally \begin{equation}\label{tarara} \left<
\left[q^{}_{ab}\right]\right.\left|\right[q^{\prime}_{ab}\left]\right>
=\int \limits_{\left[g\right]} {\cal D}[g]\ \ e^{i S([g])},
\end{equation}
where the integration on the right is performed over all space-time
metrics up to $4$-diffeomorphisms $\left[g \right]\in G/{\vani
Diff({\cal M})}$ with fixed boundary values up to
$3$-diffeomorphisms $\left[q_{ab}\right]$,
$\left[q^{\prime}_{ab}\right]$, respectively.

There are various difficulties associated with (\ref{tarara}).
Technically there is the problem of defining the functional
integration over $\left[g\right]$ on the RHS. This is partially
because of the difficulties in defining infinite dimensional
functional integration beyond the perturbative framework. In
addition, there is the issue of having to deal with the space
$G/{\vani Diff({\cal M})}$, i.e., how to characterize the
diffeomorphism invariant information in the metric. This gauge
problem ($3$-diffeomorphisms) is also present in the definition of
the boundary data. There is no well defined notion of kinematical
state $\left|\left[q_{ab}\right]\right>$ as the notion of kinematical Hilbert
space in standard metric variables has never been defined.

\begin{figure}[h]\!\!\!\!\!\!
\centerline{\hspace{0.5cm} \(\begin{array}{c}
\includegraphics[height=4.5cm]{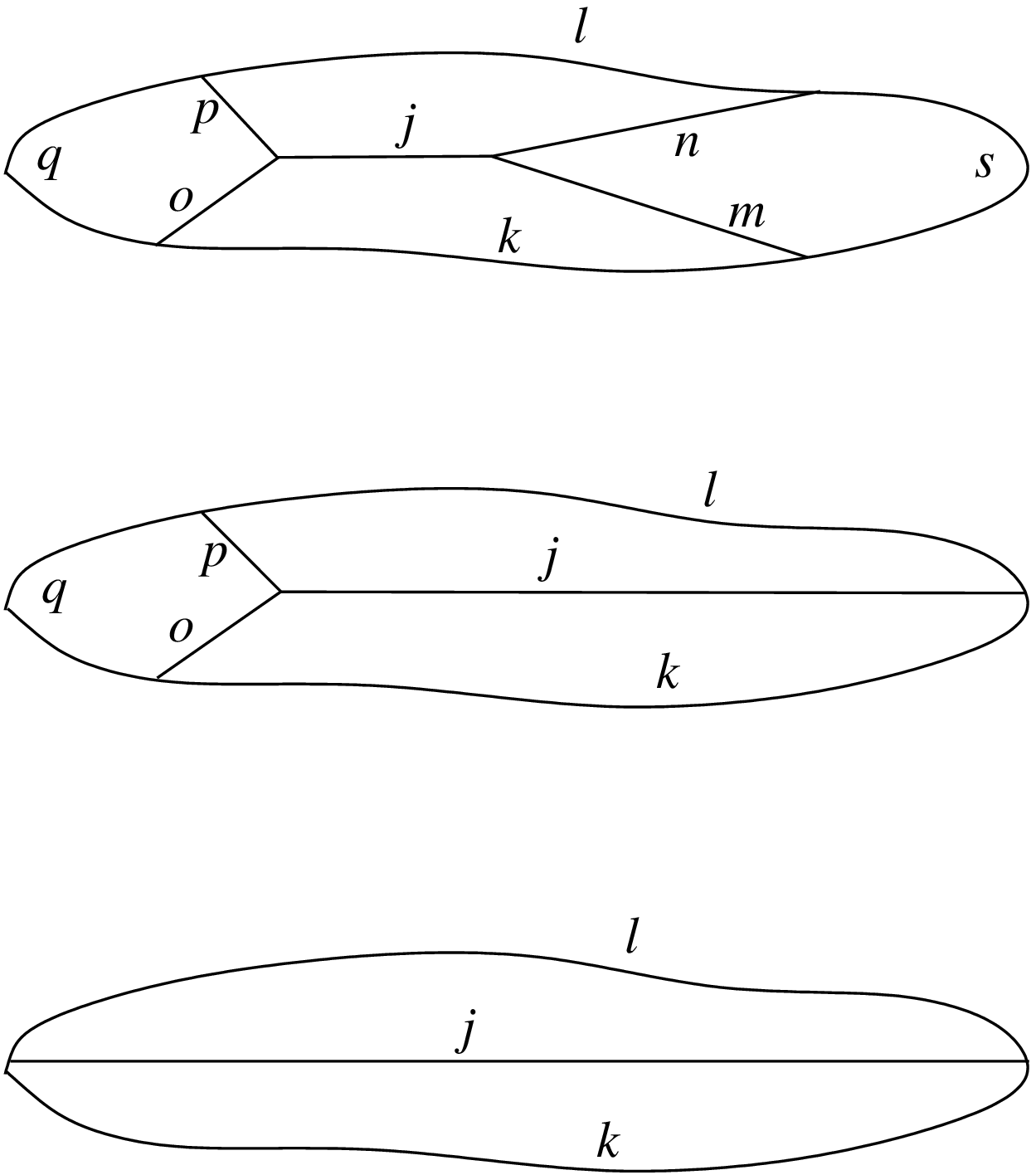}
\end{array}\ \   \rightarrow \ \
\begin{array}{c}
\includegraphics[height=4.5cm]{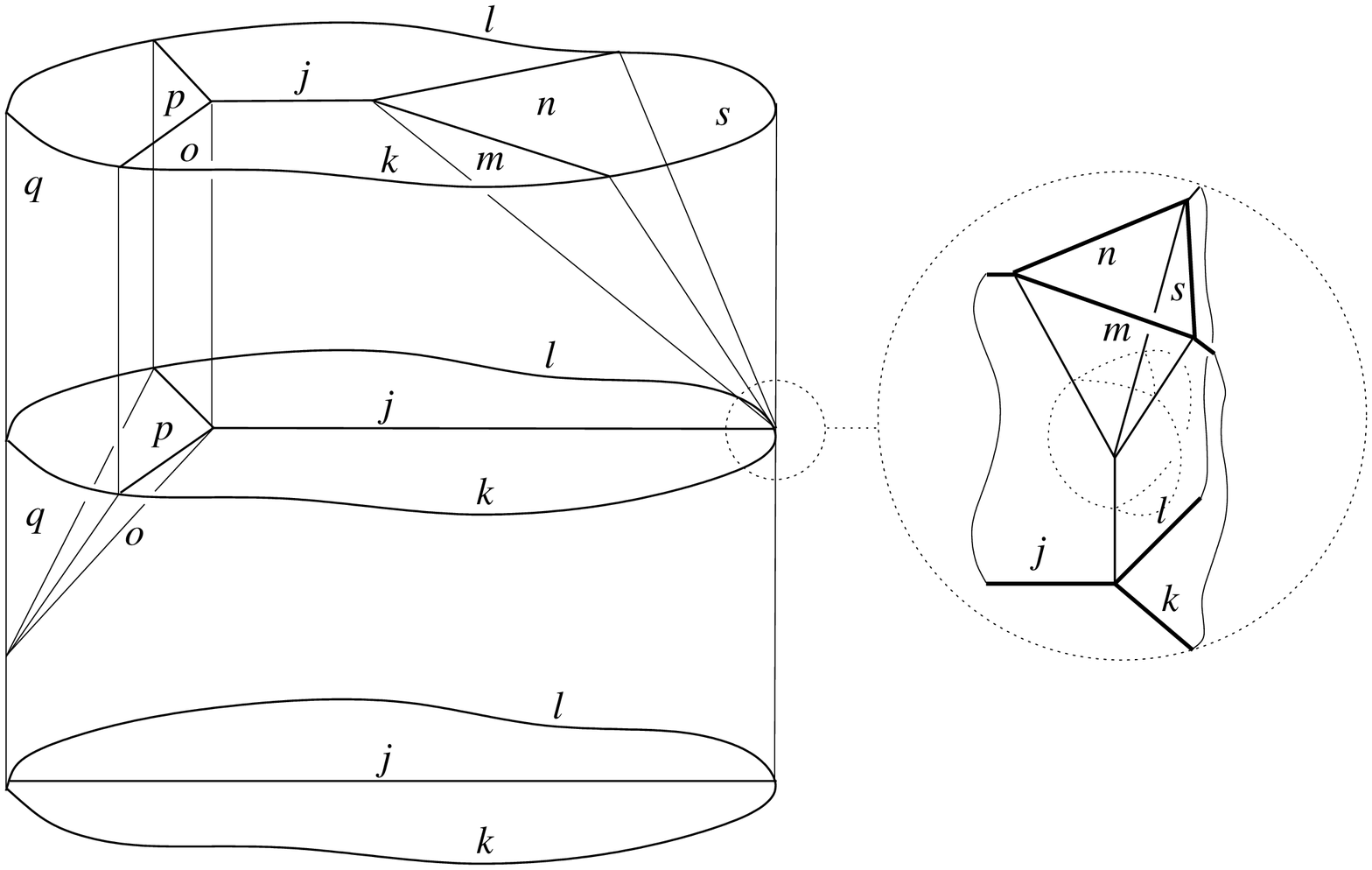}
\end{array} \) }
\caption{A typical path in a path integral version of loop quantum
gravity is given by a series of transitions through different
spin-network states representing a state of $3$-geometries.
Nodes and links in the spin network evolve
into 1-dimensional edges and faces. New links are created and
spins are reassigned at vertexes (emphasized on the right). The
`topological' structure is provided  by the underlying $2$-complex
while the geometric degrees of freedom are encoded in the labeling of its elements with
irreducible representations and intertwiners.} \label{3g}
\end{figure}

We can be more optimistic in the framework of loop quantum
gravity. The notion of quantum state of $3$-geometry is rigorously
defined in terms of spin-network states. They carry the
diff-invariant information of the Riemannian structure of
$\Sigma$. In addition, and very importantly, these states are
intrinsically discrete (colored graphs on $\Sigma$) suggesting a
possible solution to the functional measure problem, i.e., the
possibility of constructing a notion of Feynman `path integral' in
a combinatorial manner involving sums over spin network
world sheets amplitudes. Heuristically, `$4$-geometries' are to be
represented by `histories' of quantum states of $3$-geometries or
spin network states. These `histories' involve a series of
transitions between spin network states (Figure \ \ref{3g}), and
define a foam-like structure (a `$2$-graph' or $2$-complex) whose 
components inherit the spin representations from the
underlying spin networks. These spin network world sheets are the
so-called {\em spin foams}.

The precise definition of spin foams was introduced by Baez in
\cite{baez7} emphasizing their role as morphisms in the category
defined by spin networks\footnote{The role of category theory for
quantum gravity had been emphasized by Crane in 
\cite{crane3,crane4,crane0}.}. 
A spin foam ${\cal F}: s\rightarrow
s^{\prime}$, representing a transition from the spin-network $s=(
\gamma, \{ j_{\ell}\}, \{ \iota_n \})$ into $s^{\prime}=(
\gamma^{\prime}, \{ j_{\ell^{\prime}}\}, \{ \iota_{n^{\prime}}\}
)$, is defined by a $2$-complex \footnote{In most of the paper we
use the concept of piecewise linear $2$-complexes as in
\cite{baez7}; in Section \ref{sec:gft-sf} we shall study a
formulation of spin foam in terms of certain combinatorial
$2$-complexes.} ${\cal J}$ bordered by the graphs of $\gamma$ and
$\gamma^{\prime}$ respectively, a collection of spins $\{ j_f\}$
associated with faces $f \in {\cal J}$ and a collection of
intertwiners ${ \{ \iota_e}\}$ associated to edges $e\in {\cal
J}$. Both spins and intertwiners of exterior faces and edges match
the boundary values defined by the spin networks $s$ and
$s^{\prime}$ respectively. Spin foams ${\cal F}: s\rightarrow
s^{\prime}$ and ${\cal F}^{\prime}: s^{\prime}\rightarrow
s^{\prime\prime}$ can be composed into ${\cal FF}^{\prime}: s
\rightarrow s^{\prime\prime}$ by gluing together the two
corresponding 2-complexes at $s^{\prime}$. A spin foam model is an
assignment of amplitudes $A[{\cal F}]$ which is consistent with
this composition rule in the sense that
\begin{equation}\label{cobordism}
A[{\cal F F}^{\prime}]=A[{\cal F}]A[{\cal F}^{\prime}].
\end{equation}
Transition amplitudes between spin network states are defined by
\begin{equation}
\left<s,s^{\prime}\right>_{phys}=\sum_{{\cal F}: s\rightarrow
s^{\prime}} A[{\cal F}],
\end{equation}
where the notation anticipates the interpretation of such
amplitudes as defining the physical scalar product. The domain of
the previous sum is left unspecified at this stage. We shall
discuss this question further in Section \ref{sci}. This last
equation is the spin foam counterpart of equation (\ref{tarara}).
This definition remains formal until we specify what the set of
allowed spin foams in the sum are and define the corresponding
amplitudes.
\vskip1cm

In standard quantum mechanics the path integral is used to compute
the matrix elements of the evolution operator $U(t)$. It provides
in this way the solution for dynamics since for any kinematical
state $\Psi$ the state $U(t)\Psi$ is a solution to Schr\"odinger's
equation. Analogously, in a generally covariant theory the path
integral provides a device for constructing solutions to the
quantum constraints. Transition amplitudes represent the matrix
elements of the so-called generalized `projection' operator $P$
(Sections \ref{sec:ext} and \ref{psp}) such that $P\Psi$ is a
physical state for any kinematical state $\Psi$. As in the case of
the vector constraint the solutions of the scalar constraint
correspond to distributional states (zero is in the continuum part
of its spectrum). Therefore, ${\cal H}_{phys}$ is not a proper
subspace of $\cal H$ and the operator $P$ is not a projector
($P^2$ is ill defined)\footnote{In the notation of the previous
section states in ${\cal H}_{phys}$ are elements of $Cyl^*$.}. In Section
\ref{sfm3d} we give an explicit example of this
construction.

The background-independent character of spin foams is manifest. The
$2$-complex can be thought of as representing `space-time' while
the boundary graphs as representing `space'. They do not carry
any geometrical information in contrast with the standard concept
of a lattice. Geometry is encoded in the spin labelings which
represent the degrees of freedom of the gravitational field.

\subsection{Spin foams and the projection operator into ${\cal H}_{\va phys}$}\label{sec:ext}

Spin foams naturally arise in the formal definition of the
exponentiation of the scalar constraint as studied by Reisenberger
and Rovelli in \cite{reis5} and Rovelli \cite{c2}. The basic idea
consists of constructing the `projection' operator $P$ providing a
definition of the formal expression
\begin{equation}\label{PE}
P=\prod_{x\in\Sigma} \delta(\hat {\cal S}(x))=\int {\cal D}[N]\
e^{i \hat {\cal S}[N]},
\end{equation}
where $\hat {\cal S}[N]=\int dx^3 N(x) \hat {\cal S}(x)$, with $N(x)$ 
the lapse function. $P$ defines the {\em physical} scalar product
according to
\begin{equation}\label{fisi}
\left<s,s^{\prime}\right>_{\va phys}=\left<sP,s^{\prime}\right>,
\end{equation}
where the RHS is defined using the kinematical scalar product.
Reisenberger and Rovelli make progress toward a definition of
(\ref{PE}) by constructing a truncated version $P_\Lambda$ (where
$\Lambda$ can be regarded as an infrared cutoff). One of the main
ingredients is Rovelli's definition of a diffeomorphism invariant
measure ${\cal D}[N]$ generalizing the techniques of \cite{th3}.

The starting point is the  expansion of  the exponential in
(\ref{PE}) in powers \begin{equation} \left<sP_{\Lambda},
s^{\prime}\right>=\int \limits_{|N(x)| \le \Lambda} {\cal D}[N]
\left<s\sum \limits^{\infty}_{n=0} \frac{i^{n}}{n!}({\cal
S}[N])^n, s^{\prime}\right>.
\end{equation} The construction works for a generic form of quantum scalar
constraint as long as it acts locally on spin network nodes both
creating and destroying links (this local action generates a
vertex of the type emphasized in Figure \ref{branch}).
The action of $\hat {\cal S}[N]$ depends on the value of the lapse
at nodes. Integration over the lapse can be performed and the
final result is given by a power series in the cutoff $\Lambda$, namely
\begin{eqnarray}
\nonumber \left<sP_{\Lambda}, s^{\prime}\right>&=&\sum
\limits^{\infty}_{n=0} \ \frac{i^n \Lambda^{n}}{n!} \sum
\limits_{{\cal F}_n:s \rightarrow s^{\prime}} A[{\cal
F}_n]\\&=&\sum \limits^{\infty}_{n=0} \ \frac{i^n \Lambda^{n}}{n!}
\sum \limits_{{\cal F}_n:s \rightarrow s^{\prime}} \prod_{v}
A_v(\rho_v,\iota_v),
\end{eqnarray}
where ${\cal F}_n:s \rightarrow s^{\prime}$ are spin foams
generated by $n$ actions of the scalar constraint, i.e., spin
foams with $n$ vertices. The spin foam amplitude $A[{\cal F}_n]$
factorizes in a product of vertex contributions
$A_v(\rho_v,\iota_v)$ depending of the spins $j_v$ and $\iota_v$
neighboring faces and edges. The spin foam shown in Figure \ref{3g}
corresponds in this context to two actions of $\cal S$ and would contribute
to the amplitude in the order $\Lambda^2$.

Physical observables can be constructed out of kinematical
operators using $P$. If $O_{\va kin}$ represents an operator
commuting with all but the scalar constraint then $O_{\va
phys}=PO_{\va kin}P$ defines a physical observable. Its
expectation value is
\begin{equation}\label{ela}
\left<O_{\va
phys} \right>=\frac{\left<sPO_{\va kin}P,s\right>}{\left<sP,s\right>}:=\lim_{\Lambda
\rightarrow
\infty}\frac{\left<sP_{\Lambda}O_{\va kin}P_{\Lambda},s\right>}{\left<sP_{\Lambda},s\right>};
\end{equation}
where the limit of the ratio of truncated quantities is expected
to converge for suitable operators $O_{\va kin}$. Issues of
convergence have not been studied and they would clearly be
regularization dependent.

\subsection{Spin foams from lattice gravity}\label{sflg}

Spin foam models naturally arise in lattice-discretizations of the
path integral of gravity and generally covariant gauge theories.
This was originally studied by Reisenberger \cite{reis8}. The
space-time manifold is replaced by a lattice given by a cellular
complex. The discretization allows for the definition of the
functional measure reducing the number of degrees of freedom to
finitely many. The formulation is similar to that of standard
lattice gauge theory. However, the nature of this truncation is
fundamentally different: background independence implies that it
cannot be simply interpreted as a UV regulator (we will be more
explicit in the sequel).

We present a brief outline of the formulation for details see
\cite{reis8,reis6,reis4}. Start from the action of gravity in some
first order formulation ($S(e,A)$). The formal path integral takes
the form
\begin{equation}\label{ludlud}
{Z}=\int {\cal D}[e]\ {\cal D}[A]\  e^{i S(e,A)}=\int {\cal D}[A]\
e^{i S_{\va \rm eff}(A)},
\end{equation}
where in the second line we have formally integrated over the
tetrad $e$ obtaining an effective action $S_{\va \rm
eff}$\footnote{This is a simplifying assumption in the derivation.
One could put the full action in the lattice \cite{reis6} and then
integrate over the discrete $e$ to obtain the discretized version
of the quantity on the right of (\ref{ludlud}). This is what we
will do in Section \ref{sfm3d}.}. From this point on, the
derivation is analogous to that of generally covariant gauge
theory. The next step is to define the previous equation on a
`lattice'.

As for Wilson's action for standard lattice gauge theory the
relevant structure for the discretization is a 2-complex ${\cal
J}$. We assume the $2$-complex to be defined in terms of the dual
$2$-skeleton ${\cal J}_{\Delta}$ of a simplicial complex $\Delta$.
Denoting the edges $e \in {\cal J}_{\Delta}$ and the plaquettes or
faces $f \in {\cal J}_{\Delta}$ one discretizes the connection by
assigning a group element $g_e$ to edges
\[A\rightarrow \{g_e\}.\]
The Haar measure on the group is used to represent the connection
integration:
\[{\cal D}[A]\rightarrow \prod_{e\in{\cal J}_{\Delta}} dg_e.\]
The action of gravity depends on the connection $A$ only through
the curvature $F(A)$ so that upon discretization the action is
expressed as a function of the holonomy around faces $g_f$
corresponding to the product of the $g_e$'s which we denote
$g_f=g^{\va (1) }_e\cdots g_e^{\va (n)}$:
\[F(A)\rightarrow \{g_f\}.\]
In this way, $S_{\va \rm eff}(A)\rightarrow S_{\va \rm
eff}(\{g_f\})$. Thus the lattice path integral becomes:
\begin{equation}\label{discri}
{Z}= \int \ \prod_{e\in{\cal J}_{\Delta}}\  dg_{e}\  {\rm
exp}\left[{i S_{\rm eff}(\{g_f\})}\right].
\end{equation}

\begin{figure}[h]
\centering{\includegraphics[width=5cm]{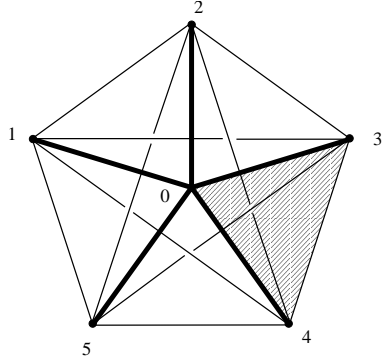}}
\caption{A fundamental {\em atom} is defined by the intersection
of a dual vertex in ${\cal J}_{\Delta}$ (corresponding to a
4-simplex in $\Delta$) with a 3-sphere. The thick lines represent
the internal edges while the thin lines the intersections of the
internal faces with the boundary. They define the boundary graph denoted
$\gamma_5$ below. One of the faces has been
emphasized.}\label{chunk}
\end{figure}

Reisenberger assumes that $S_{\va \rm eff}(\{g_f\})$ is local in
the following sense: the amplitude of any piece of the $2$-complex
obtained as its intersection with a ball depends only on the
value of the connection on the corresponding boundary. Degrees of freedom communicate
through the lattice connection on the boundary. One can compute
amplitudes of pieces of ${\cal J}_{\Delta}$ (at fixed boundary
data) and then obtain the full ${\cal J}_{\Delta}$ amplitude by
gluing the pieces together and integrating out the mutual boundary
connections along common boundaries. The boundary of a portion of
${\cal J}_{\Delta}$ is a graph. The boundary value is an
assignment of group elements to its links. The amplitude is a
function of the boundary connection, i.e., an element of $Cyl$. In
the case of a cellular 2-complex there is a maximal splitting
corresponding to cutting out a neighborhood around each vertex. If
the discretization is based on the dual of a triangulation these
elementary building blocks are all alike and denoted {\em atoms}.
Such an atom in four dimensions is represented in Figure
\ref{chunk}.

The atom amplitude depends on the boundary data given by the value
of the holonomies on the ten links of the pentagonal boundary
graph $\gamma_5$ shown in the figure. This amplitude can be
represented by a function
\begin{equation}\label{atomamp}
{\cal V}(\alpha_{ij})\ \ \  {\rm for} \ \ \ \alpha_{ij}\in G\ \ \ {\rm
and}\ \ \ i\not=j=1,\dots,5
\end{equation}
where $\alpha_{ij}$ represents the boundary lattice connection
along the link $ij$ in Figure \ref{chunk}. Gauge invariance
($V(\alpha_{ij})=V(g_i\alpha_{ij}g^{-1}_j)$) implies that the
function can be spanned in terms of spin networks functions
$\Psi_{\gamma_5,\{\rho_{ij}\}, \{\iota_{i}\}}(\alpha_{ij})$ based
on the pentagonal graph $\gamma_5$, namely
\begin{equation}
{\cal V}(\alpha_{ij})=\sum_{\rho_{ij}} \sum_{\iota_i} {\tilde {\cal V}
}(\{\rho_{ij}\}, \{\iota_{i}\})\ \Psi_{\gamma_5,\{\rho_{ij}\},
\{\iota_{i}\}}(\alpha_{ij})
\end{equation}
where $\tilde {\cal V}(\{\rho_{ij}\}, \{\iota_{i}\})$ is the atom
amplitude in `momentum' space depending on ten spins $\rho_{ij}$
labeling the faces and five intertwiners $\iota_{i}$ labeling
the edges. Gluing the atoms together the integral over common
boundaries is replaced by the sum over common values of spin
labels and intertwiners\footnote{\label{lejos} This is a
consequence of the basic orthogonality of unitary representations,
namely
\[\int {j[g]}_{\alpha \beta} k[g]_{\gamma \mu} dg =
\frac{\delta_{jk^*}}{2j+1}\delta_{\alpha \gamma} \delta_{\beta \mu}.\]}. The total amplitude becomes
\begin{equation}\label{sixteen}
Z[{\cal J}_{\Delta}]= \sum_{{\cal C}_f:\{f\} \rightarrow \{
\rho_f\}} \sum_{{\cal C}_e:\{e\} \rightarrow \{ \iota_e\}}
\prod_{f\in {{\cal J}_{\Delta}}} A_{f}(\rho_{f},\iota_f) \
\prod_{v\in {{\cal J}_{\Delta}}} \tilde {\cal V} (\{\rho_v\}, \{\iota_v\}),
\end{equation}
where ${\cal C}_e:\{e\} \rightarrow \{ \iota_e\}$  denotes the
assignment of intertwiners to edges, ${\cal C}_f:\{\rho_f\}
\rightarrow \{ f\}$ the assignment of spins $\rho_f$ to faces, and
$A_{f}(\rho_{f},\iota_f)$ is the face amplitude arising in the
integration over the lattice connection. The lattice definition of
the path integral for gravity and covariant gauge theories becomes
a discrete sum of spin foam amplitudes!

\subsection{Spin foams for gravity from BF theory}\label{sfbf}

The integration over the tetrad we formally performed in
(\ref{ludlud}) is not always possible in practice. There is
however a type of generally covariant theory for which the analog
integration is trivial. This is the case of a class of theories
called BF theory. General relativity in three dimensions is of
this type. Consequently, BF theory can be quantized along the
lines of the previous section. BF spin foam amplitudes are simply
given by certain invariants in the representation theory of the
gauge group. We shall study in some detail the case of
$3$-dimensional Riemannian quantum gravity in the next section.

The relevance of BF theory is its close relation to general
relativity in four dimensions. In fact, general relativity can be
described by certain BF theory action plus Lagrange multiplier
terms imposing certain algebraic constraints on the
fields \cite{pleb}. This is the starting point for the definition
of several of the models we will present in this article: a spin
foam model for gravity can be defined by imposing restrictions on
the spin foams that enter in the partition function of the BF
theory. These restrictions are essentially the translation into
representation theory of the constraints that reduce BF theory to
general relativity.

\subsection{Spin foams as Feynman diagrams}\label{sffd}

As already pointed out in \cite{baez7} spin foams can be
interpreted in close analogy to Feynman diagrams. Standard Feynman
graphs are generalized to $2$-complexes and the labeling of
propagators by momenta to the assignment of spins to faces.
Finally, momentum conservation at vertices in standard
Feynmanology is now represented by spin-conservation at edges,
ensured by the assignment of the corresponding intertwiners. In
spin foam models the non-trivial content of amplitudes is
contained in the vertex amplitude as pointed out in Sections
\ref{sec:ext} and \ref{sflg} which in the language of Feynman
diagrams can be interpreted as an interaction. We shall see that
this analogy is indeed realized in the formulation of spin foam
models in terms of a group field theory (GFT) \cite{reis1,reis2}.

\section{Spin foams for $3$-dimensional gravity}\label{sfm3d}

Three dimensional gravity is an example of BF theory for which the
spin foam approach can be implemented in a rather simple way.
Despite its simplicity the theory allows for the study of many of
the conceptual issues to be addressed in four dimensions. In
addition, as we mentioned in Section \ref{sfbf}, spin foams for BF
theory are the basic building block of $4$-dimensional gravity
models. For a beautiful presentation of BF theory and its relation
to spin foams see  \cite{baez5}. For simplicity we study the
Riemannian theory; the Lorentzian generalization of the
results of this section have been studied in \cite{fre1}.

\subsection{The classical theory}

Riemannian gravity in $3$ dimensions is a theory with no local
degrees of freedom, i.e., no gravitons. Its action is given by
\begin{equation}\label{bfaction} S_{}(e,A)=\int
\limits_{\cal M} Tr(e\wedge F(A)),
\end{equation}
where the field $A$ is an $SU(2)$-connection and the triad $e$ is
a Lie algebra valued $1$-form. The local symmetries of the action
are $SU(2)$ gauge transformations
\begin{equation}\label{gauge1}
\delta e = \left[e,\omega \right], \ \ \ \ \ \ \ \ \ \delta A =
d_{A} \omega,
\end{equation}
where $\omega$ is a ${{\tt su(2)}}$-valued $0$-form, and
`topological' gauge transformation
\begin{equation}\label{gauge2}
\delta e = d_{\va A} \eta, \ \ \ \ \ \ \ \ \ \delta A = 0,
\end{equation}
where $d_{\va A}$ denotes the covariant exterior derivative and
$\eta$ is a ${\tt su(2)}$-valued $0$-form. The first invariance is
manifest from the form of the action, while the second is a
consequence of the Bianchi identity, $d_{\va A}F=0$. The gauge
freedom is so big that the theory has only global degrees of
freedom. This can be checked directly by writing the equations of
motion
\begin{equation}\label{soluto}
F(A)=d_{\va A}A=0, \ \ \ \ \ \ \ \ \ d_{\va A}e=0.
\end{equation}
The first implies that the connection is flat which in turn means
that it is locally gauge ($A=d_{\va A}\omega$). The solutions of
the second equation are also locally gauge as any closed form is
locally exact ($e=d_{\va A}\eta$)
\footnote{\label{dibf} One can easily check that the infinitesimal
diffeomorphism gauge action $\delta e={\cal L}_v e$, and $\delta
A={\cal L}_v A$, where ${\cal L}_v$ is the Lie derivative in the
$v$ direction, is a combination of (\ref{gauge1}) and
(\ref{gauge2}) for $\omega=v^aA_a$ and $\eta=v^a e_{a}$,
respectively, acting on the space of solutions, i.e. when
(\ref{soluto}) holds.}
.  This very simple theory can be quantized in a direct manner.

\subsection{Canonical quantization}

Assuming ${\cal M}=\Sigma \times \R$ where $\Sigma$ is a Riemann
surface representing space, the phase space of the theory is
parameterized by the spatial connection $A$ (for simplicity we use the same notation
as for the space-time connection) and its conjugate
momentum $E$. The constraints that result from the gauge freedoms
(\ref{gauge1}) and (\ref{gauge2}) are
\begin{equation}
{\cal D}_a E^a_i=0, \ \ \ \ \ F(A)=0.
\end{equation}
The first is the familiar Gauss constraint (in the notation of equation (\ref{constro})), and 
we call the second the {\em curvature constraint}. There are $6$ independent constraints for
the $6$ components of the connection, i.e., no local degrees of
freedom. The kinematical setting is analogous to that of
$4$-dimensional gravity and the quantum theory is defined along
similar lines. The kinematical Hilbert space (quantum geometry) is
spanned by $SU(2)$ spin-network states on $\Sigma$ which
automatically solve the Gauss constraint. ``Dynamics'' is governed
by the curvature constraint. The physical Hilbert space is
obtained by restricting the connection to be flat and the physical
scalar product is defined by a natural measure in the space of
flat connections \cite{baez5}. The distributional character of the
solutions of the curvature constraint is manifest here. Different
spin network states are physically equivalent when they differ by
a null state (states with vanishing physical scalar product with
all spin network states). This happens when the spin networks are
related by certain skein relations. One can reconstruct ${\cal
H}_{phys}$ directly from the skein relations which in turn can be
found by studying the covariant spin foam formulation of the
theory.

\subsection{Spin foam quantization of 3d gravity}\label{sec:qbf}

Here we apply the general framework of Section \ref{sflg}. This
has been studied by Iwasaki in  \cite{iwa3,iwa4}. The partition
function, ${\cal Z}$, is formally given by\footnote{We are dealing with Riemannian
$3$-dimensional gravity. This should not be confused with the approach of
Euclidean quantum gravity formally obtained by a Wick rotation of 
Lorentzian gravity. Notice the imaginary unit
in front of the action. The theory of Riemannian quantum gravity should be regarded as a toy
model with no obvious connection to the Lorentzian sector.}
\begin{equation}\label{zbf}
{\cal Z}=\int  {\cal D}[e] {\cal D}[A]\ \ e^{i \int_{\va \cal M}
{\rm Tr}[e\wedge F(A)]},
\end{equation}
where for the moment we assume ${\cal M}$ to be a compact and
orientable. Integrating over the $e$ field in (\ref{zbf}) we
obtain
\begin{equation}\label{VA}
{\cal Z}=\int {\cal D}[A] \ \ \delta \left(F(A)\right).
\end{equation}
The partition function ${\cal Z}$ corresponds to the `volume' of
the space of flat connections on $\cal M$.

In order to give a meaning to the formal expressions above, we
replace the $3$-dimensional manifold ${\cal M}$ with an arbitrary
cellular decomposition $\Delta$. We also need the notion of the
associated dual 2-complex of $\Delta$ denoted by ${\cal
J}_{\Delta}$. The dual 2-complex ${\cal J}_{\va \Delta}$ is a
combinatorial object defined by a set of vertices $v\in {\cal
J}_{\va \Delta}$ (dual to 3-cells in $\Delta$) edges $e\in {\cal
J}_{\va \Delta}$ (dual to 2-cells in $\Delta$) and faces $f\in {\cal
J}_{\va \Delta}$ (dual to $1$-cells in $\Delta$).

The fields $e$ and $A$ have support on these discrete structures.
The $su(2)$-valued $1$-form field $e$ is represented by the
assignment of an $e \in {su(2)}$ to each $1$-cell in $\Delta$. The
connection field $A$ is represented by the assignment of group
elements $g_e \in SU(2)$ to each edge in ${\cal J}_{\va \Delta}$.


The partition function is defined by
\begin{equation}\label{Zdiscrete}
{\cal Z}(\Delta)=\int \prod_{f \in {\cal J}_{\va \Delta}} de_f \
\prod_{e \in {\cal J}_{\va \Delta}} dg_e  \ e^{i {\rm Tr}
\left[e_f U_f\right]},
\end{equation}
where $de_f$ is the regular Lebesgue measure on $\R^3$,
$dg_e$ is the Haar measure on $SU(2)$, and $U_f$
denotes the holonomy around faces, i.e., $U_f=g^1_e\dots g^{\va
N}_e$ for $N$ being the number of edges bounding the corresponding
face. Since $U_f \in SU(2)$ we can write it as $U_f=u^0_f\ {\mathbbm{1}} + F_f$ 
where $u^0_f\in \C$ and $F_f \in su(2)$. $F_f$ is interpreted as 
the discrete curvature around the face $f$. Clearly ${\rm Tr}[e_f U_f]={\rm Tr}[e_f F_f]$.
An arbitrary orientation is assigned to faces when computing
$U_f$. We use the fact that faces in ${\cal J}_{\va \Delta}$ are
in one-to-one correspondence with $1$-cells in $\Delta$ and label
$e_f$ with a face subindex.

Integrating over $e_f$, we obtain
\begin{equation}\label{Zdiscrete0}
{\cal Z}(\Delta)=\int \ \prod_{e \in {\cal J}_{\va \Delta}} dg_e \
\prod_{f \in {\cal J}_{\va \Delta}}{\huge \delta}(g^1_e\dots
g^{\va N}_e),
\end{equation}
where $\delta$ corresponds to the delta distribution defined on
${\cal L}^2(SU(2))$. Notice that the previous equation corresponds
to the discrete version of equation (\ref{VA}).

The integration over the discrete connection ($\prod_e dg_e$) can
be performed expanding first the delta function in the previous
equation using the Peter-Weyl decomposition \cite{vilenkin}
\begin{equation}\label{deltarep}
\delta(g)=\sum \limits_{j \in {\rm irrep}(SU(2))} \Delta_{j} \
{\rm Tr} \left[ j(g)\right],
\end{equation}
where $\Delta_{j}=2j+1$ denotes the dimension of the unitary
representation $j$, and $j(g)$ is the corresponding representation
matrix. Using equation (\ref{deltarep}), the partition function
(\ref{Zdiscrete0}) becomes
\begin{equation}\label{coloring}
{\cal Z}(\Delta)=\sum \limits_{{\cal C}:\{j\} \rightarrow \{ f\}}
\int \ \prod_{e \in {\cal J}_{\va \Delta}} dg_e \ \prod_{f \in
{\cal J}_{\va \Delta}} \Delta_{j_f} \ {\rm Tr}\left[j_f(g^1_e\dots
g^{\va N}_e)\right],
\end{equation}
where the sum is over coloring of faces in the notation of
(\ref{sixteen}).

Going from equation (\ref{Zdiscrete}) to (\ref{coloring}) we have
replaced the continuous integration over the $e$'s by the sum over
representations of $SU(2)$. Roughly speaking, the degrees of
freedom of $e$ are now encoded in the representation being summed
over in (\ref{coloring}).

Now it remains to integrate over the lattice connection $\{g_e\}$.
If an edge $e\in {\cal J}_{\va \Delta}$ bounds $n$ faces there are
$n$ traces of the form ${\rm Tr}[j_f(\cdots g_e\cdots)]$ in
(\ref{coloring}) containing $g_e$ in the argument. The relevant
formula is
\begin{equation}\label{3dp}
P^{n}_{inv}:= \int dg\ {j_1(g)}\otimes j_2(g) \otimes \cdots \otimes j_n(g)=
\sum_{\iota} {C^{\va \iota}_{\va j_1 j_2 \cdots j_n} \ C^{*{\va
\iota}}_{\va j_1 j_2 \cdots j_n}},
\end{equation}
where $P^{n}_{inv}$ is the projector onto ${\rm Inv}[j_1\otimes j_2 \otimes \cdots
\otimes j_n]$. On the RHS we have chosen an orthonormal basis of
invariant vectors (intertwiners) to express the projector. Notice
that the assignment of intertwiners to edges is a consequence of
the integration over the connection. This is not a particularity
of this example but rather a general property of local spin foams
as pointed out in Section \ref{sflg}. Finally (\ref{Zdiscrete0})
can be written as a sum over spin foam amplitudes
\begin{equation}\label{statesum}
{\cal Z}(\Delta)=\sum \limits_{ {\cal C}:\{j\} \rightarrow \{ f\}
} \ \sum \limits_{ {\cal C}:\{\iota\} \rightarrow \{ e\} }\
\prod_{f \in {\cal J}_{\va \Delta}} \Delta_{j_f} \prod_{v\in {\cal
J}_{\va \Delta}} A_v(\iota_v,j_v),
\end{equation}
where $A_v(\iota_v,j_v)$ is given by the appropriate trace of the
intertwiners $\iota_v$ corresponding to the edges bounded by the
vertex and $j_v$ are the corresponding representations. This
amplitude is given in general by an $SU(2)$ $3Nj$-symbol
corresponding to the flat evaluation of the spin network defined
by the intersection of the corresponding vertex with a $2$-sphere.
When $\Delta$ is a simplicial complex all the edges in ${\cal
J}_{\Delta}$ are $3$-valent and vertices are $4$-valent (one such
vertex is emphasized in Figure \ref{3g}, the intersection with the
surrounding $S^2$ is shown in dotted lines). Consequently, the
vertex amplitude is given by the contraction of the corresponding
four $3$-valent intertwiners, i.e., a $6j$-symbol. In that case
the partition function takes the familiar
Ponzano-Regge \cite{ponza} form
\begin{equation}\label{statesum}
{\cal Z}(\Delta)=\sum \limits_{ {\cal C}:\{j\} \rightarrow \{ f\}
} \ \prod_{f \in {\cal J}_{\va \Delta}} \Delta_{j_f} \prod_{v\in
{\cal J}_{\va \Delta}} \begin{array}{c}
\includegraphics[width=3cm]{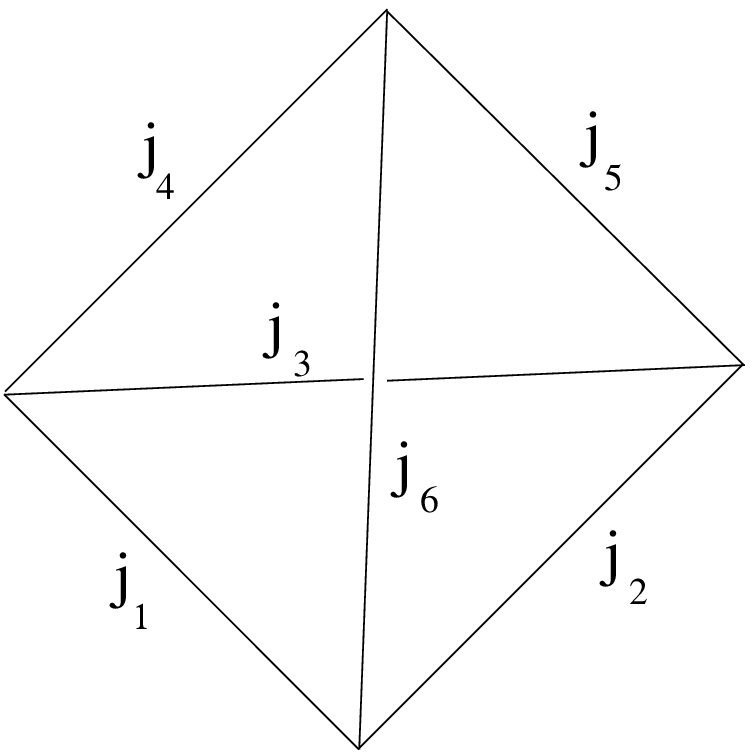}\end{array},
\end{equation}
were the sum over intertwiners disappears since ${\rm dim}({\rm
Inv}[j_1\otimes j_2 \otimes j_3])=1$ for $SU(2)$ and there is only
one term in (\ref{3dp}). Ponzano and Regge originally defined the
amplitude (\ref{statesum}) from the study of the asymptotic
properties of the $6j$-symbol.

\subsubsection{Discretization independence}\label{disind}

A crucial property of the partition function (and transition
amplitudes in general) is that it does not depend on the
discretization $\Delta$. Given two different cellular
decompositions $\Delta$ and $\Delta^{\prime}$ (not necessarily
simplicial) \begin{equation}\label{rucu} \tau^{-n_0}
Z(\Delta)=\tau^{-n^{\prime}_0} Z(\Delta^{\prime}),
\end{equation}
where $n_0$ is the number of 0-simplexes in $\Delta$ (hence the
number of bubbles in ${\cal J}_{\Delta}$), and $\tau=\sum_j (2j+1)$
is clearly divergent which makes discretization independence a
formal statement without a suitable regularization.

The sum over spins in (\ref{statesum}) is typically divergent, as
indicated by the previous equation. Divergences occur due to
infinite volume factors corresponding to the topological gauge
freedom (\ref{gauge2})(see \cite{fre7})\footnote{\label{u1} For simplicity we
concentrate on the Abelian case $G=U(1)$. The analysis can be
extended to the non-Abelian case.  Writing $g\in U(1)$ as
$g=e^{i\theta}$ the analog of the gravity simplicial action is
\begin{equation}
S(\Delta, \{e_f\},\{\theta_e\})=\sum_{f \in {\cal J}_{\va \Delta}}
e_f F_f(\{\theta_e\}),
\end{equation}
where $F_f(\{\theta_e\}) = \sum_{e\in f} \theta_{e}$. Gauge
transformations corresponding to (\ref{gauge1}) act at the end
points of edges $e\in {\cal J}_{\Delta}$ by the action of group
elements $\{\beta\}$ in the following way
\begin{eqnarray}
\nonumber && B_f \rightarrow B_f,\\
&& \theta_{e} \rightarrow \theta_{e}+\beta_{s}-\beta_{t},
\end{eqnarray}
where the sub-index $s$ (respectively $t$) labels the source
vertex (respectively target vertex) according to the orientation
of the edge. The gauge invariance of the simplicial action is
manifest. The gauge transformation corresponding to (\ref{gauge2})
acts on vertices of the triangulation $\Delta$ and is given by
\begin{eqnarray}
\nonumber && B_f \rightarrow B_f + \eta_s -\eta_t,\\
&& \theta_{e} \rightarrow \theta_{e}.
\end{eqnarray}
According to the discrete analog of Stokes theorem  \[\sum_{f \in
Bubble} F_f(\{\theta_e\})=0,\] which implies the invariance of the
action under the transformation above. The divergence of the
corresponding spin foam amplitudes is due to this last freedom.
Alternatively, one can understand it from the fact that Stokes
theorem implies a redundant delta function in (\ref{Zdiscrete0})
per bubble in ${\cal J}_{\Delta}$  \cite{fre7}. }. The factor
$\tau$ in (\ref{rucu}) represents such volume factor. It can also
be interpreted as a $\delta(0)$ coming from the existence of a
redundant delta function in (\ref{Zdiscrete0}). One can partially
gauge fix this freedom at the level of the discretization. This
has the effect of eliminating bubbles from the 2-complex.

In the case of simply connected $\Sigma$ the gauge fixing is
complete. One can eliminate bubbles and compute finite transition
amplitudes. The result is equivalent to the physical scalar
product defined in the canonical picture in terms of the delta
measure\footnote{If ${\cal M}=S^2\times [0,1]$ one can construct a
cellular decomposition interpolating any two graphs on the
boundaries without having internal bubbles and hence no
divergences.}.

In the case of gravity with cosmological constant the state-sum
generalizes to the Turaev-Viro model \cite{TV} defined in terms of $SU_q(2)$
with $q^n=1$ where the representations are finitely many.
Heuristically, the presence of the cosmological constant
introduces a physical infrared cutoff. Equation (\ref{rucu}) has been proved in this case
for the case of simplicial decompositions in  \cite{TV}, see also  \cite{tur,kau}.
The generalization for arbitrary cellular decomposition was obtained in  \cite{a1}.

\subsubsection{Transition amplitudes}

Transition amplitudes can be defined along similar lines using a
manifold with boundaries. Given $\Delta$, ${\cal J}_{\Delta}$ then
defines graphs on the boundaries. Consequently, spin foams induce
spin networks on the boundaries. The amplitudes have to be
modified concerning the boundaries to have the correct composition
property (\ref{cobordism}). This is achieved by changing the face
amplitude from $(\Delta_{j_{f}})$ to $(\Delta_{j_{\ell}})^{1/2}$
on external faces.

The crucial property of this spin foam model is that the
amplitudes are independent of the chosen cellular decomposition
 \cite{tur,a1}. This allows for computing transition amplitudes
between any spin network states $s=(\gamma, \{j\},\{\iota\})$ and
$s^{\prime}=(\gamma, \{j^{\prime}\},\{\iota^{\prime}\})$ according
to the following rules\footnote{Here we are ignoring various
technical issues in order to emphasize the relevant ideas. The
most delicate is that of the divergences due to gauge factors
mentioned above. For a more rigorous treatment see  \cite{za1}.}:

\begin{itemize}
\item Given ${\cal M}=\Sigma \times [0,1]$ (piecewise linear) and
spin network states $s=(\gamma, \{j\},\{\iota\})$ and
$s^{\prime}=(\gamma, \{j^{\prime}\},\{\iota^{\prime}\})$ on the
boundaries---for $\gamma$ and $\gamma^{\prime}$ piecewise linear
graphs in $\Sigma$---choose any cellular decomposition $\Delta$
such that the dual $2$-complex ${\cal J}_{\Delta}$ is bordered by
the corresponding graphs $\gamma$ and $\gamma^{\prime}$
respectively (existence can be shown easily).
%
%
\item Compute the transition amplitude between $s$ and $s^{\prime}$
by summing over all spin foam amplitudes (rescaled as in (\ref{rucu}))
for the spin foams $F:s\rightarrow s^{\prime}$ defined on the $2$-complex ${\cal
J}_{\Delta}$.
\end{itemize}

\subsubsection{The generalized projector}\label{gp}

We can compute the transition amplitudes between any element of
the kinematical Hilbert space ${\cal H}$\footnote{The sense in
which this is achieved should be apparent from our previous
definition of transition amplitudes. For a rigorous statement see
 \cite{za1}.}. Transition amplitudes define the physical scalar
product by reproducing the skein relations of the canonical
analysis. We can construct the physical Hilbert space by
considering equivalence classes under states with zero transition
amplitude with all the elements of ${\cal H}$, i.e., null states.
\begin{figure}[h]
\centering {\includegraphics[width=2cm]{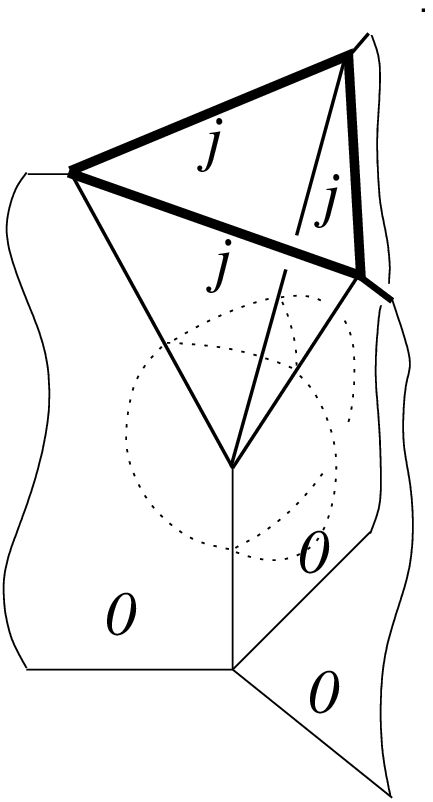}\ \ \ \
\ \ \includegraphics[width=5cm]{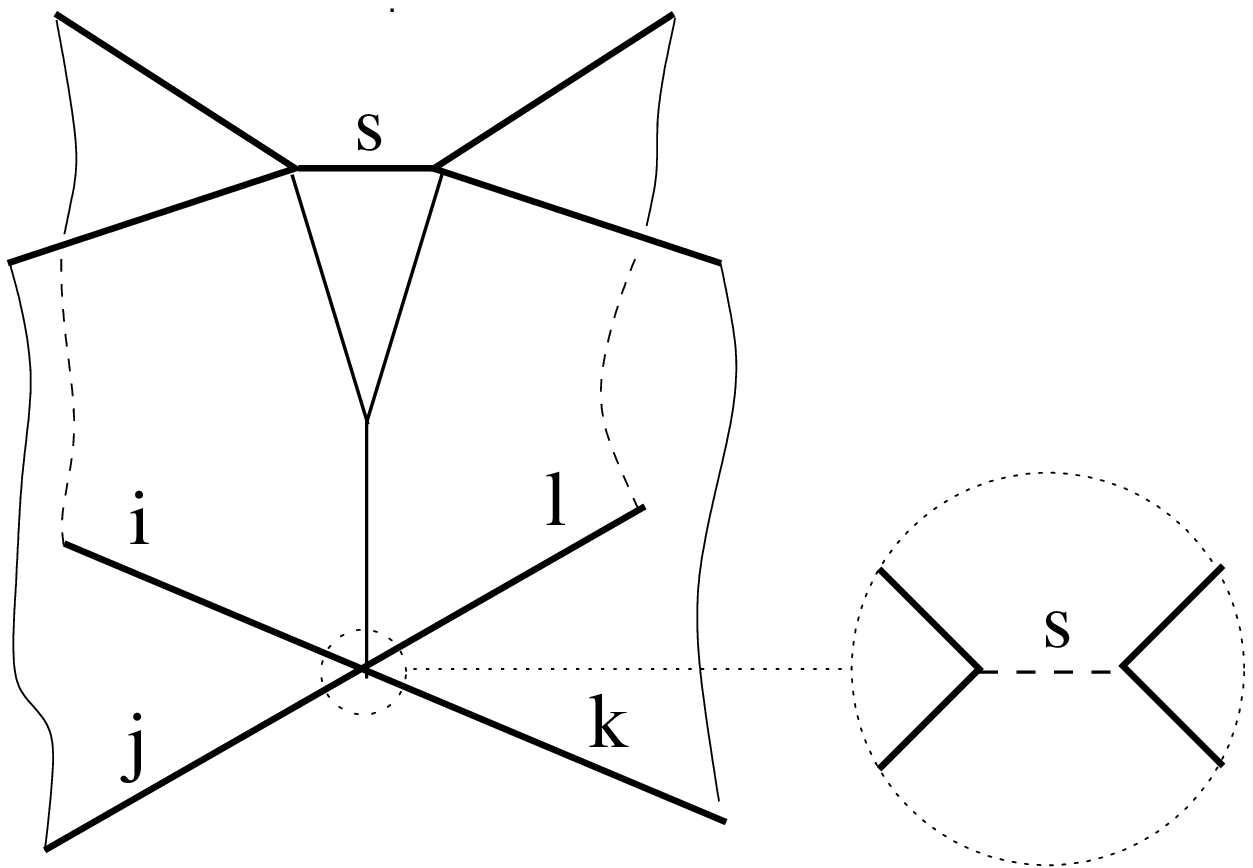}\ \ \
\includegraphics[width=3cm]{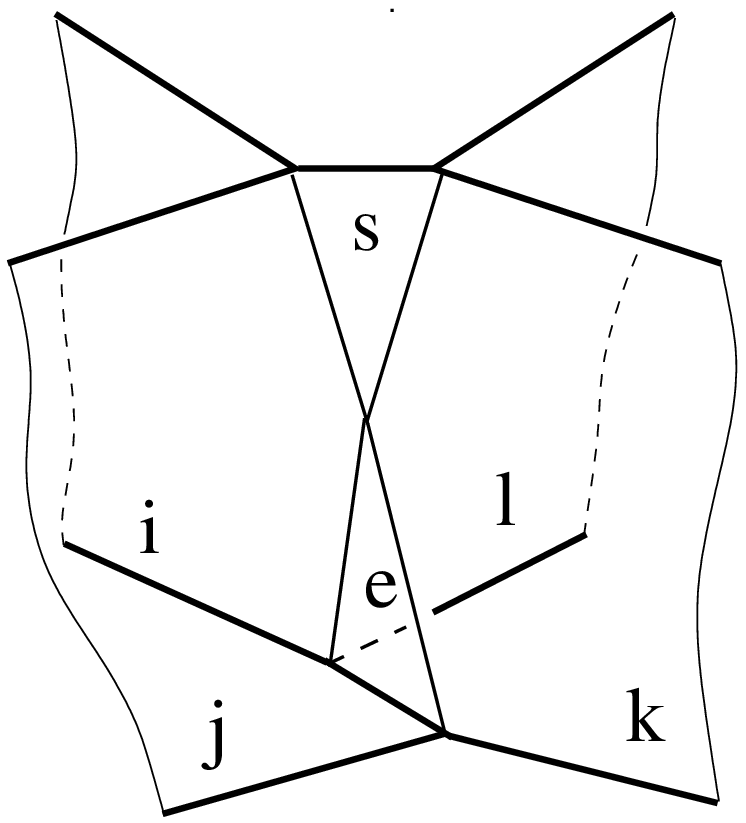}}
\caption{Elementary spin foams used to prove skein
relations.}\label{deltasf}
\end{figure}

Here we explicitly construct a few examples of null states. For
any contractible Wilson loop in the $j$ representation the state
\begin{equation} \psi=(2j+1)\ s - \begin{array}{c}
\includegraphics[width=1cm]{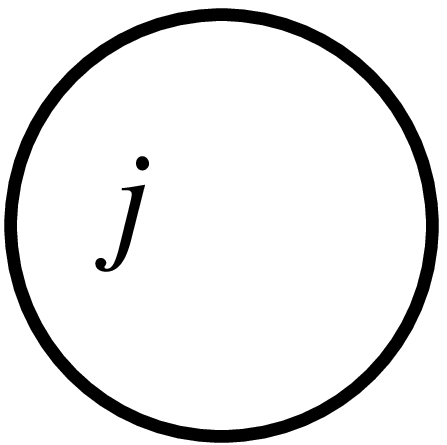}\end{array} \otimes s \phys 0,
\end{equation}
for any spin network state $s$, has vanishing transition amplitude
with any element of ${\cal H}$. This can be easily checked by
using the rules stated above and the portion of spin foam
illustrated in Figure \ref{deltasf} to show that the two terms in
the previous equation have the same transition amplitude (with
opposite sign) for any spin-network state in ${\cal H}$. Using the
second elementary spin foam in Figure \ref{deltasf} one can
similarly show that
\begin{eqnarray} \begin{array}{ccc}
\includegraphics[width=2cm]{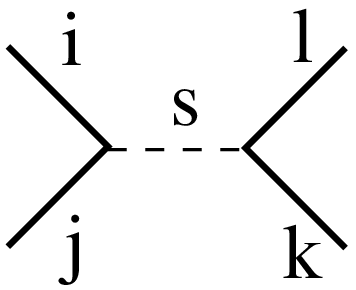}\end{array}
 - \begin{array}{c}
\includegraphics[width=2cm]{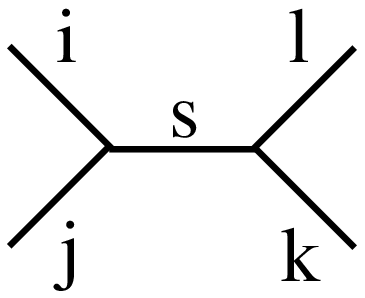}\end{array} \phys 0,
\end{eqnarray}
or the re-coupling identity using the elementary spin foam on the right of Figure \ref{deltasf}
\begin{eqnarray}\label{42} \begin{array}{ccc}
\includegraphics[width=2cm]{intu1.eps}\end{array}
 - \sum \limits_e \sqrt{2s+1}\sqrt{2e+1} \left\{\begin{array}{ccc}i\ \ j \ \ s\\ k\ \  l\ \  e  \end{array}\right\}
\begin{array}{c}
\includegraphics[height=2cm]{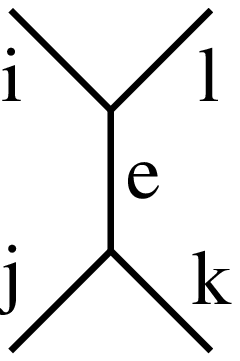}\end{array} \phys 0,
\end{eqnarray}
where the quantity in brackets represents an $SU(2)$ $6j$-symbol.
All skein relations can be found in this way. The transition
amplitudes imply the skein relations that define the physical
Hilbert space! The spin foam quantization is equivalent to the
canonical one.

\subsubsection{The continuum limit}

Recently Zapata  \cite{za1} formalized the idea of a continuum spin
foam description of $3$-dimensional gravity using projective
techniques inspired by those utilized in the canonical picture
 \cite{ash3}. The heuristic idea is that due to the discretization
invariance one can define the model in an `infinitely' refined
cellular decomposition that contains any possible spin network
state on the boundary (this intuition is implicit in our rules for
computing transition amplitudes above). Zapata concentrates on the
case with non-vanishing cosmological constant and constructs the
continuum extension of the Turaev-Viro model.

\subsection{Conclusion}

We have illustrated the general notion of the spin foam
quantization in the simple case of $3$ dimensional Riemannian
gravity (for the generalization to the Lorentzian case see
 \cite{fre1}). The main goal of the approach is to provide a
definition of the physical Hilbert space. The example of this
section sets the guiding principles of what one would like to
realize in four dimensions. However, as should be expected,
there are various new issues that make the task by far more
involved.

\section{Spin foams for $4$-dimensional quantum gravity}\label{sfm4d}

In this section we briefly describe the various spin foam models
for quantum gravity in the literature.

\subsection{The Reisenberger model}\label{Reise}

According to Plebanski  \cite{pleb} the action of self dual Riemannian
gravity can be written as a constrained $SU(2)$ BF theory
\begin{equation}\label{pleb00}
S(B,A)=\int {\rm Tr}\left[B\wedge F(A)\right]-\psi_{ij}\left[
B^i\wedge B^j-\frac{1}{3}\delta^{ij} B^{k}\wedge B_k\right],
\end{equation}
where variations with respect to the symmetric 
(Lagrange multiplier) tensor $\psi_{ij}$ imposes the
constraints
\begin{equation}\label{forty}
\Omega^{ij}=B^i\wedge B^j-\frac{1}{3}\delta^{ij} B^{k}\wedge
B_k=0.
\end{equation}
When $B$ is non degenerate the constraints are satisfied if and
only if $B^i=\pm(e^0\wedge e^i+{\vani \frac{1}{2}}\epsilon^i_{\
jk}e^{j}\wedge e^k)$ which reduces the previous action to that of
self-dual general relativity. Reisenberger studied the simplicial
discretization of this action in  \cite{reis6} as a preliminary
step toward the definition of the corresponding spin foam model.
The consistency of the simplicial action is argued by showing that
the simplicial theory converges to the continuum formulation when
the triangulation is refined: both the action and its variations
(equations of motion) converge to those of the continuum theory.

In reference  \cite{reis4} Reisenberger constructs a spin foam
model for this simplicial theory by imposing the constraints
$\Omega^{ij}$ directly on the $SU(2)$ BF amplitudes. The spin foam
path integral for BF theory is obtained as in Section \ref{sfm3d}.
The constraints are imposed by promoting the $B^i$ to operators
${\chi}^i$ (combinations of left/right invariant vector fields)
acting on the discrete connection \footnote{Notice that (for
example) the right invariant vector field ${\cal J}^{i}(U)=
{\sigma}^{i\ A }_{\ B}U^{B}_{\  C} {\partial}/{\partial U^{A}_{\
C}}$ has a well defined action at the level of equation
(\ref{coloring}) and acts as a B operator at the level of
(\ref{Zdiscrete}) since
\begin{eqnarray}\label{RIV}
-i{\chi}^{i}(U)\left[ e^{i{\rm Tr}[BU]}\right]|_{U\sim 1} = {\rm
Tr}[\sigma^{i}UB] e^{i{\rm Tr}[BU]}|_{U\sim 1}\sim B^{i}e^{i{\rm
Tr}[BU]},
\end{eqnarray}
where $\sigma^{i}$ are Pauli matrices.}. The model is defined as
\begin{equation}
Z_{GR}={\underbrace{\int \ \ \prod_{e \in {\cal J}_{\va \Delta}}
dg_e}_{\int {\cal D}[A]} } \ \overbrace{\delta ( \hat \Omega^{ij} )}^{\int {\cal D}[\psi] \  e^{i \psi_{ij}\Omega^{ij} }}\!\!
\underbrace{\sum \limits_{{\cal C}:\{j\} \rightarrow \{ f\}}
\prod_{f \in {\cal J}_{\va \Delta}} \Delta_{j_f} \ {\rm
Tr}\left[j_f(g^1_e\dots g^{\va N}_e)\right]}_{\int {\cal D}[B] \
e^{i\int {\rm Tr}[B\wedge F(A)]}},
\end{equation}
where $\hat \Omega={\cal J}^i\wedge {\cal
J}^j-\frac{1}{3}\delta^{ij} {\cal J}^{k}\wedge {\cal J}_k $ and we
have indicated the correspondence of the different terms with the
continuum formulation. The preceding equation is rather formal; for
the rigorous implementation see  \cite{reis4}. Reisenberger uses
locality so that constraints are implemented on a single
$4$-simplex amplitude. There is however a difficulty with the
this procedure: the algebra of operators $\hat \Omega^{ij}$ do
not close so that imposing the constraints sharply becomes a too
strong condition on the BF configurations\footnote{This difficulty also arises
in the Barrett-Crane model as we shall see in Section \ref{BCM}.}. 
In order to avoid this, Reisenberger defines a one-parameter 
family of models by inserting the operator
\begin{equation}\label{sw}
e^{-\frac{1}{2z^2}\hat \Omega^2}
\end{equation}
instead of the delta function above. In the limit $z\rightarrow
\infty$ the constraints are sharply imposed. This introduces an
extra parameter to the model. The properties of the kernel of
$\hat \Omega$ have not been studied in detail.

\subsection{The Freidel-Krasnov prescription}\label{Freide}

In reference  \cite{fre5} Freidel and Krasnov define a general framework
to construct spin foam models corresponding to theories whose action
has the general form
\begin{equation}
S(B,A)=\int {\rm Tr}\left[B\wedge F(A)\right] + \Phi(B),
\end{equation}
where the first term is the BF action while $\Phi(B)$ is a certain
polynomial function of the $B$ field. The formulation is
constructed for compact internal groups. The definition is based on
the formal equation
\begin{equation}
\int {\cal D}[{B}]{\cal D}[{ A}]\  e^{i\int {\rm Tr}\left[B\wedge F(A)\right] + \Phi(B)}:=
\left.e^{i\int \Phi(\frac {\delta}{\delta J})}   Z[J]\right|_{J=0},
\end{equation}
where the {\em generating functional} $Z[J]$ is defined as
\begin{equation}
Z[J]:=\int {\cal D}[{B}]{\cal D}[{A}] \ e^{i\int {\rm Tr}\left[B\wedge F(A)\right] + {\rm Tr}\left[B\wedge J\right]},
\end{equation}
where $J$ is an algebra valued $2$-form field. They provide a
rigorous definition of the generating functional by introducing a
discretization of $\cal M$ in the same spirit of the other spin
foam models discussed here. Their formulation can be used to
describe various theories of interest such as BF theories with
cosmological terms, Yang-Mills theories (in $2$ dimensions) and
Riemannian self-dual gravity. In the case of self dual gravity $B$
and $A$ are valued in $su(2)$, while
\begin{equation}
\Phi(B)= \int \psi_{ij}\left[ B^i \wedge B^j - \frac{1}{3} \delta^{ij} B^k \wedge B_k\right]
\end{equation}
according to equation (\ref{pleb00}). The model obtained in this way
is very similar to Reisenberger's one. There are however some
technical differences. One of the more obvious one is that the
non-commutative invariant vector fields ${\cal J}^i$ representing $B^i$ are replaced
here by the commutative functional derivatives $\delta/\delta
J^{i}$. The explicit properties of these models have not
been studied further.

\subsection{The Iwasaki model}

Iwasaki defines a spin foam model of self dual Riemannian gravity
\footnote{Iwasaki defines another model involving multiple
cellular complexes to provide a simpler representation of wedge
products in the continuum action. A more detail presentation of
this model would require the introduction of various
technicalities at this stage so we refer the reader to
 \cite{iwa2}.} by a direct lattice discretization of the continuous
Ashtekar formulation of general relativity. This model constitutes
an example of the general prescription of Section \ref{sflg}. The
action is
\begin{equation}
S(e,A)=\int dx^4\ \epsilon^{\mu \nu \lambda \sigma}\left[2\
e^{0}_{[\mu} e_{\nu ] i} + \epsilon^{0}_{\ ijk}
e^{j}_{\mu}e^{k}_{\nu} \right]\left[ 2\ \partial_{[\lambda}
A_{\sigma ]}^i + \epsilon^{0i}_{\ \ lm}
A^{l}_{\lambda}A^{m}_{\sigma}\right],
\end{equation}
where $A^i_a$ is an $SU(2)$ connection. The fundamental
observation of  \cite{iwa1} is that one can write the discrete
action in a very compact form if we encode part of the degrees of
freedom of the tetrad in an $SU(2)$ group element. More precisely,
if we take $g_{\mu\mu}=e^{i}_{\mu}e^{j}_{\mu}\delta_{ij}=1$ we can
define ${\mathbf e}_{\mu}:=e_{\mu}^0+i\sigma_i e^i_{\mu}\in SU(2)$
where $\sigma_i$ are the Pauli matrices. In this parameterization
of the `angular' components of the tetrad and using a hypercubic
lattice the discrete action becomes
\begin{equation}
S_{\Delta}=-\beta \sum_{v\in \Delta}
\epsilon^{\mu\nu\lambda\sigma} {r_{\mu} r_{\nu}} \ {\rm
Tr}\left[{\mathbf e}^{\dagger}_{\mu}{\mathbf e}_{\nu} U_{\lambda
\sigma} \right],
\end{equation}
where $r_{\mu}:=(\beta^{1/2} \ell_p)^{-1}\epsilon
\sqrt{g_{\mu\mu}}$, $U_{\mu \nu}$ is the holonomy around the
$\mu\nu$-plaquette, $\epsilon$ the lattice constant and $\beta$ is
a cutoff for $r_{\mu}$ used as a regulator ($r_{\mu}\le
\beta^{1/2}\ell_p \epsilon^{-1}$). The lattice path integral is
defined by using the Haar measure both for the connection and the
`spherical' part of the tetrad $\mathbf e$'s and the radial part
$dr_{\mu}:= dr_{\mu} r^{3}_{\mu}$. The key formula to obtain an
expression involving spin foams is
\begin{equation}
e^{i{x}{\rm Tr}[U]}=\sum_j (2j+1) \frac{J_{2j+1}(2x)}{x} \chi_j(U).
\end{equation}
Iwasaki writes down an expression for the spin foam amplitudes in
which the integration over the connection and the $\mathbf e$'s
can be computed explicitly. Unfortunately, the integration over
the radial variables $r$ involves products of Bessel functions and
its behavior is not analyzed in detail. In $3$ dimensions the
radial integration can be done and the corresponding amplitudes
coincide with the results of Section (\ref{sec:qbf}).

\subsection{The Barrett-Crane model}

The appealing feature of the previous models is the clear
connection to loop quantum gravity, since they are defined directly using
the self dual formulation of gravity (boundary states are
$SU(2)$-spin networks). The drawback is the lack of closed simple
expressions for the amplitudes which complicates their analysis.
There is however a simple model that can be obtained as a systematic
quantization of simplicial $SO(4)$ Plebanski's
action. This model was introduced by Barrett
and Crane in  \cite{BC2} and further motivated by Baez in
 \cite{baez7}. The basic idea behind the definition was that of the
{\em quantum tetrahedron} introduced by Barbieri in  \cite{barb2}
and generalized to 4d in  \cite{baez6}. The beauty of the model
resides in its remarkable simplicity. This has stimulated a great
deal of explorations and produced many interesting results. We
will review most of these in Section {\ref{BCM}}.

\subsection{Markopoulou-Smolin causal spin networks}\label{fotin}

Using the kinematical setting of LQG with the assumption of the
existence of a micro-local (in the sense of Planck scale) causal
structure Markopoulou and Smolin define a general class of
(causal) spin foam models for gravity  \cite{fot1,fot2} (see also
 \cite{fot3}). The elementary transition amplitude
$A_{s_I\rightarrow s_{I+1}}$ from an initial spin network $s_{I}$
to another spin network $s_{I+1}$ is defined by a set of simple combinatorial rules
based on a definition of causal propagation of the information at
nodes. The rules and amplitudes have to satisfy certain causal
restrictions (motivated by the standard concepts in classical
Lorentzian physics). These rules generate surface-like excitations
of the same kind we encounter in the more standard spin foam model
but endow the foam with a notion of causality. Spin foams ${\cal
F}^{N}_{s_i\rightarrow s_{f}}$ are labeled by the number of times
these elementary transitions take place. Transition amplitudes are
defined as
\begin{equation}
\left<s_i,s_f\right>=\sum_{N} \prod \limits^{N-1}_{I=0} A_{s_I\rightarrow s_{I+1}}.
\end{equation}
The models are not related to any continuum action. The only
guiding principles are the restrictions imposed by causality,
simplicity and the requirement of the existence of a non-trivial
critical behavior that would reproduce general relativity at large
scales. Some indirect evidence of a possible non trivial continuum
limit has been obtained in some versions of the model in $1+1$
dimensions.

\subsection{Gambini-Pullin model}

In reference  \cite{pul2} Gambini and Pullin introduced a very
simple model obtained by modification of the $BF$ theory skein
relations. As we argued in Section \ref{sfm3d} skein relations
defining the physical Hilbert space of BF theory are implied by
the spin foam transition amplitudes. These relations reduce the
large kinematical Hilbert space of $BF$ theory (analogous to that of
quantum gravity) to a physical Hilbert space corresponding to the
quantization of a finite number of degrees of freedom. Gambini and
Pullin define a model by modifying these amplitudes so that some
of the skein relations are now forbidden. This simple modification
frees local excitations of a field theory. A remarkable feature
is that the corresponding physical states are (in a certain sense)
solutions to various regularizations of the scalar constraint for
(Riemannian) LQG. The fact that physical states of BF theory solve
the scalar constraint is well known  \cite{th6}, since roughly
$F(A)=0$ implies $EEF(A)=0$. The situation here is of a similar
nature, and---as the authors argue---one should interpret this
result as an indication that some `degenerate' sector of quantum
gravity might be represented by this model. The definition of this
spin foam model is not explicit since the theory is directly
defined by the physical skein relations.

\subsection{Capovilla-Dell-Jacobson theory on the lattice}

The main technical difficulty that we gain in going from
$3$-dimensional general relativity to the $4$-dimensional one is
that the integration over the $e$'s becomes intricate. In the
Capovilla-Dell-Jacobson \cite{cap1,cap2} formulation of general
relativity this `integration' is partially performed at the
continuum level. The action is
\begin{equation}
S(\eta,A)=\int \eta {\rm Tr}\left[\epsilon \cdot F(A) \wedge F(A) \epsilon \cdot F(A)\wedge F(A) \right],
\end{equation}
where $\epsilon \cdot F \wedge F:=\epsilon^{abcd}F_{ab} F_{cd}$.
Integration over $\eta$ can be formally performed in the path integral and we obtain
\begin{equation}
Z=\int \prod_x \delta\left({\rm Tr}\left[\epsilon \cdot F(A) \wedge F(A) \epsilon \cdot F(A)\wedge F(A) \right]\right),
\end{equation}
This last expression is of the form (\ref{ludlud}) and can be
easily discretized along the lines of Section \ref{sflg}. The
final expression (after integrating over the lattice connection)
involves a sum over spin configurations with no implicit
integrations. One serious problem of this formulation is that it 
corresponds to a sector of gravity where the Weyl tensor satisfy 
certain algebraic requirements. In particular flat geometries are 
not contained in this sector. 

\section{Some conceptual issues}\label{sci}

\subsection{Anomalies and gauge fixing}\label{anom}

As we mentioned before and illustrated with the
example of Section \ref{sfm3d}, the spin foam path integral is meant
to provide a definition of the physical Hilbert space.
Spin foam transition amplitudes are not interpreted as defining
propagation in time but rather as defining the physical scalar
product. This interpretation of spin foam models is the only one
consistent with general covariance. However, in the path integral
formulation, this property relies on the gauge invariance 
of the path integral measure. If the measure meets this property we say it is 
{\em anomaly free}. It is well known that in addition to the invariance of the measure,
one must provide appropriate gauge fixing conditions for 
the amplitudes to be well defined. In this section we analyze 
these issues in the context of the spin foam approach.

Since we are interested in gravity in the first order formalism,
in addition to diffeomorphism invariance one has to deal with the
gauge transformations in the internal space.
Let us first describe the situation for the latter. If this
gauge group is compact then anomaly free measures are defined using
appropriate variables and invariant measures.
In this case gauge fixing is not necessary for the amplitudes to 
be well defined. Examples where this happens are:
the models of Riemannian gravity considered in this paper (the
internal gauge group being $SO(4)$ or $SU(2)$), and standard
lattice gauge theory. In these cases, one represents the connection in terms
of group elements (holonomies) and uses the (normalized) Haar
measure in the integration. In the Lorentzian sector (internal
gauge group $SL(2,\C)$) the internal gauge orbits have infinite
volume and the lattice path integral would diverge without an
appropriate gauge fixing condition. These conditions
generally exist in spin foam models and we
will study an example in Section \ref{BCGFT} (for a general treatment
see  \cite{frei8}).

The remaining gauge freedom is diffeomorphism invariance. 
It is generally assumed that spin foam encode diff-invariant information
about $4$-geometries. Thus, no gauge fixing would be necessary as 
one would be already summing over physical configurations. 
To illustrate this perspective we concentrate 
on the case of a model defined on a fixed discretization $\Delta$ 
as described in Section \ref{sflg}.

Let us start by considering the spin network states, at $\partial \Delta$: boundary of $\Delta$,
for which we want to define the transition amplitudes. According to 
what we have learned from the canonical approach, $3$-diffeomorphism 
invariance is implemented by considering (diffeomorphism) equivalence classes of 
spin-network states. In the context of spin foams, the underlying discretization $\Delta$
restricts the graphs on the boundary to be contained on the dual $1$-skeleton of the boundary 
complex $\partial\Delta$. These states are regarded as representative elements of 
the corresponding $3$-diffeomorphism equivalence class. The discretization can be interpreted, 
in this way, as a gauge fixing of $3$-diffeomorphisms on the boundary.
This gauge fixing is partial in the sense that, generically, there will remain
a discrete symmetry remnant given by the discrete symmetries of
the spin network. This remaining symmetry has to be factored out when computing
transition amplitudes (in fact this also plays a role in the definition of the
kinematical Hilbert space of Section \ref{lqg}).

The standard view point (consistent with LQG and quantum geometry) 
is that this should naturally generalize to
$4$-diffeomorphisms for spin foams. The underlying $2$-complex ${\cal J}_{\Delta}$ on which
spin foams are defined represents a partial gauge fixing for the configurations 
(spin foams) entering in the path integral.
The remaining symmetry, to be factored out in the computation of transition amplitudes,
corresponds simply to the finite group of discrete symmetries of the
corresponding spin foams\footnote{Baez \cite{baez7} points out this
equivalence relation between spin foams as a necessary condition
for the definition of the {\em category of spin foams}.}. 
This factorization is well defined since the number of equivalent spin foams 
can be characterized in a fully combinatorial manner, and is finite for any spin foam
defined on a finite discretization. In addition, a spin foam model is 
anomaly free if the amplitudes are invariant under this discrete symmetry. 
This requirement is met by all the spin foam models
we considered in this article.

\begin{figure}[h]
\centerline{\hspace{0.5cm} \(\begin{array}{c}
\includegraphics[height=2.5cm]{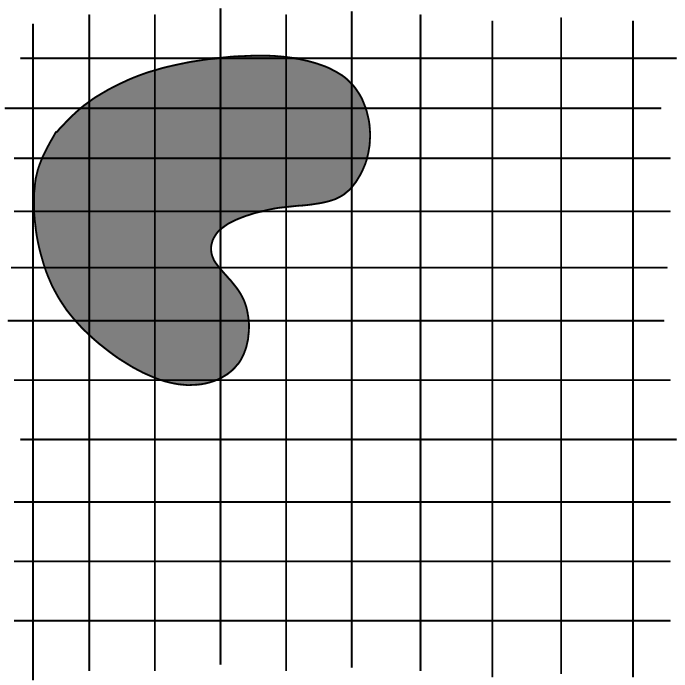}
\end{array}\ \   \begin{array}{ccc}\rightarrow \\ (a)\end{array} \ \
\begin{array}{c}
\includegraphics[height=2.5cm]{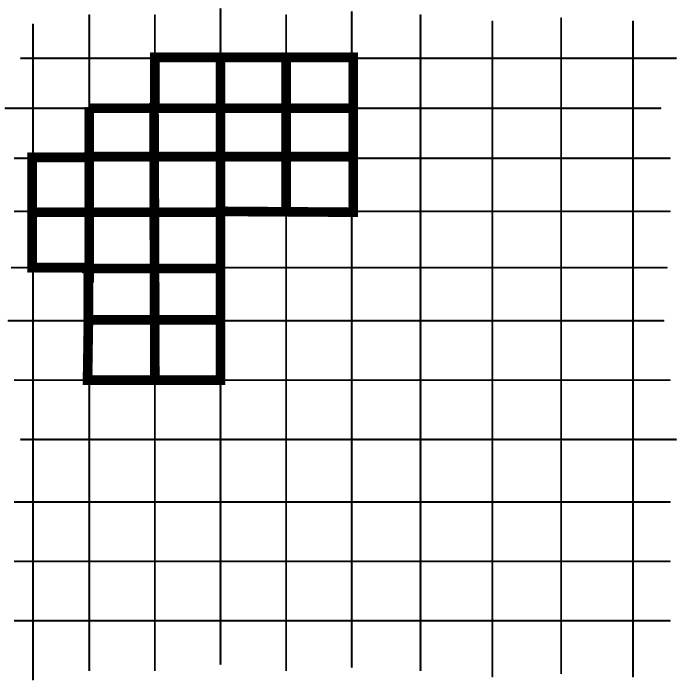}
\end{array}\ \ \ \ \ \ \
\begin{array}{c}
\includegraphics[height=2.5cm]{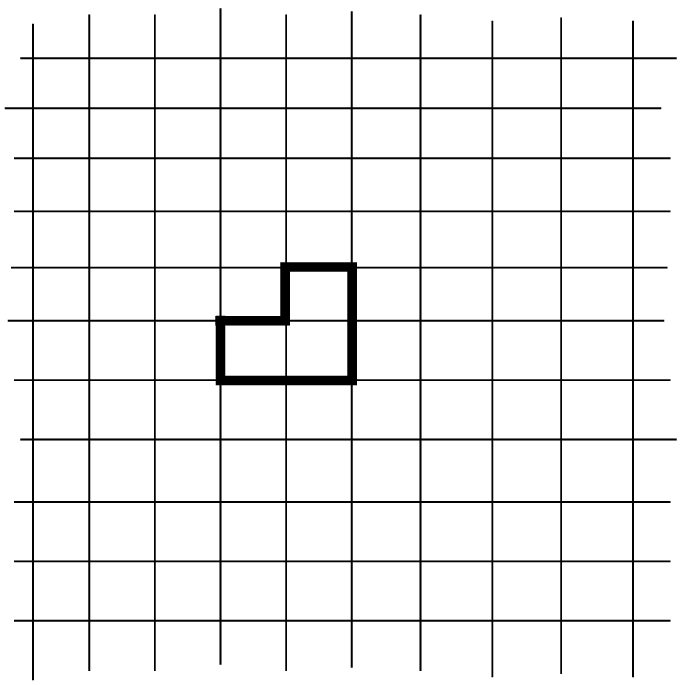}
\end{array}\ \   \begin{array}{ccc}\rightarrow \\ (b)\end{array} \ \
\begin{array}{c}
\includegraphics[height=2.5cm]{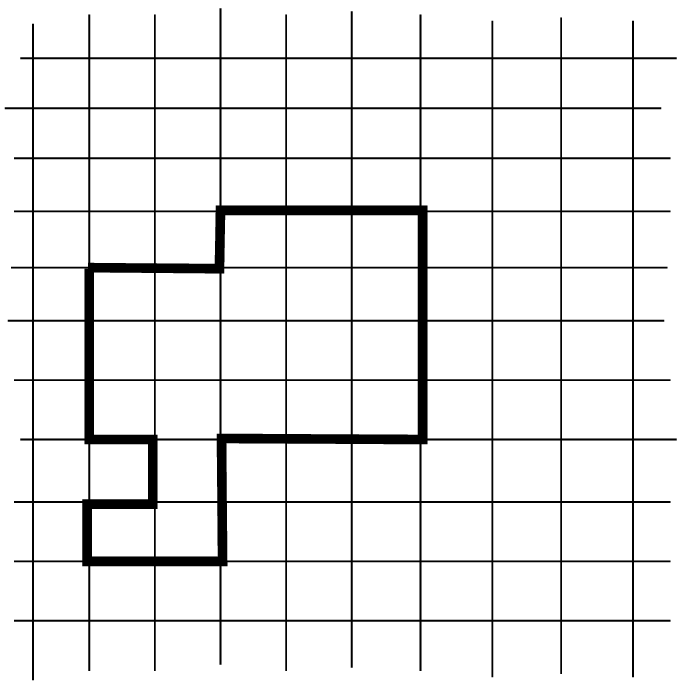}
\end{array}  \) }
\caption{Diffeomorphisms are replaced by discrete symmetries in spin foams.
In this simplified picture there are two possible colorings of $0$ or $1$, the latter
represented by the darkening of the corresponding lattice element. The two
configurations on $(b)$ are regarded as physically equivalent in the
background independent context.}
\label{grid}
\end{figure}

We illustrate the intuitive idea of the previous paragraph in Figure \ref{grid}.
On the diagram $(a)$ a continuum configuration is represented by a discrete (spin foam)
configuration on the lattice. On $(b)$ two configurations are shown.
In the background dependent context (e.g., lattice gauge theory)
these two configurations would be physically inequivalent as the lattice carries metric 
information (length of the edges).
In the context of spin foam models there is no geometric information encoded in the
discretization and in an anomaly free spin foam model the two configurations should 
be regarded as equivalent.

The discretization of the manifold $\Delta$ is seen as a regulator introduced
to define the spin foam model. Even when the regulator (or the discretization dependence) 
eventually has to be removed (see next subsection), the theory remains 
discrete at the fundamental level.  The smooth manifold diffeomorphism invariant 
description is regarded in this context as a derived concept in the (low energy) continuum limit.
Fundamental excitations are intrinsically discrete. From this viewpoint, the precise meaning 
of  the gauge symmetries of such a theory would have to be formulated directly 
at the discrete level. We have seen that this can be achieved in the case of $3$-dimensional
gravity (recall Section \ref{disind}). From this perspective, the discrete symmetries of
colored $2$-complexes (spin foams) represent the fundamental `gauge' freedom. This symmetry 
would manifest itself as diffeomorphisms only in the continuum limit. 

The viewpoint stated above is consistent with recent results obtained
by  Gambini and Pullin \cite{pul0,pul00}. They study the canonical
formulation of theories defined on a lattice from the onset.
This provides a way to analyze the meaning of gauge symmetries
directly \`a la Dirac. Their results indicate that diff-invariance
is indeed broken by the discretization in the sense that there is
no infinitesimal generator of diffeomorphism. This is consistent
with our covariant picture of discrete symmetries above. In their
formulation the canonical equations of motion fix the value of
what were Lagrange multipliers in the continuum (e.g. lapse and
shift). This is interpreted as a breaking of diffeomorphism
invariance; however, the solutions of the multiplier equations are
highly non unique. The ambiguity in selecting a particular solution
corresponds to the remnant diffeomorphism invariance of the
discrete theory.

However, the notion of anomaly freeness stated at the beginning of this section should 
be strengthened. In fact according to our tentative definition,
an anomaly free measure can be multiplied by any gauge invariant function and
yield a new anomaly free measure. This kind of ambiguity is not what 
we want; however, it is in fact present in most of the spin foam models defined so far.
In standard QFT theory, the formal (phase space) path integral measure in the 
continuum has a unique meaning (up to a constant normalization) emerging 
from the canonical formulation. Provided an appropriate gauge fixing, 
the corresponding Dirac bracket determines the formal measure 
on the gauge fixed constraint surface.
As we will see later, there is a certain degree of ambiguity in the definition
of various spin foam models in four dimensions. This ambiguity concerns the 
evaluation of lower dimensional simplexes and is directly
related to the definition of measure in the spin foam sum. 
One would expect that a strengthened definition of anomaly freeness
should resolve these ambiguities. This possibility
is studied in \cite{myo}.

Finally let us briefly recall the situation in $3$-dimensional
gravity. In three dimensions the discrete action is invariant
under transformations that are in correspondence with the
continuum gauge freedoms (\ref{gauge1}) and (\ref{gauge2}). As we
mentioned in Section \ref{disind} spin foam amplitudes diverge due
to (\ref{gauge2}). This means that spin foams do not fix (\ref{gauge2}) up to
a `finite volume' discrete symmetry, which seems to be in conflict 
with the argument stated above as diffeomorphism can be 
obtained combining the transformations (\ref{gauge1})
and (\ref{gauge2}). However, the topological gauge symmetry (\ref{gauge2}) involves 
more than just diffeomorphisms. Indeed, only on shell one can express diffeomorphisms as a combination
of (\ref{gauge1}) and (\ref{gauge2}). This representation of diffeomorphisms
turns out to be field dependent (recall Footnote \ref{dibf}).
At the canonical level we have seen that the topological symmetry
(\ref{gauge2}) acts in away that is totally different from diffeomorphisms.
In particular spin network states defined on graphs which
which differ in the number of edges and vertices can be 
physically equivalent (recall Section \ref{gp}).
Thus $3$-dimensional gravity is a degenerate example and 
extrapolation of its gauge properties to four 
dimensions seems misleading.

In Section \ref{BF} we will study the spin foam quantization of BF
theory---a topological theory which corresponds to a generalization of
$3$-dimensional gravity to four dimensions. The spin foam amplitudes
are also divergent in this case due to the analog of the gauge symmetry (\ref{gauge2}). 
Some of the spin foam models in Section \ref{sfm4d} are defined
from BF theory by implementing constraints that reduce the topological theory to
general relativity. The implementation of the 
constraints breaks the topological symmetry (\ref{gauge2}) and the
resulting model is no longer topological invariant. 
At the continuum level it is clear that the remnant gauge symmetry is 
diffeomorphism invariance. Whether in the the resulting   
spin foam model for gravity remnant is larger than
the discrete spin foam symmetries advocated above has
not been studied in detail.

The action of diffeomorphism is far from understood at a rigorous level in the context of
spin foam models. Here we have presented an account of some of the ideas under
consideration, and we tried to point out the relevance of an issue that certainly
deserves detailed investigation.

\subsection{Discretization dependence}\label{dd}

The spin foam models we have introduced so far are defined on a
fixed cellular decomposition of $\cal M$. This is to be
interpreted as an intermediate step toward the definition of the
theory. The discretization reduces the infinite dimensional
functional integral to a multiple integration over a finite number
of variables. This cutoff is reflected by the fact that only a
restrictive set of spin foams (spin network histories) is allowed
in the path integral: those that can be obtained by all possible
coloring of the underlying $2$-complex. In addition it restricts
the number of possible 3-geometry states (spin network states) on
the boundary by fixing a finite underlying boundary graph. This
represents a truncation in the allowed fluctuations and the set of
states of the theory that can be interpreted as a regulator.
However, the nature of this regulator is fundamentally different
from the standard concept in the background independent framework:
since geometry is encoded in the coloring (that can take any spin
values) the configurations involve fluctuations all the way to
Plank scale\footnote{Changing the label of a face from $j$ to
$j+1$ amounts to changing an area eigenvalue by an amount of the
order of Planck length squared according to (\ref{aarreeaa}).}.
This scenario is different in lattice gauge theories where the
lattice introduces an effective UV cutoff given by the lattice
spacing. Transition amplitudes are however discretization
dependent now. A consistent definition of the path integral using
spin foams should include a prescription to eliminate this
discretization dependence.

A special case is that of topological theories such as gravity in
3 dimensions. In this case, one can define the sum over spin foams
with the aid of a fixed cellular decomposition $\Delta$  of the
manifold. Since the theory has no local excitations (no
gravitons), the result is independent of the chosen cellular
decomposition. A single discretization suffices to capture the
degrees of freedom of the topological theory.

In lattice gauge theory the solution to the problem is implemented
through the so-called continuum limit. In this case the existence
of a background geometry is crucial, since it allows one to define the
limit when the lattice constant (length of links) goes to zero. In
addition the possibility of working in the Euclidean regime allows
the implementation of statistical mechanical methods.

None of these structures are available in the background
independent context. The lattice (triangulation) contains only
topological information and there is no geometrical meaning
associated to its components. As we mentioned above this has the
novel consequence that the truncation can not be regarded as an UV
cutoff as in the background dependent context. This in turn
represents a conceptual obstacle to the implementation of standard
techniques. Moreover, no Euclidean formulation seems meaningful in
a background independent scenario. New means to eliminate the
truncation introduced by the lattice have to be developed.

This is a major issue where concrete results have not been
obtained so far beyond the topological case. Here we explain the
two main approaches to recover general covariance corresponding to
the realization of the notion of `summing over discretizations' of
 \cite{c2}.

\begin{itemize}
\item {\em Refinement of the discretization:}

According to this idea topology is fixed by the simplicial
decomposition. The truncation in the number of degrees of freedom
should be removed by considering triangulations of increasing
number of simplexes for that fixed topology. The flow in the space
of possible triangulations is controlled by the Pachner moves. The
formal idea is to take a limit in which the number of four
simplexes goes to infinity together with the number of tetrahedra
on the boundary. Given a $2$-complex ${\cal J}_2$ which is a
refinement of a  $2$-complex ${\cal J}_1$ then the set of all
possible spin foams defined on ${\cal J}_1$ is naturally contained
in those defined on ${\cal J}_2$ (taking into account the
equivalence relations for spin foams mentioned in the previous
section). The refinement process should also enlarge the space of
possible 3-geometry states (spin networks) on the boundary
recovering the full kinematical sector in the limit of infinite
refinements. An example where this procedure is well defined is 
Zapata's treatment of the Turaev-Viro model
 \cite{za1}. The key point in this case is that amplitudes are
independent of the discretization (due to the topological
character of the theory) so that the refinement limit is trivial.
In the general case there is a great deal of ambiguity involved in
the definition of refinement\footnote{It is not difficult to
define the refinement in the case of a hypercubic lattice. In the
case of a simplicial complex a tentative definition can be
attempted using Pachner moves. To illustrate we can concentrate on the 
simple $2$-dimensional case. Given an initial
triangulation of a surface with boundary $\Delta_1$ define
$\Delta^{\prime}_1$ by implementing a $1-3$ Pachner move to each
triangle in $\Delta_1$. $\Delta^{\prime}_1$ is a homogeneous
refinement of $\Delta_1$; however, the boundary triangulation
remain unchanged so that they will support the same space of
boundary data or spin networks. As mentioned above, in the
refinement process one also wants to refine the boundary
triangulation so that the corresponding dual graph will get
refined and the space of possible boundary data will become bigger
(it should involve all of $Cyl$ in the limit). In order to achieve
this we define $\Delta_2$ by erasing from $\Delta^{\prime}_{1}$
triangles sharing a $1$-simplex with the boundary. This
amounts to carrying out a $1-2$ Pachner move on the boundary. This
completes the refinement $\Delta_1\rightarrow\Delta_2$. This
refinement procedure seems however not fully satisfactory as
the refinement keeps the memory of the initial triangulation. 
This can be easily visualized in the dual $2$-complex.
An improvement of this prescription could involve
the random implementation of $2-2$ Pachner moves between 
refinements.
}. The hope is that the nature of the transition amplitudes 
would be such that these ambiguities will not affect the final result. 
The Turaev-Viro model is an example where this prescription works.

If the refinement limit is well defined one would expect that
working with a `sufficiently refined' but fixed discretization
would serve as an approximation that could be used to extract
physical information with some quantifiable precision.

\item{\em Spin foams as Feynman diagrams:}

This idea has been motivated by the generalized matrix models of
Boulatov and Ooguri  \cite{bu,oo}. The fundamental observation is
that spin foams admit a dual formulation in terms of a field
theory over a group manifold  \cite{fre2,reis1,reis2}. The duality
holds in the sense that spin foam amplitudes correspond to Feynman
diagram amplitudes of the GFT. The perturbative Feynman expansion
of the GFT (expansion in a fiducial coupling constant $\lambda$)
provides a definition of {\em sum over} discretizations which is
fully combinatorial and hence independent of any manifold
structure\footnote{This is more than a `sum over topologies' as
many of the 2-complex appearing in the perturbative expansion
cannot be associated to any manifold  \cite{pietri1}.}. The latter
is most appealing feature of this approach. 

However, the convergence issues clearly become  more involved. The perturbative series
are generically divergent. This is not necessarily a definite obstruction as
divergent series can often be given an asymptotic meaning and provide physical 
information. Moreover, there are standard techniques that can allow to `re-sum' a
divergent series in order to obtain non perturbative information.
Recently, Freidel and Louapre \cite{fre10} have shown that this is indeed
possible for certain GFT's in three dimensions.
Other possibilities have been proposed in \cite{reis1}. 

It is not clear how the notion of diffeomorphism would be addressed in
this framework. Diffeomorphism equivalent configurations (in the
discrete sense described above) appear at all orders in the
perturbation series\footnote{The GFT formulation is clearly non
trivial already in the case of topological theories. There has
been attempts to make sense of the GFT formulation dual to BF
theories in lower dimensions \cite{a5}.}. From this perspective
(and leaving aside the issue of convergence) the sum of different
order amplitudes corresponding to equivalent spin foams should be interpreted 
as the definition of the physical amplitude. The discussion of the
previous section does not apply in the GFT formulation, i.e.,
there is no need for gauge fixing.

The GFT formulation resolves by definition the two fundamental conceptual
problems of the spin foam approach: diffeomorphism gauge symmetry and
discretization dependence. The difficulties are shifted to the question of the
physical role of $\lambda$ and the convergence of the corresponding perturbative series.
The GFT formulation has also been very useful in
the definition of a simple normalization of the Barret-Crane model
and has simplified its generalizations to the Lorentzian sector.
We will study this formulation in detail in Section
\ref{sec:gft-sf}.

\end{itemize}

\subsection{Physical scalar product revisited}\label{psp}

In this subsection we study the properties of the generalized
projection operator $P$ introduced before. The generalized
projection operator is a linear map from the kinematical Hilbert
space into the physical Hilbert space (states annihilated by the
scalar constraint). These states are not contained in the
kinematical Hilbert space but are rather elements of the dual
space $Cyl^*$. For this reason the operator $P^2$ is not well
defined. We have briefly mentioned the construction of the
generalized projection operator in the particular context of
Rovelli's model of Section (\ref{sec:ext}). After the above
discussion of gauge and discretization dependence we want to
revisit the construction of the $P$ operator. We consider the case
in which discretization independence is obtained by a refinement
procedure.

The number of 4-simplexes $N$ in the triangulation plays the role
of cutoff. The matrix elements of the $P$ operator is expected to
be defined by the refinement limit of fixed discretization
transition amplitudes ($\left<sP_{N},s^{\prime}\right>$), namely
\begin{equation}\label{milo}
\left<s,s^{\prime}\right>_{phys}=\left<sP,s^{\prime}\right>=\lim \limits_{N\rightarrow \infty} \left<sP_{N},s^{\prime}\right>.
\end{equation}

If we have a well defined operator $P$ (formally corresponding to
$\delta[\hat{\cal S}]$) the reconstruction of the physical Hilbert
space goes along the lines of the GNS construction \cite{haag}
where the algebra is given by $Cyl$ and the state $\omega(\Psi)$
for $\Psi\in Cyl$ is defined by $P$ as
\[\omega(\Psi)=\frac{\left<0 P,\Psi\right>}{\left<0 P,0\right>},\]
In the spin foam language this corresponds to the transition
amplitude from the state $\Psi$ to the `vacuum'. The
representation of this algebra implied by the GNS construction has
been used in
 \cite{a3} to define observables in the theory.

In Section \ref{sec:ext} we mentioned another way to obtain observables
according to the definition
\begin{equation}\label{pilo}
\left<s O_{\va
phys},s^{\prime}\right>={\left<sPO_{\va kin}P,s^{\prime}\right>},
\end{equation}
where $O_{\va kin}$ is a kinematical observable. This construction
is expected to be well defined only for some suitable kinematical
operators $O_{\va kin}$ (e.g., the previous equation is clearly
divergent for the kinematical identity). There are kinematical
operators which can be used as regulators
\footnote{ \label{quelo} Consider the following simple example.
Take the torus $S^1\times S^1$ represented  by
$[0,2\pi]\times[0,2\pi]\in \R^2$ with periodic boundary
conditions. An orthonormal basis of states is given by the (`spin
network') wave functions $\left<xy,nm\right>=\frac{1}{2 \pi}
e^{inx}e^{imy}$ for $n,m\in\Z$. Take the scalar constraint to be
given by $\hat X$ so that the generalized projection $P$ simply
becomes
\[\left<n m,\hat P\right|=\sum \limits_{\alpha}\left<\alpha  m \right|. \]
The RHS is no longer in ${\cal H}_{kin}$ since is not
normalizable: it is meaningful as a distributional state. The
physical scalar product is defined as in (\ref{milo}), namely
\[\left< n m,n^{\prime}m^{\prime}\right>_{phys} =\left< n m P,n^{\prime}m^{\prime}\right> =\sum \limits_{\alpha}
\left< \alpha m,n^{\prime} m^{\prime}\right>=\delta_{m
m^{\prime}}.\] Therefore we have that $\left|nm\right> \phys
\left|km\right> \forall k$. We can define physical observables by
means of the formula
\[\left<m, \hat O_{ph}\right|\left.m^{\prime}\right>=\left<n m ,\hat P\hat O\hat P\right|\left.n^{\prime} m^{\prime}\right>,\]
for suitable kinematic observables $O$. If for example we take
$O=1$ the previous equation would involve $\hat P^2$ which is not
defined. However, take the following family of kinematical
observables
\[\left|O_k, n m\right>=\delta_{n,k} \left|k m\right>, \]
then $\left<m, \hat
O_{ph}\right|\left.m^{\prime}\right>=\delta_{m,m^{\prime}}$, i.e.,
$\hat O_{ph}$ is just the physical identity operator for any value
of $k$. The whole family of kinematic operators define a single
physical operator. The key of the convergence of the previous
definition is in the fact that $O_k$ is a projection operator into
an eigenstate of the operator $\hat P_x$ canonically conjugate to
the constraint $\hat X$. In particular any operator that decays
sufficiently fast along going to $\pm \infty$ in the spectrum of
$\hat p_x$ will do. This illustrates what kind of operators will
be suitable for the above definition to work: they have to be
sufficiently localized along the gauge orbit, hence `sufficiently
localized in time' in the case of the scalar constraint. For
example, one may also use the Gaussian $\hat O_k\left|nm\right>=
e^{\frac{-(k-n)^2}{\epsilon}}\left|nm\right>$ which also projects
down to a multiple of the physical identity. Moreover,
availability of such a kinematical operator can serve as a
regulator of non suitable operators, for instance
\[\left<m, \hat O_{ph}\right|\left.m^{\prime}\right>=\left<n m ,\hat P\hat O_k \hat O\hat O_k \hat P\right|\left.n^{\prime} m^{\prime}\right>,\]
will converge even for $\hat O=1$. The operator $\hat O_k$ can be
regarded as a quantum gauge fixing: it selects a point along the
gauge orbit generated by the constraint.}. They corresponds to
Rovelli's `sufficiently localized in time' operators  \cite{c2}.
In  \cite{baez1} it is argued that in posing a physical question one
is led to conditioning the path integral and in this way improving
the convergence properties. Conditioning is represented here by
the regulator operators (see previous footnote).

\subsection{Contact with the low energy world}\label{clew}

A basic test to any theory of quantum gravity is the existence of
a well defined classical limit corresponding to general
relativity. In the case of the spin foam approach this should be
accomplished simultaneously with a limiting procedure that bridges
the fundamental discrete theory with the smooth description of
classical physics. That operation is sometimes referred to as the
{\em continuum limit} but it should not be confused with the
issues analyzed in the previous subsection.

There is debate on how one would actually setup the definition of the
`low energy limit' in the background independent context.
The general strategy---in fact motivated by our experience
in background physics---is to setup a renormalization scheme (\`a la
Wilson) where microscopic degrees of freedom are summed out
to obtain a coarse grained effective description. The previous
strategy is to be regarded as heuristic until a clear-cut
definition is found which encompasses all the issues discussed
in this section. It should be emphasized that even when various spin foam
models for quantum gravity have been defined, none of them has been show
to reproduce gravity at `low energies'. This is a major open question that
deserves all our efforts.

Interesting ideas on how to define the renormalization program for
spin foams have been proposed by Markopoulou \cite{fot4,fot5}.
They make use of novel mathematical techniques shown to be useful
in dealing with Feynman diagrams in standard QFT \cite{kre,kre1}.
In this approach Pachner moves on the underlying $2$-complex are
regarded as radiative corrections in standard QFT\footnote{this
interpretation becomes completely precise in the GFT formulation
of spin foams presented in Section \ref{BCGFT}.}. In order to
understand the `scaling' properties of the theory one has to study
the behavior of amplitudes under the action of these elementary
moves. Amplitudes for this elementary moves in the Barrett-Crane
model have been computed in  \cite{a10}. A full implementation of
Markopoulou's proposal to a concrete model remains to be studied.

Recently Oeckl \cite{oeckl} has studied the issue of 
renormalization in the context of spin foam models containing 
a coupling parameter. These models include generalized covariant gauge theories \cite{oeckl2,oeckl3}, 
the Reisenberger model, and the so called {\em interpolating model} 
(defined by Oeckl). The latter is given by a one-parameter family of  models that interpolate 
between the trivial BF topological model and the Barrett-Crane model according to the 
value of a `coupling constant'. Qualitative aspects of the renormalization 
groupoid flow of the couplings are studied in the various models.

\part{The Barrett-Crane model}

\section{SO(4) Plebanski's action and the Barrett-Crane model}\label{BCM}

The Barrett-Crane model is one of the most extensively studied
spin foam models for quantum gravity. 
In this section we concentrate on the definition of
the model in the Riemannian sector and sketch its generalization to
the Lorentzian sector.

The Barrett-Crane model can be viewed as a spin foam
quantization of $SO(4)$ Plebanski's formulation of general
relativity. The general idea is analogous to the one
implemented in the model of  Section \ref{Reise}.
We introduce the model from this perspective in the Riemannian sector
so we need to start with the study
of $Spin(4)$ BF theory.

\subsection{Quantum $Spin(4)$ BF theory}\label{BF}

Classical ($Spin(4)$) BF theory  is defined by the action
\begin{equation}
S[B,A]=\int \limits_{\cal M}{\rm Tr}\left[B\wedge F(A) \right],
\end{equation}
where $B^{IJ}_{ab}$ is a $Spin(4)$ Lie-algebra valued 2-form,
$A^{IJ}_a$ is a connection on a $Spin(4)$ principal bundle over
$\cal M$. The theory has no local excitations. Its properties are
very much analogous to the case of $3$-dimensional gravity studied
in Section \ref{sfm3d}.

A discretization $\Delta$ of $\cal M$ can be introduced along the same
lines presented in Section \ref{sfm3d}. For simplicity we concentrate in the 
case when $\Delta$ is a triangulation. The field $B$ is associated
with Lie algebra elements $B_t$ assigned to triangles $t\in \Delta$. 
In four dimensions triangles $t\in \Delta$
are dual to faces $f\in {\cal J}_{\Delta}$. This one-to-one
correspondence allows us to denote the discrete $B$ by either a
face ($B_f$) or a triangle ($B_t$) subindex respectively.
$B_f$ can be interpreted as the `smearing' of the continuous 2-form $B$ on
triangles in $\Delta$.
The connection $A$ is discretized by the assignment of group elements $g_e \in Spin(4)$
to edges $e\in {\cal J}_{\Delta}$.
The path integral becomes
\begin{equation}\label{papart}Z(\Delta)= \int \prod_{e \in {\cal J}_{\Delta}}
dg_e \prod_{f \in {\cal J}_{\Delta}} dB_f \ e^{iB_f U_f} = \int \prod_{e \in {\cal J}_{\Delta}}
dg_e \prod_{f \in {\cal J}_{\Delta}} \delta(g_{e_1} \cdots g_{e_n}),
\end{equation}
where the first equality is the analog of (\ref{Zdiscrete}) while
the second is the result of the $B$ integration  \cite{thesis}. Using
Peter-Weyl's theorem as in (\ref{coloring}) one obtains
\begin{equation}\label{coloring4}
{\cal Z}(\Delta)=\sum \limits_{{\cal C}:\{\rho\} \rightarrow \{
f\}} \int \ \prod_{e \in {\cal J}_{\va \Delta}} dg_e \ \prod_{f
\in {\cal J}_{\va \Delta}} \Delta_{\rho_f} \ {\rm
Tr}\left[\rho_f(g^1_e\dots g^{\va N}_e)\right],
\end{equation}
where $\rho$ are $Spin(4)$ irreducible representations. Integration over the
connection can be performed as in the case of $3$-dimensional gravity. 
In a triangulation $\Delta$, the edges $e\in {\cal J}_{\Delta}$ bound
precisely four different faces; therefore, the $g_e$'s in (\ref{coloring4})
appear in four different traces. The relevant
formula is
\begin{equation}\label{4dp}
P^{4}_{inv}:= \int dg\ {\rho_1(g)}\otimes \rho_2(g) \otimes \cdots \otimes \rho_4(g)=
\sum_{\iota} {C^{\va \iota}_{\va \rho_1 \rho_2 \cdots \rho_4} \ C^{*{\va
\iota}}_{\va \rho_1 \rho_2 \cdots \rho_4}},
\end{equation}
where $P^{4}_{inv}$ is the projector onto ${\rm Inv}[\rho_1\otimes \rho_2 \otimes \cdots
\otimes \rho_n]$ and on the RHS we have expressed the projector in terms of normalized
intertwiners as in (\ref{3dp}). Finally, integration over the connection yields
\begin{eqnarray}\label{bf4} Z_{BF}(\Delta)=\sum \limits_{ {\cal
C}_f:\{f\} \rightarrow \rho_f }  \sum \limits_{{\cal C}_e:\{e\}
\rightarrow \{ \iota_e \}} \ \prod_{f \in {\cal J}_{\Delta}} \Delta_{\rho}
\prod_{v \in {{\cal J}_{\Delta}}}
\begin{array}{c}
\includegraphics[width=3cm]{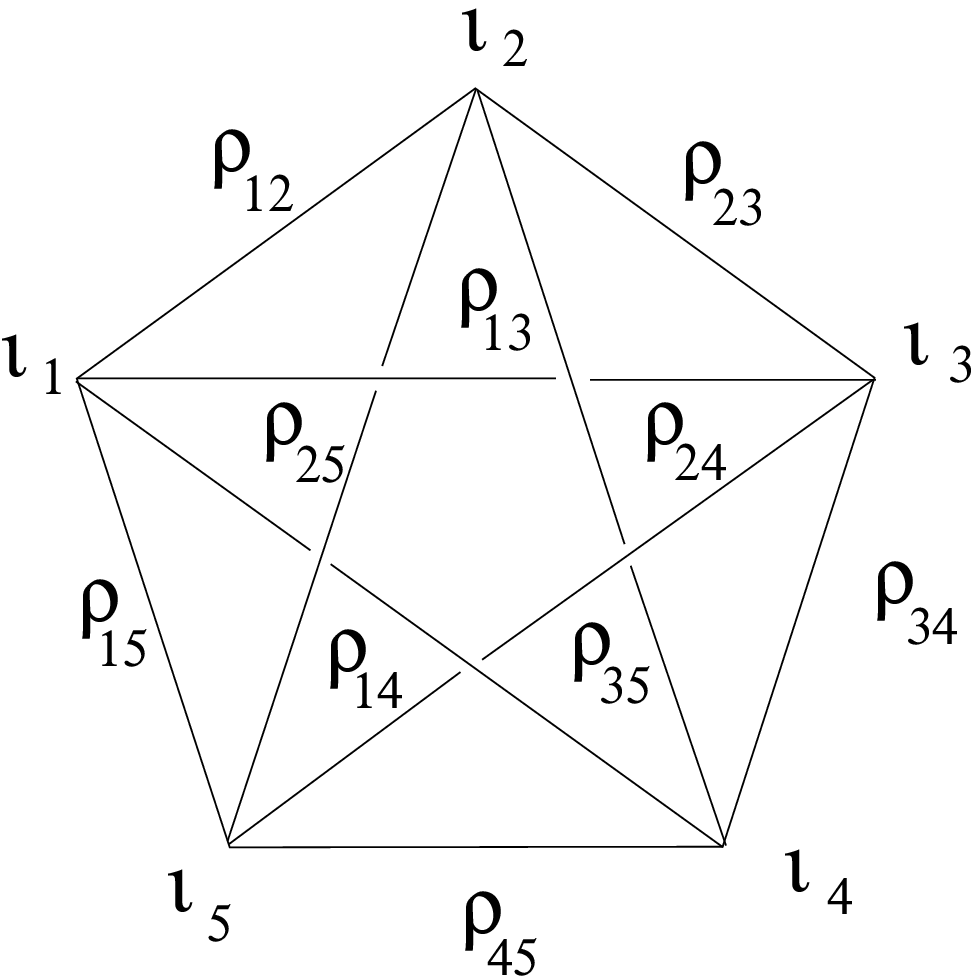}\end{array},
\end{eqnarray}
where the pentagonal diagram representing the vertex amplitude
denotes the trace of the product of five normalized $4$-intertwiners $C^{\va
\iota}_{\va \rho_1 \rho_2 \rho_3 \rho_4}$. As in the model for
$3$-dimensional gravity, the vertex amplitude corresponds to the
flat evaluation of the spin network state defined by the
pentagonal diagram in (\ref{bf4}), a $15j$-symbol. Vertices $v\in
{\cal J}_{\Delta}$ are in one-to-one correspondence to 4-simplexes
in the triangulation $\Delta$. In addition we have ${\cal
C}_e:\{e\} \rightarrow \{ \iota_e \}$ representing the assignment 
of intertwiners to edges coming from (\ref{4dp}). 

The state sum (\ref{bf4}) is generically divergent (due to the
gauge freedom analogous to (\ref{gauge2})). A regularized version
defined in terms of $SU_q(2)\times SU_q(2)$ was introduced by
Crane and Yetter  \cite{crane0, crane00}. As in three dimensions
(\ref{bf4}) is topologically invariant and the spin foam path
integral is discretization independent.

\subsection{Classical $SO(4)$ Plebanski action}\label{plebe}

Plebanski's Riemannian action can be thought of as $Spin(4)$
BF theory plus constraints on the $B$-field.
It depends on an $so(4)$ connection
$A$, a Lie-algebra-valued 2-form $B$ and Lagrange multiplier
fields $\lambda$ and $\mu$. Writing explicitly the Lie-algebra
indexes, the action is given by
\begin{equation}\label{pleb}
S[B,A,\lambda,\mu]=\int \left[B^{IJ}\wedge F_{IJ}(A) + \lambda_{IJKL}
\ B^{IJ} \wedge B^{KL} +\mu \epsilon^{IJKL}\lambda_{IJKL} \right],
\end{equation}
where $\mu$ is a 4-form and
$\lambda_{IJKL}=-\lambda_{JIKL}=-\lambda_{IJLK}=\lambda_{KLIJ}$ is
a tensor in the internal space. Variation with respect to $\mu$
imposes the constraint $\epsilon^{IJKL}\lambda_{IJKL}=0$ on
$\lambda_{IJKL}$. The Lagrange multiplier tensor $\lambda_{IJKL}$
has then $20$ independent components. Variation with respect to
$\lambda$ imposes $20$ algebraic equations on the $36$ components
of $B$. Solving for $\mu$ they are
\begin{equation}\label{ito}
 \epsilon^{\mu\nu\rho\sigma} B^{IJ}_{\mu\nu}B^{KL}_{\rho\sigma}=e \ \epsilon^{IJKL}
\end{equation}
which is equivalent to
\begin{equation}\label{constraints}
\epsilon_{IJKL} B^{IJ}_{\mu\nu}B^{KL}_{\rho\sigma}=e \
 \epsilon_{\mu\nu\rho\sigma},
\end{equation}
for $e\not=0$ where
$e=\frac{1}{4!}\epsilon_{OPQR}B^{OP}_{\mu\nu}B^{QR}_{\rho\sigma}\epsilon^{\mu\nu\rho\sigma}$
\cite{fre6}. The  solutions to these equations are
\begin{equation}\label{ambi}
B=\pm {}^*( e \wedge e), \ \ \ {\rm and}\ \ \ B=\pm e\wedge e,
\end{equation}
in terms of the $16$ remaining degrees of freedom of the tetrad
field $e^I_a$. If one substitutes the first solution into the
original action one obtains Palatini's formulation of general
relativity
\begin{equation}\label{pala}
S[e,A]=\int {\rm Tr}\left[e\wedge e \wedge {}^*F(A)\right].
\end{equation}
This property is the key to the definition of 
the spin foam model for gravity of the next section.

\subsection{Discretized Plebanski's constraints}\label{dpc}

In Section \ref{BF} we have derived the spin foam model for
$Spin(4)$ BF theory. We have seen that the sum over
$B$-configurations is encoded in the sum over unitary irreducible
representations of $Spin(4)$. Can we restrict the spin foam
configurations to those that satisfy Plebanski's constraints? To
answer this question one first needs to translate the constraints
of Plebanski's formulation into the simplicial framework. There
are two ways in which we can approach the problem. One is by
looking at the constraints in the form (\ref{pleb}) the other is
by using (\ref{constraints}).

The first version of the constraints is of the type of
$\Omega^{ij}$ of Section \ref{Reise}. This possibility has not
been explored in the literature. A simple representation of the constraints
is obtained if we choose to discretize (\ref{constraints}), 
for which there are no free algebra indexes.

As explained in Section \ref{BF} in the discretization of the $B$
field we replace the space-time indexes of the $2$-form by a
triangle:
\begin{equation}\label{rule}
\mu \nu \ \rightarrow \ \ t={\rm triangle}
\end{equation}
Using this it is easy to translate (\ref{constraints}) into the simplicial
framework. The RHS of (\ref{constraints}) vanishes when there are repeated space-time
indexes. At the discrete level, this means that whenever the triangles $t$ and 
$t^{\prime}$ belong to the same tetrahedron $T$, $t,t^{\prime}\in {T}$, (i.e., either
$t=t^{\prime}$  or  $t$ and $t^{\prime}$ share an edge) we have
\begin{equation}\label{opc1}
\epsilon_{IJKL} B^{IJ}_t B^{KL}_{t^{\prime}} =0,
\end{equation}
The remaining constraints correspond to the situation when $t$ and
$t^{\prime}$ do not lie on the same tetrahedron (no `repeated space-time indexes'). 
We can write these constraints as follows.
If we label a vertex in a given $4$-simplex, $0\cdots 4$ we can label the six triangles $0jk$ containing
the vertex $0$ simply as $jk$ ($j,k=1\cdots 4$). Only these are
needed to express the independent constraints
\begin{equation}\label{cdos}
\epsilon_{IJKL}B^{IJ}_{12}B^{KL}_{34}=\epsilon_{IJKL}B^{IJ}_{13} B^{KL}_{42}
=\epsilon_{IJKL}B^{IJ}_{14}B^{KL}_{23}\propto e.
\end{equation}

Now we want to implement the constraints.
The strategy is to impose them directly on the BF spin foam sum, (\ref{coloring4}), after the
$B$-integration has been performed. 
Formally one would associate the discrete $B_f$
to the differential operator $-i \partial / \partial U_f$ in (\ref{papart}). More precisely,
the observation is that the $Spin(4)$ left invariant vector
field $-i{\cal X}^{IJ}(U):= U^{\mu}_{\ \
\nu} X^{IJ\ \nu }_{\ \ \ \ \ \sigma}\frac{\partial}{\partial
U^{\mu}_{\ \ \sigma}}$  acts as a quantum $B^{IJ}$ on (\ref{papart}) since
\begin{eqnarray}\label{RIV}
\nonumber& &-i{\cal X}^{IJ}(U)\left( e^{i{\rm
Tr}[BU]}\right)|_{U\sim 1} = U^{\mu}_{\ \
\nu} X^{IJ\ \nu }_{\ \ \ \ \ \sigma}\frac{\partial}{\partial
U^{\mu}_{\ \ \sigma}} e^{i{\rm Tr}[BU]}|_{U\sim 1}= \\
&& ={\rm Tr}[U X^{IJ}B] e^{i{\rm Tr}[BU]}|_{U\sim 1}\sim B^{IJ}e^{i{\rm Tr}[BU]},
\end{eqnarray}
where $X^{IJ}$ are elements of an orthonormal basis in the $SO(4)$
Lie-algebra. The evaluation at $U=1$ is motivated by the fact that
configurations in the BF partition function (\ref{papart}) have
support on flat connections. This approximation is made in order to motivate our
definition but it plays no role in the implementation of the
constraints.

The constraints (\ref{constraints}) are quadratic in the $B$'s.
We have to worry about
cross terms; the nontrivial case corresponds to:
\begin{eqnarray}\label{RIV2}&& \nonumber
\epsilon_{IJKL}{\cal X}^{IJ}(U){\cal X}^{KL}(U)\left( e^{i{\rm
Tr}[BU]}\right)|_{U\sim 1}\\ \nonumber && =
-\epsilon_{IJKL} \left({\rm Tr}[X^{IJ}UB]{\rm Tr}[X^{KL}UB] e^{i{\rm Tr}[BU]}
+i{\rm Tr}[X^{IJ}X^{KL}UB]e^{i{\rm Tr}[BU]}\right)|_{U\sim 1}\\ &&
\sim \epsilon_{IJKL}B^{IJ}B^{KL}e^{i{\rm Tr}[BU]},
\end{eqnarray}
where the second term on the second line can be dropped using that
$\epsilon_{IJKL} X^{IJ}X^{KL}\propto 1$ (one of the two $SO(4)$
Casimir operators) and $U\sim 1$. Therefore, we define the $B_f$
field associated to a face at the level of equation
(\ref{coloring}) as the appropriate left invariant vector field
$-i{\cal X}^{IJ}(U_f)$ acting on the corresponding discrete
holonomy $U_f$, namely
\begin{equation}\label{B_f}
B_f^{IJ}\rightarrow -i{\cal X}^{IJ}(U_f).
\end{equation}
Gauge invariance of the BF partition function implies that
for every tetrahedron 
\begin{equation}\label{gauss}
\sum_{t \in T} B^{IJ}_t=0
\end{equation}
where $t \in T$ denotes the triangles in the corresponding tetrahedron.

In order to implement (\ref{opc1}) we concentrate on a single
4-simplex amplitude, i.e., the fundamental {\em atom} of Figure
\ref{chunk}. Once the
constrained 4-simplex amplitude is constructed any spin foam
amplitude can be obtained by gluing atoms together along faces by
integration over common boundary data as in Section \ref{sflg}.
The BF 4-simplex wave function is obtained using (\ref{papart}) on
the dual 2-complex with boundary defined by the intersection of
the dual of a single 4-simplex with a 3-sphere, see Figure
\ref{chunk}.

The amplitude of the fundamental {\em atom} is a (cylindrical) function
depending on the boundary values of the connection on the 
boundary graph $\gamma_5$. We denote as $h_{ij}\in Spin(4)$ ($i\not=j$, $i,j=1\cdots 5$ and
$h_{ij}=h^{-1}_{ji}$) the corresponding 10 boundary variables
(associated to thin boundary edges in Figure
\ref{chunk})\footnote{Strictly speaking, the boundary connections
$h_{ij}$ are defined as the product $h^{\prime}_{ij} h^{''}_{ij}$
where $h^{\prime}$ and $h^{''}$ are associated to half paths as
follows: take the edge $ij$ for simplicity and assume it is
oriented from $i$ to $j$. Then $h^{\prime}_{ij}$ is the discrete
holonomy from $i$ to some point in the center of the path and
$h^{''}_{ij}$ is the holonomy from that center point to $j$. This
splitting of variables is necessary when matching different atoms
to reconstruct the simplicial amplitude
 \cite{reis4,reis6}.}
and $g_i\in Spin(4)$ ($i=1,\cdots, 5$) the internal connection (corresponding
to the thick edges in Figure \ref{chunk}). According to (\ref{papart}) the 4-simplex BF amplitude
$4SIM_{BF}(h_{ij})$ is given by
\begin{equation}\label{pito}
4SIM_{BF}(h_{ij})=\int \prod_i dg_i  \prod_{i<j} \delta(g_ih_{ij}g_j),
\end{equation}
where $U_{ij}=g_ih_{ij}g_j$ is the holonomy around the triangular
face $0ij$.
With the definition of the $B$ fields given in (\ref{B_f}) the
constrained amplitude, $4SIM_{const}(h_{ij})$, formally becomes
\begin{equation}
4SIM_{const}(h_{ij})=\int \prod_i dg_i \delta\left[{\rm Constraints}(-i{\cal X}(U_{ij}))\right]\prod_{i<j} \delta(g_ih_{ij}g_j).
\end{equation}
It is easy to verify, using an equation analogous to
(\ref{RIV}) and the invariance of $\epsilon_{IJKL}$, that one can
define the $B$'s by simply acting with the left invariant vector
fields on the boundary connection $h_{ij}$. Therefore, the
previous equation is equivalent to
\begin{equation}\label{4sgr}
4SIM_{const}(h_{ij})= \delta\left[{\rm Constraints}(-i{\cal X}(h_{ij}))\right] \int \prod_i dg_i \prod_{i<j} \delta(g_ih_{ij}g_j),
\end{equation}
where we have taken the delta function out of the integral. The
quantity on which the formal delta distribution acts is simply
$4SIM_{BF}(h_{ij})$. The amplitude $4SIM_{BF}(h_{ij})$ can be
expressed in a more convenient way if we expand the delta
functions in modes as in (\ref{coloring4}) and then integrate over
the internal connection $g_i$. The integration is analogous to the
one in (\ref{4dp}), for example integration over $g_1$ yields
\begin{equation}\label{nodo} P^{4}_{inv} \ \rho_{12}(h_{12})\otimes\rho_{13}(h_{13})\otimes\rho_{14}(h_{14})\otimes \rho_{15}(h_{15})
=\sum_{\iota} \Delta_{\iota}
\begin{array}{c}{\includegraphics[width=5cm]{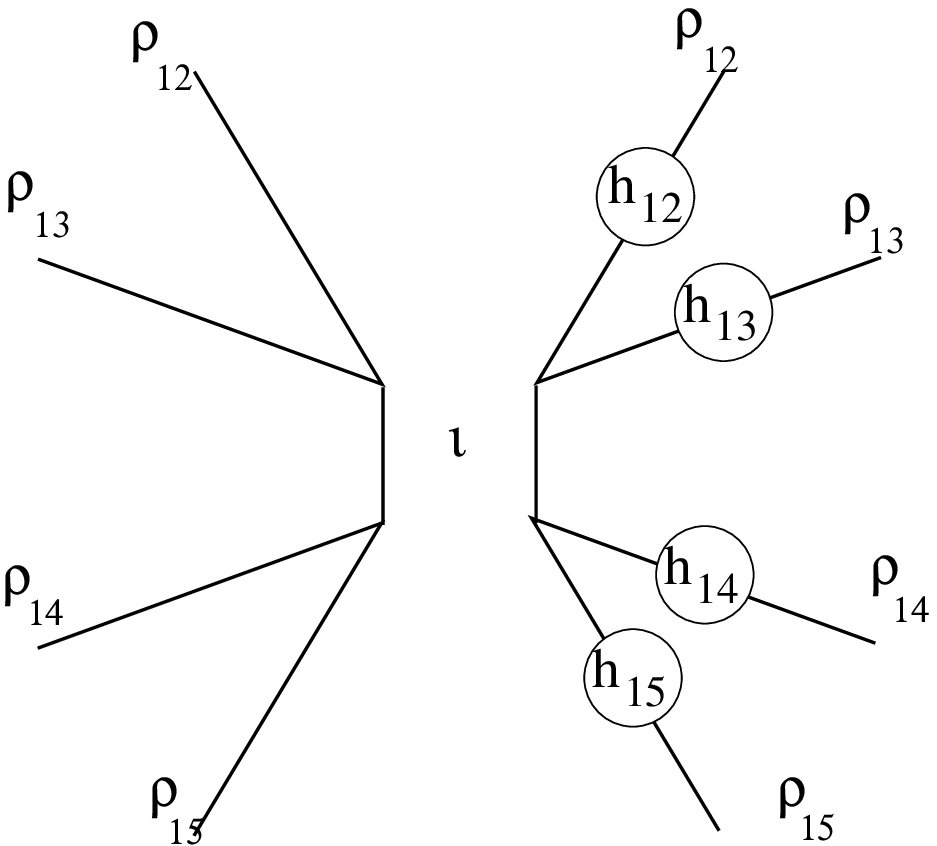}}\end{array},
\end{equation}
where on the RHS we have chosen a particular basis of 4-intertwiners to span the
projector $P^{4}_{inv}$ into ${\rm Inv}\left[\rho_{12}\cdots \rho_{15}\right]$.
The $4$-intertwiners are explicitly given as contractions of normalized 
intertwiners using the graphical notation introduced in Section \ref{sfm3d}.
The circles represent the corresponding $\rho$-representation
matrices evaluated on the the corresponding boundary connection $h$.
The $4$-simplex amplitude becomes
\begin{equation}\label{bfwf}
\!\!
4SIM_{BF}(h_{ij})=\!\! \sum \limits_{\rho_{ij},\iota_i }\! \ \ \prod \limits_{i<j}\  \Delta_{\rho_{ij}}\!\!
\begin{array}{c}
\includegraphics[width=4.5cm]{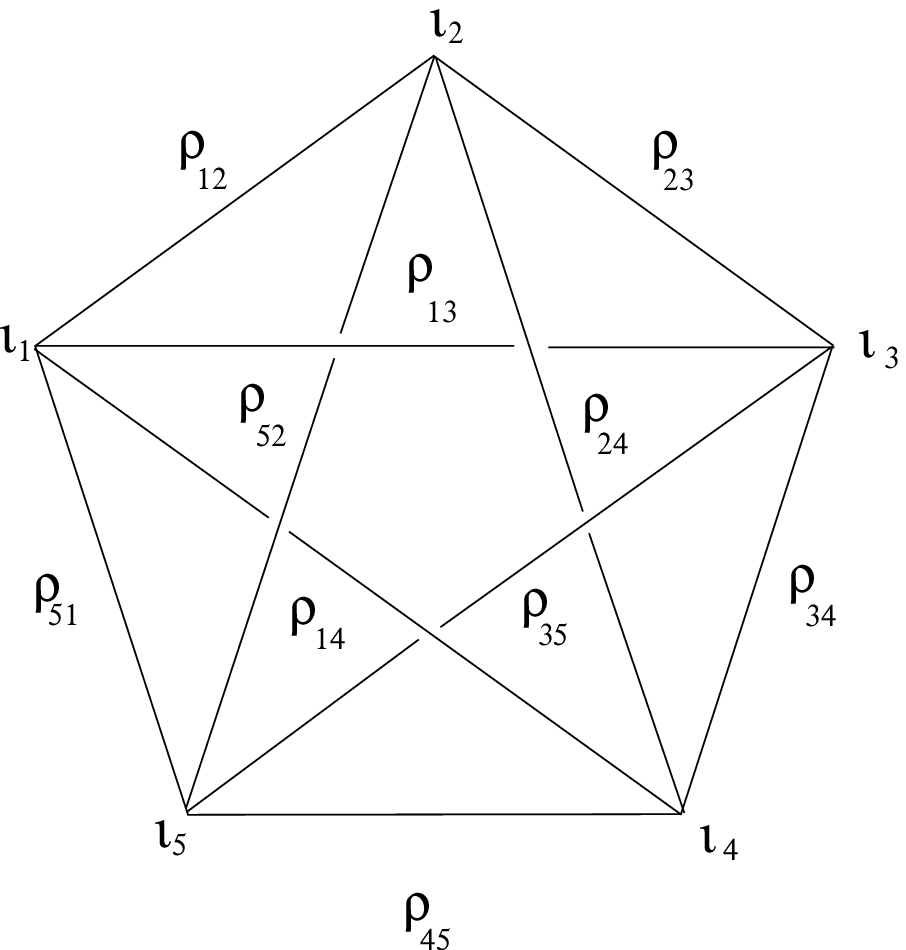}\end{array}
\begin{array}{c}
\includegraphics[width=4.5cm]{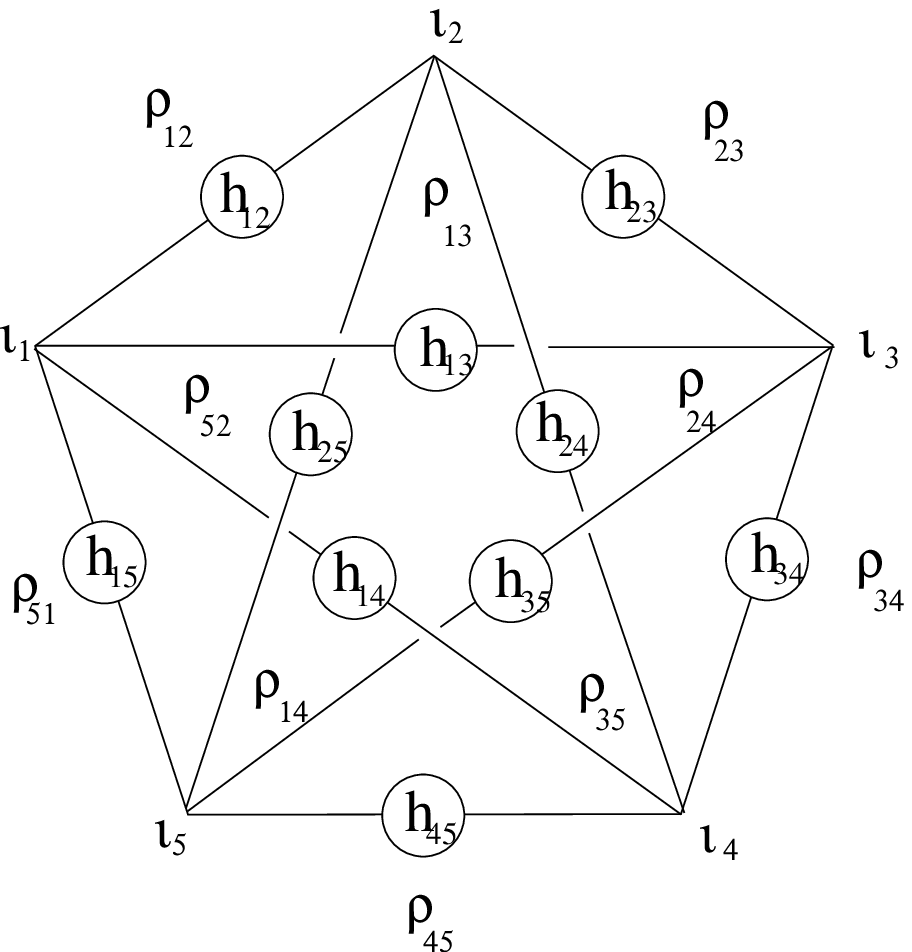}\end{array}.
\end{equation}
In the previous equation $4$-valent nodes denote normalized $4$-intertwiners
and the tree decomposition is left implicit (the factors $\Delta_{\iota}$ in (\ref{nodo})
have been absorbed into the notation).
The term on the left is a $15j$-symbol as in (\ref{bf4}) while the term
on the right is the trace of five $4$-intertwiners with the respective
boundary connection insertions.

The 4-simplex amplitude for the constrained spin foam model is then defined as the
restriction of $4SIM_{BF}(h_{ij})$ imposed by the quantum version
of the constraints (\ref{opc1}) defined by the
following set of differential equations
\begin{equation}\label{XX}
\hat C_{ij,\ ik}\ 4SIM_{const}(h_{ij})=\epsilon_{IJKL}{\cal
X}^{IJ}(h_{ij}){\cal X}^{KL}(h_{ik})\ 4SIM_{const}(h_{ij})= 0 \ \
\ \forall \ j,k,
\end{equation}
and where the index $i=1,\cdots, 5$ is held fixed and $\hat
C_{ij,\ ik}$ denotes the corresponding constraint operator.

There are six independent constraints (\ref{XX}) for each value of
$i=1,\cdots,5$. If we consider all the equations for the 4-simplex
amplitude then some of them are redundant due to (\ref{gauss}). 
The total number of independent conditions is 20, in agreement 
with the number of classical constraints (\ref{constraints}). For a given $i$ in (\ref{XX})
(i.e., a given tetrahedron) and for $j=k$ the equation becomes
\begin{eqnarray}\label{XX1} \nonumber &&
\epsilon_{IJKL}{\cal X}^{IJ}(h_{ij}){\cal X}^{KL}(h_{ij})\
4SIM_{const}(h_{ij})= \\ &&\ \ \ \ \ \ \ \ \ \ \ \ \ \ \ \ \ \
\nonumber = \delta_{ik} \left[ J^{i}(h^{\ell}_{ij}) J^{k}(h^{\ell}_{ij})- 
J^{i}(h^{r}_{ij}) J^{k}(h^{r}_{ij})\right]\
4SIM_{const}(h^{\ell}_{ij},h^{r}_{ij})=  \\ &&\ \ \ \ \ \ \ \ \ \ \ \ \ \ \ \ \ \
=\left[j_{ij}^{\ell}(j_{ij}^{\ell}+1)-j_{ij}^r(j_{ij}^r+1)\right]
4SIM_{const}(h_{ij})=0,
\end{eqnarray}
where we have used that: $Spin(4)=SU(2)\times SU(2)$ so that for $h\in Spin(4)$,
$h^{\ell}, h^r\in SU(2)$ denote its right and left components, irreducible representation
can be expressed as $\rho=j^{\ell}\otimes j^{r}$ for $j^{\ell},j^{r} \in {\rm Irrep}[SU(2)]$, 
and the left invariant vector field
\begin{equation}\label{r-l}
{\cal X}^{IJ}(h)=P^{+IJ}_i J^i(h^{\ell})+ P^{-IJ}_i J^i(h^{r}),\end{equation}
for $\epsilon_{IJ}^{\ \ \ KL}P^{\pm IJ}=\pm P^{\pm KL}$ and $J^i$'s
being left invariant vector fields on the corresponding left and right
$SU(2)$ copies of $Spin(4)$.
The previous constraints are solved
by requiring the corresponding representation $\rho_{ij}$ to be
simple, i.e.,
\begin{equation}
\rho_{ij}=j_{ij} \otimes j^*_{ij} \ \ \ {\rm or}\ \ \  \rho_{ij}=j_{ij} \otimes j_{ij}.
\end{equation}
This ambiguity is analogous to the classical one in (\ref{ambi}).
We take $\rho_{ij}=j_{ij} \otimes j^*_{ij}$ in correspondence to the choice
${}^*( e \wedge e)$ that produces the gravity sector
\footnote{In a
detailed analysis Baez and Barrett \cite{baez6} justify this
choice in a rigorous way. This restriction (imposed by the
so-called chirality constraint, Section \ref{arvol}) implies that
the `fake tetrahedron' configurations---corresponding to solutions
of the constraints on the right of equation (\ref{ambi})---are
dropped from the state sum.}. This solves 10 of the 20 equations.
The next non-trivial conditions imposed by (\ref{XX}) is when
$j\not=k$. In this case we have
\begin{eqnarray}\label{XX2} \nonumber &&
2 \epsilon_{IJKL}{\cal X}^{IJ}(h_{ij}){\cal X}^{KL}(h_{ik})\
4SIM_{const}(h_{ij})\\ && \nonumber =\epsilon_{IJKL} \left({\cal
X}^{IJ}(h_{ij})+{\cal X}^{IJ}(h_{ik})\right) \left({\cal
X}^{KL}(h_{ij})+{\cal X}^{KL}(h_{ik})\right)4SIM_{const}(h_{ij})\\
&&\nonumber
=\left[\iota^{\ell}(\iota^{\ell}+1)-\iota^r(\iota^r+1)\right]
4SIM_{const}(h_{ij})\\ && =0,
\end{eqnarray}
where in the second line we used the fact that we have already solved (\ref{XX1}). 
In the third line we have used the gauge invariance (or the analog of (\ref{gauss}) for the
3-valent node in the tree decomposition that pairs the
representation $\rho_{ij}$ with the $\rho_{ik}$) in order to express
the sum of invariant vector fields as the invariant vector filed acting 
on the virtual link labeled by $\iota$. 
This choice of tree decomposition in the case $ij=12$ and $ik=13$ is the one used in
equation (\ref{nodo}). The solution is clearly
$\iota=\iota\otimes\iota^*$.

What happens now to any of the two remaining conditions (only one
is independent), for example, $\hat C_{ij^{\prime},\ ik^{\prime}}$
for $k\not=k^{\prime}$, $j\not=j^{\prime}$ and
$j^{\prime}\not=k^{\prime}$? At first sight it looks like this
equations cannot be (generically) satisfied because an
intertwiner that has simple $\iota$ in one tree decomposition does not have
only simple $\iota^{\prime}$ components in a different tree
decomposition as a consequence of the recoupling identity
(\ref{42}). There is however a linear combination of intertwiners
found by Barrett and Crane in  \cite{BC2} which is simple in any
tree decomposition, namely
\begin{equation}\label{ii}
\left|\Psi_{BC}\right>=\sum_{simple \ \iota} C^{\iota}_{\rho_1,\cdots, \rho_4}.
\end{equation}
$C^{\iota}_{\rho_1,\cdots, \rho_4}$ is a normalized 4-intertwiner
and the summation is over simple  $\iota$ (i.e.
$\iota=\iota\otimes \iota^*$) and the $\rho_i$ are also simple
($\rho_i=j_i\otimes j^*_i$ for $i=1..4$). This is clearly a
solution to all the constraints and has been shown to be the
unique one (up to an overall factor) by Reisenberger in
\cite{reis3}. $\left|\Psi_{BC}\right>$ defines the so-called
Barrett-Crane intertwiner. Now the projector $P^{4}_{inv}$ in
(\ref{nodo})---the building block of the BF amplitude---can be
written as
\begin{equation} P^{4}_{inv}=\left|\Psi_{BC}\right> \left<\Psi_{BC}\right| \ \ + \ \ orthogonal \ \
terms,
\end{equation}
using the standard Gram-Schmidt construction of a basis in ${\rm Inv}\left[\rho_{1}\cdots \rho_{4}\right]$.
In other words $P^{4}_{inv}$ is the sum of 1-dimensional projector to the solutions of the
constraints (\ref{XX}) plus the orthogonal complement. The
solution to (\ref{XX}) is then unique (up to scaling) and can be written as
\begin{equation}\label{grave}
\!\! 4SIM_{const}(h_{ij})=
\!\! \sum \limits_{j_{ij},\iota_i }\! \ \ \prod \limits_{i<j}\  \Delta_{\rho_{ij}}\!\!
\begin{array}{c}
\includegraphics[width=4.3cm]{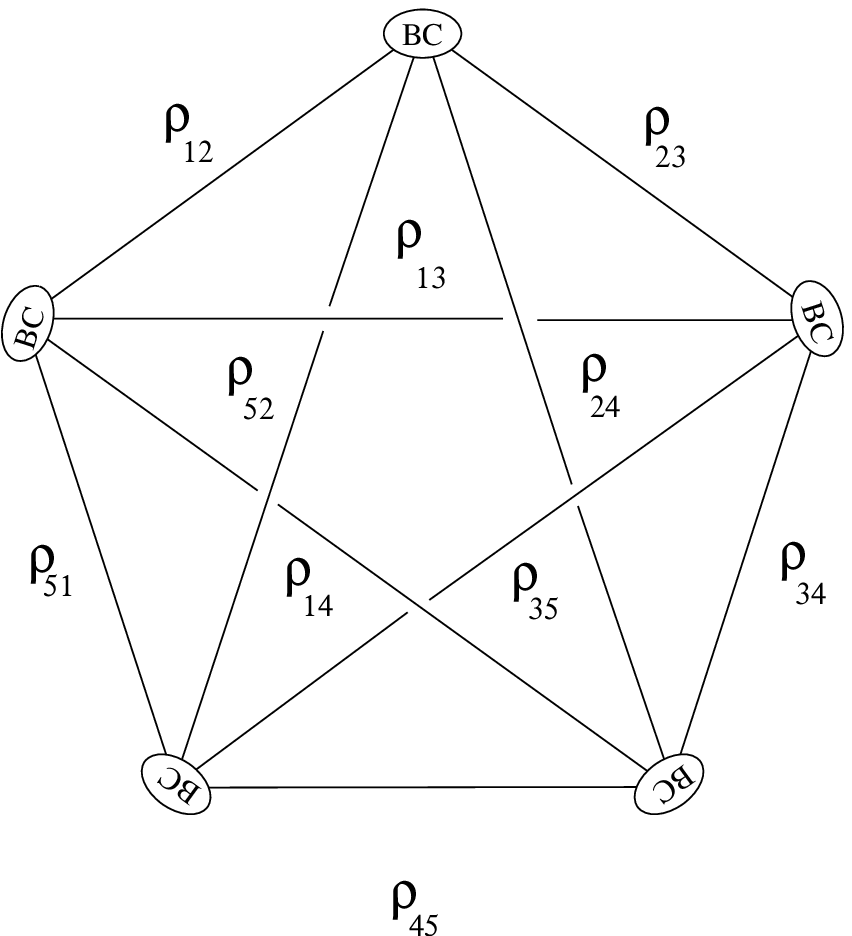}\end{array}
\begin{array}{c}
\includegraphics[width=4.5cm]{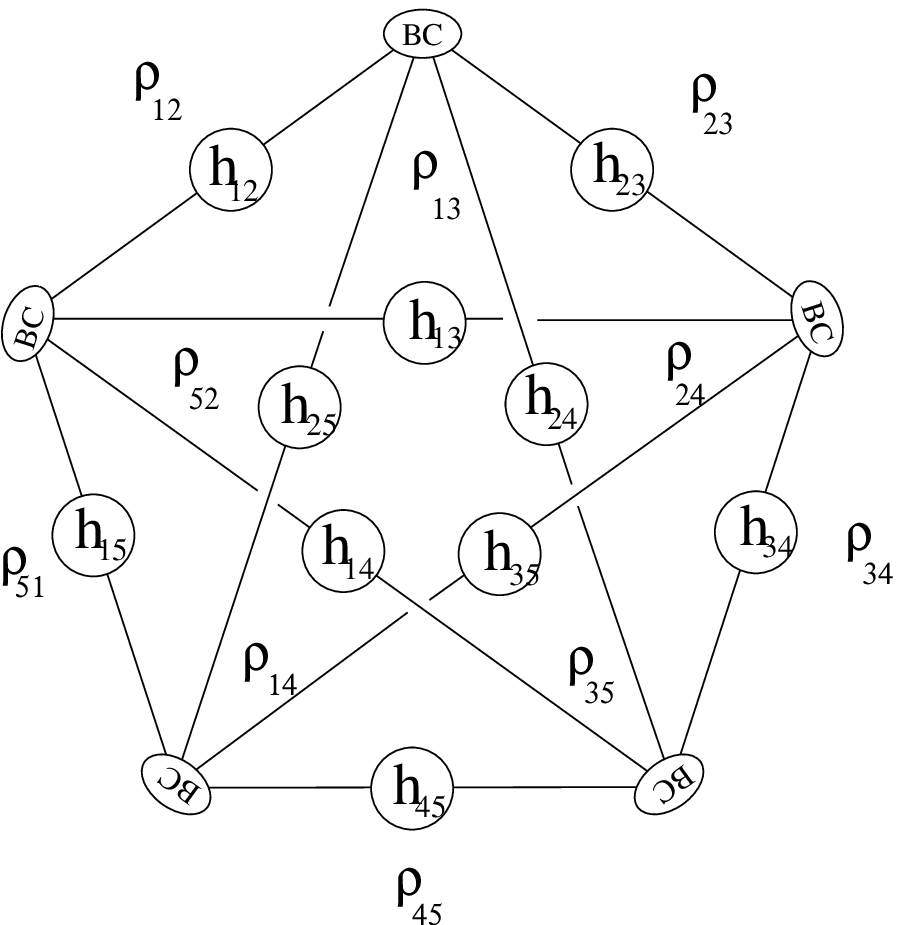}\end{array},
\end{equation}
where the $\rho_{ij}=j_{ij} \otimes j^*_{ij}$ and we graphically represent $\left|\Psi_{BC}\right>$
by $\includegraphics[width=.7cm]{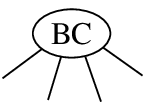}$.

The amplitude of an arbitrary simplicial complex is computed by
putting together 4-simplexes with consistent boundary connections
and gluing them by integration over boundary data. If we do that
we obtain the Barrett-Crane state sum on a fixed triangulation
\begin{eqnarray}\label{BC4} Z_{BC}({\cal J}_{\Delta})=\sum \limits_{ {\cal
C}_f:\{f\} \rightarrow j_f }   \ \prod_{f \in {\cal J}_{\Delta}}
(2j_f+1)^2 \ \prod_{e \in {\cal J}_{\Delta}} A_{e} \prod_{v \in {{\cal
J}_{\Delta}}} \begin{array}{c}
\includegraphics[width=3.2cm]{forobc2.eps}\end{array},
\end{eqnarray}
where we have made the replacement $\Delta_{\rho}=(2j+1)^2$ for $\rho=j\otimes j^*$.
The vertex amplitude depends on the ten representations labeling the ten faces in a
4-simplex and it is referred to as $10j$-symbol. Using the definition (\ref{ii}) of the Barrett-Crane 
intertwiner,
$\includegraphics[width=.7cm]{chi.eps}$, the $10j$-symbol 
can be written explicitly in terms of $15j$-symbols as
\begin{eqnarray}\label{VBC} \begin{array}{c}
\includegraphics[width=3.2cm]{forobc2.eps}\end{array}.
= \sum \limits_{\iota_1\cdots \iota_5}
\begin{array}{c}
\includegraphics[width=7cm]{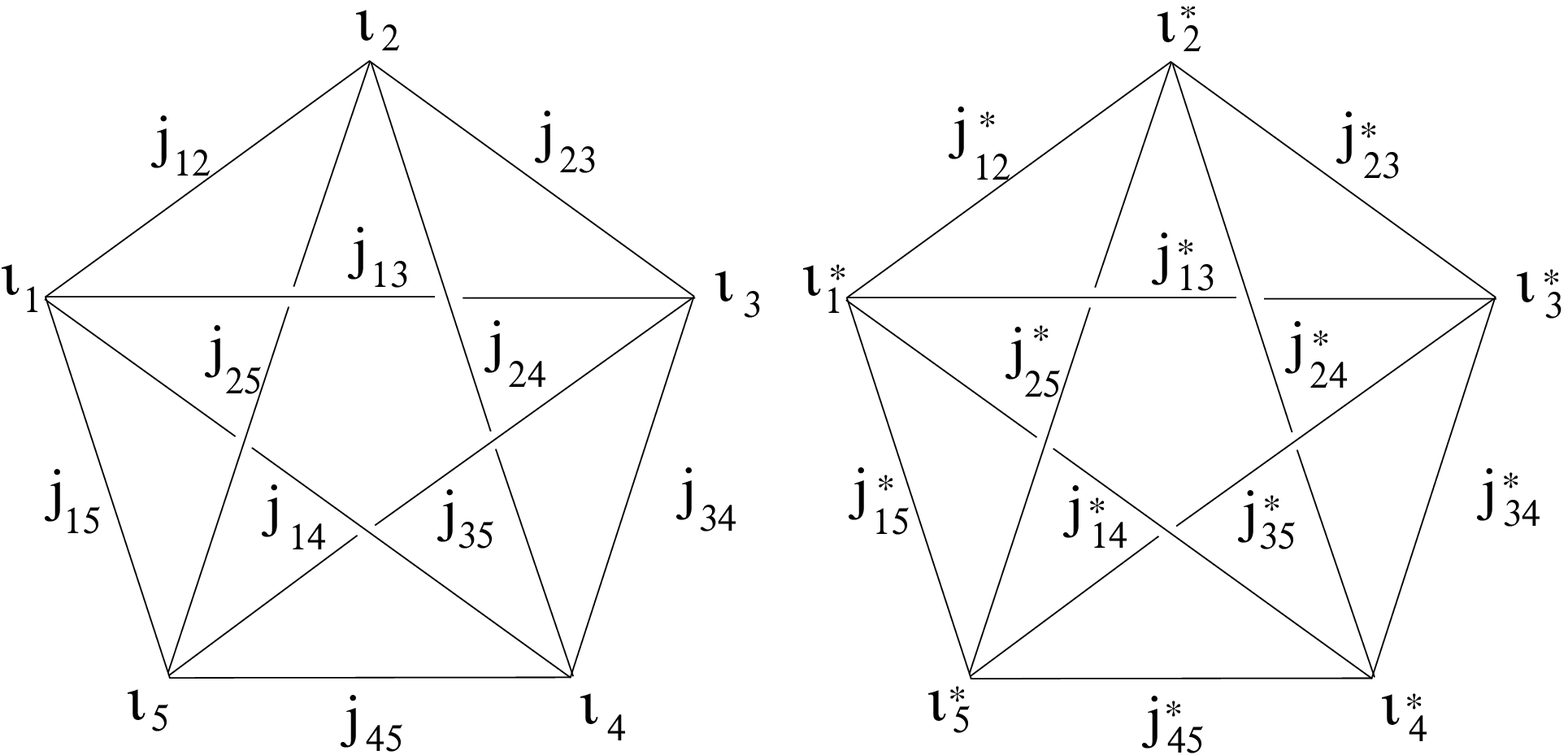}\end{array},
\end{eqnarray}
where we represent the normalized $SO(4)$-intertwiners
$C^{\iota}_{\rho_1\cdots \rho_4}$ in (\ref{ii}) in terms of the
corresponding pair of left-right $SU(2)$-intertwiners. 

$A_e$ denotes a possible edge amplitude which is not determined by
our argument. The implementation of constraints in the path
integral should be supplemented with the appropriate modification
of the measure. This should affect the values of lower dimensional
simplexes such as face and edge amplitudes. Constraints (\ref{XX})
act on each edge (tetrahedron) separately; heuristically one would
expect a Jacobian factor to modify and so determine the value of
the edge amplitude $A_e$. As we pointed out 
in Section \ref{anom}, the resolution of this ambiguity is related to the
question of how to define the correct anomaly-free measure.
In Section \ref{BCGFT} we will see how the GFT formulation provides a natural definition
of $A_e$.

Constraints that involve different tetrahedra in a given 4-simplex---corresponding to (\ref{cdos})---are
automatically satisfied as operator equations on the Barrett-Crane
solutions. This can be checked using (\ref{XX}) and (\ref{gauss}).

\subsection{The quantum tetrahedron in 4d and the Barrett-Crane intertwiner}

We have re-derived the Barrett-Crane model directly from a
simplicial formulation of Plebanski's action. The original
definition makes use of the concept of  the `quantum tetrahedron'
\cite{baez6,baez7}. In that context, the analog of our $B$'s is
given by the bivectors associated with the triangles of a
classical tetrahedron. A Hilbert space is defined using geometric
quantization and the classical triangle bivectors are promoted to
operators. This Hilbert space is reduced by implementation of
constraints---quantization of the geometric constraints satisfied
by the classical tetrahedron---to the Hilbert space of the
so-called `quantum tetrahedron'. These constraints are precisely of the 
form (\ref{cdos}).
For fixed triangle quantum
numbers the state of the quantum tetrahedron is defined by the
Barrett-Crane intertwiner (\ref{ii}). There is however no
systematic prescription for the construction of the state sum and
even the single $4$-simplex amplitude has to be given as an, although natural, 
ad hoc definition.

As we have shown in the previous sections all this can be obtained
from the path integral approach applied to (\ref{pleb}). In
addition to simplifying the derivation---no additional quantization
principle is required---our framework provides a direct and
systematic definition of the state sum. In
 \cite{baez6,baez7} one can only access the constraint operator
algebra of a single tetrahedron and conclude that this sub-algebra
is closed. The full constraint algebra of (\ref{XX}) can only be
studied in the formalism presented here. 
Finally, since our prescription is directly derived from an action
principle it is conceivable that a rigorous derivation of the
undetermined amplitudes (such as $A_e$ in (\ref{BC4})) could
exist. These important questions remain to be investigated.

\subsection{An integral expression for the $10j$-symbol}

In Section \ref{BCGFT} we will present a discretization
independent formulation of the Barrett-Crane model based on a GFT.
The realization by Barrett  \cite{baba} that the vertex amplitude
(\ref{VBC}) admits an integral representation plays a key role in
the construction. The integral formula is also important in the
computation of the $10j$-symbol asymptotics that we briefly
describe in the next subsection.

The basic observation in the construction of an
integral expression for (\ref{VBC}) is that equation (\ref{ii})
has precisely the form (\ref{3dp}) (for $n=4$) if we write the $Spin(4)$
intertwiners as tensor products
of $SU(2)$ ones. Therefore,
\begin{equation}\label{bcint1}
\left|\Psi_{BC}\right>=\int \limits_{SU(2)} du\ {j_1(u)}\otimes j_2(u) \otimes j_3(u) \otimes
j_4(u),
\end{equation}
where $j(u)$ denotes $SU(2)$ representation matrices in the
representation $j$. Each one of the five pairs of intertwiners in
(\ref{VBC}) can be obtained as an integral (\ref{bcint1}) over $SU(2)$. Each of
the ten representation matrices $j_{ik}$ ($i\not=k=1\cdots 5$)
appears in two integrals corresponding to the intertwiners at the
node $i$ and $j$ respectively. Contracting the matrix indexes according
to (\ref{VBC}) these two representation matrices combine into a
trace ${\rm Tr}\left[j_{ik}(u_iu^{-1}_k)\right]$ ($u_i \in SU(2)$). Parameterizing
$SU(2)$ with spherical coordinates on $S^3$
\begin{equation} \label{kernito} {\rm
Tr}\left[j_{ik}(u_iu^{-1}_k)\right]=\frac{{\rm
sin}(2j_{ik}+1)\psi_{ik}}{{\rm sin}(\psi_{ik})} :=(2j_{ik}+1)
K_{j_{ik}}(y_i,y_k),\end{equation} where $\psi_{ik}$ is the
azimuthal angle between the points $y_i,y_k$ on the sphere
corresponding to $u_i$ and $u_k$ respectively. We have also
introduced the definition of the kernel $K_{j_{ik}}(y_i,y_k)$ in
terms of which the Barret-Crane vertex amplitude (\ref{VBC})
becomes
\begin{eqnarray}\label{intform}
A_v(j_{ik}) =  \int \limits_{(S^3)^5} \prod \limits_{i=1}^5 dy_i \
\prod \limits_{i<k}(2j_{ik}+1) K_{\va j_{ik}}(y_i,y_k).
\end{eqnarray} Each of the five integration variables in $S^3$ can be regarded
as a unit vector in $\R^4$. They are interpreted as unit normal
vectors to the 3-dimensional hyperplanes spanned by the
corresponding five tetrahedra. The angles $\psi_{ik}$ is defined
by ${\rm cos}\ \psi_{ik}=y_i\cdot y_k$ and corresponds to the
exterior angle between two hyperplanes (analogous to the dihedral
angles of Regge calculus). These normals determine a 4-simplex in
$\R^4$ up to translations and scaling \cite{baba}.

\subsection{The asymptotics for the vertex amplitude}\label{asy}

The large spin behavior of the spin foam amplitudes provides
information about the low `energy' or semi-classical limit of the
model\cite{major} in the naive sense $\hbar \rightarrow \infty$
while geometric quantities such as the area (\ref{aarreeaa})
are held fixed.  Evidence showing a connection between the
asymptotics of the Barrett-Crane vertex and the action of general
relativity was found by Crane and Yetter in
 \cite{crane}.

A computation of the asymptotic (large $j$) expression of
the Barrett-Crane vertex amplitude for non-degenerate
configurations was obtained by Barrett and Williams in
\cite{bawi}. They computed $A_v(j_{ik})$ for large $j_{ik}$ by
looking at the stationary phase approximation of the oscillatory
integral (\ref{intform}). The large spin behavior of the vertex
amplitude is given by
\begin{equation}
A_v(j_{ik}) \sim \sum_{\sigma} \ P(\sigma)\ {\rm cos}\left[S_{\va
Regge}(\sigma)+\kappa \frac{\pi}{4}\right]+D
\end{equation}
where the sum is over geometric 4-simplexes $\sigma$  whose
face areas are fixed by the spins. The action in the argument of
the cosine corresponds to Regge action which in four dimensions is
defined by $S_{\va Regge}(\sigma)=\sum_{i<k} A_{ik}\
\psi_{ik}(\sigma)$ where $A_{ik}$ is the area of the
$ik$-triangle. $P(\sigma)$ is a normalization factor which does
not oscillate with the spins. $D$ is the contribution of
degenerate configurations, i.e. those for which some of the
hyperplane normals defined above coincide. However, in a recent
paper  \cite{baez8} Baez, Christensen and Egan show that the term
$D$ is in fact dominant in the previous equation, i.e. the leading
order terms are contained in the set of degenerate configurations!
This has been later confirmed by the results of Freidel and 
Louapre \cite{frei9} and Barrett and Steele \cite{bastee}.

\subsection{Area and Volume in the Barrett-Crane model}\label{arvol}

Using the representation of the discrete $B$ of Section \ref{dpc} we
can define the operator corresponding to the square of $B_t$
corresponding to any triangle $t\in \Delta$, namely
\[\hat B_{t\, IJ}\hat B_t^{IJ}=-{\cal X}_{IJ}(U_f){\cal X}^{IJ}(U_f),\]
where $f$ is the face dual to $t\in \Delta$. Acting on the modes
of the expansion (\ref{coloring4}) this is simply the Casimir
\begin{equation}\label{casi2}
\hat B_{t\, IJ}\hat
B_t^{IJ}=\left[j^{\ell}(j^{\ell}+1)+j^{r}(j^r+1)\right]\
\mathds{1}.
\end{equation}
It is easy to check that this operator commutes with all the
constraints and therefore is well defined on the space of solutions
of the constraints. Its
geometrical meaning is clear if we recall that at the classical
level $B_t$ represent the bivector associate to the triangle $t\in
\Delta$. Once the constraints are imposed (\ref{casi2}) is
proportional to square of the area of the corresponding triangle.
The simplicity constraint implies $j_{\ell}=j^*_{r}=j$ so that the area
of the triangles have discrete eigenvalues $a_j$ given by
\begin{equation}\label{super}
a_j \propto \sqrt{j(j+1)},
\end{equation}
in agreement with the result of quantum geometry
(\ref{aarreeaa})\footnote{ Given a triangulation $\Delta$ of $\cal
M$ and the induced triangulation $\Delta_{\Sigma}$ of a slice
$\Sigma \subset {\cal M}$ any spin foam defined on ${\cal
J}_{\Delta}$ induces a spin network state on a graph $\gamma_{\va
\Sigma}$ dual to $\Delta_{\Sigma}$. Links of $\gamma_{\va \Sigma}$
are dual to triangles in $\Delta_{\Sigma}$. These triangles play
the role of the surface $S$ in the definition of the area operator
(\ref{aarreeaa}).}.

Can we define the analog of volume operator of quantum geometry?
The candidate for the square of such operator are
\begin{equation}
U_{\pm}=\epsilon_{ijk}\left[J^{i}_{{\va
R}1}J^{j}_{{\va R}2}J^{k}_{{\va R}3} \pm J^{i}_{{\va
L}1}J^{j}_{{\va L}2}J^{k}_{{\va L}3}\right],
\end{equation}
where $J_1,J_2$, and $J_3$ are the operators defined in (\ref{r-l})
corresponding to three different triangles in the tetrahedron. The
operator $U_+$ vanishes identically on the solutions of the
quantum constraints. In fact it can be expressed as the commutator
of the constraints (\ref{XX}). This is the chirality constraint of
\cite{baez6}.

One would like to define the volume operator as the square root of
$U_-$; however, $U_{-}$ is not a well defined operator on the
Hilbert space of the quantum tetrahedron since it does not commute
with the simplicity constraints. In other words the action of the
volume operators map states out of the solution of the
constraints. Only when the dimension of ${\rm Inv}\left[j_1\otimes
j_2\otimes j_3\otimes j_4\right]$ is one the commutator vanishes
(this is in fact a necessary and sufficient condition). The volume
operator is well defined in this subspace but it vanishes. In
conclusion, generically the volume operator is not defined in the
Barrett-Crane model.

\subsection{Lorentzian generalization}

A Lorentzian generalization of the Riemannian Barrett-Crane vertex
amplitude was proposed in  \cite{BC1}. Using the spin foam model
GFT duality of Section \ref{sec:gft-sf} a generalization of the
full model (including face and edge amplitudes) was found in
 \cite{a9}. The relevant representations of the Lorentz group are
the unitary ones. Unitary irreducible representations of
$SL(2,\C)$ are infinite dimensional and labeled by $n\in \N$ and
$\rho\in \R^+$. The simplicity constraints select the
representations for which $n=0$. The triangle area spectrum is
given by
\begin{equation}
a_{\rho}\propto \sqrt{\rho^2+1}.
\end{equation}
There is still a minimum eigenvalue but the spectrum is
continuous. The derivation of the model can be obtained from
$SO(3,1)$ Plebanski's action by generalizing the computations of
Section \ref{BCM}. We will study this model in Section \ref{LO}
were we show that the Lorentzian extension is not unique. In fact
a new Lorentzian model can be defined following the Barrett-Crane
prescription  \cite{a8}.

\subsection{Positivity of spin foam amplitudes}

Baez and Christensen \cite{baez9} showed that spin foam amplitudes
of the (Riemannian) Barrett-Crane model are positive for any
closed $2$-complex. For open spin foams they are real and (if not
zero) their sign is given by $(-1)^{2J}$ where $J$ is the sum of
the spin labels of the edges of the boundary spin networks. This
is a rather puzzling property since (as the authors point out)
this seems to imply the absence of quantum interference.
Positivity of the Lorentzian model seems to hold according to
numerical evaluations of the vertex amplitude.
The positivity results have been generalized to generic 2-complexes (not 
necessarily dual to a triangulation) by Pfeiffer in \cite{pfei}, who
also showed that the sign $(-1)^{2J}$ is not really present if one chooses 
the correct bases of intertwiners.

\subsection{Degenerate Sectors of $Spin(4)$ Plebanski's formulation}\label{ccc}

In  \cite{reis6} Reisenberger showed that, solving the constraints
(\ref{ito}) in the degenerated sector---i.e., where
$e=\frac{1}{4!}\epsilon_{OPQR}B^{OP}_{\mu\nu}B^{QR}_{\rho\sigma}\epsilon^{\mu\nu\rho\sigma}=0$---and 
substituting the solution back into the Plebanski action, the theory
reduces to two sectors described by the actions
\begin{equation}\label{rei}
S^{\pm}_{deg}=\int B^{r}_i\wedge (F_i(A^r) \pm V_i^j
F_j(A^{\ell})),
\end{equation}
where the upper index $r$ (respectively $\ell$) denotes the
self-dual (respectively anti-self-dual) part of $B$ and $A$ in the
internal space, and $V \in SO(3)$. Let us concentrate on the
sector with the minus sign in the previous expression. Then it is
straightforward to define the discretized path integral along the
same lines as BF theory in Section \ref{BF}. The result is
\begin{equation}\label{Zdeg}
{\cal Z}(\Delta)=\int \prod_{f \in \Delta^*} dB^{r\va (3)}_f dv_f
\ \prod_{e \in \Delta^*} dg^{\ell}_e dg^{r}_e \ e^{i {\rm
Tr}\left[B^{r}_f U^{r}_f v_f U^{\ell - 1}_f v_f^{-1}\right]},
\end{equation}
where $dg^{\ell}_e$, $dg^{r}_e$, and $dv_f$ are defined in terms
of the $SU(2)$ Haar measure. Integrating over the $B$ field we
obtain
\begin{equation}\label{deg}Z(\Delta)= \int \prod_{e \in \Delta^*}
dg^{\ell}_e dg^{r}_e  \prod_{f \in \Delta^*} dv_f\
\delta^{(3)}(g^{r}_{e_1} \cdots g^{r}_{e_n}v_f(g^{\ell}_{e_1}
\cdots g^{\ell}_{e_n})^{-1}v_f^{-1}),
\end{equation}
where the delta function $\delta^{(3)}$ denotes an $SU(2)$
distribution.

Integration over $g^{\ell}_e$, $g^{r}_e$ and $v_f$ yields a spin
foam model where only face representations are constrained to be
simple while intertwiners are arbitrary. Explicitly
\begin{eqnarray}\label{grdeg}
Z_{deg}(\Delta)=\sum \limits_{ {\cal C}_f:\{f\} \rightarrow j_f }
\sum \limits_{{\cal C}_e:\{e\} \rightarrow \{ \iota_e \}} \
\prod_{v \in {\Delta^*}}
\begin{array}{c}
\includegraphics[width=3cm]{4simm.eps}\end{array},
\end{eqnarray}
where $\rho_f=j_f \otimes j_f^*$ and $j_f \in {\rm Irrep}[SU(2)]$.
This is precisely the spin foam obtained in  \cite{fre2}! This
model was obtained as a natural modification of the GFT that
defines a variant of the BC model. Here we have rediscovered the
model from the simplicial quantization of $S^{-}_{deg}$ defined in
(\ref{rei}). This establishes the relation of the model with a
classical action! It corresponds to spin foam quantization of the
`$-$' degenerate sector of $SO(4)$ Plebanski's theory.

The $+$ sector action (\ref{rei}) can be treated in a similar way.
The only modification is that of the subgroup. Instead of using
the diagonal insertion defined above one has to define $u\in
SU(2)\subset Spin(4)$ so that $ug=(ug^{\ell}, u^{-1}g^r)$. This
selects representations of the form $\rho= j \otimes j$ instead of
$\rho= j \otimes j^*$ for faces.

Notice that the allowed 4-simplex configurations of the model of
Section \ref{ccc} are fully contained in the set of 4-simplex
configurations of the model obtained here.

\section{The group field theory (GFT) ansatz}\label{sec:gft-sf}

In this section we present the motivation and main results of
 \cite{reis1,reis2} where the duality relation between spin foam
models and group field theories is established. This is the
formulation we referred to in Section \ref{dd} as one of the
possible discretization independent definitions of spin foam
models. The result is based on the generalization of matrix models
introduced by Boulatov \cite{bu} and Ooguri \cite{oo} dual to BF
theory in three and four dimensions respectively. This formulation
was first proposed for the Riemannian Barrett-Crane model in
 \cite{fre2} and then generalized to a wide class of spin foam
models  \cite{reis1,reis2}.

Given a spin foam model defined on an arbitrary $2$-complex ${\cal J}_{\Delta}$
(dual to a triangulation $\Delta$) --thus the partition function
$Z[{\cal J}_{\Delta}]$ is of the form (\ref{sixteen})-- there exists a GFT
such that the perturbative expansion of the field theory partition
function generalizes (\ref{sixteen}) to a sum over $2$-complexes
represented by the Feynman diagrams ${\cal J}$ of the field theory. These
diagrams look `locally' as dual to triangulations (vertices are
$5$-valent, edges are $4$-valent) but they are no longer tied to
any manifold structure \cite{pietri1}.

We now motivate the duality using what we know of the model
defined on a fixed simplicial decomposition. The action of the GFT
is of the form
\begin{equation}\label{pepito}
I[\phi]=I_0[\phi]+\frac{\lambda}{5!}{V}[\phi],
\end{equation}
where $I_0[\phi]$ is the `kinetic' term quadratic in the field and
${V}[\phi]$ denotes the interaction term. The field $\phi$ is
defined below. The expansion in $\lambda$ of the partition
function takes the form
\begin{equation}\label{exp0}
{\cal Z}=\int {\cal D}[\phi] e^{-I[\phi]}=\sum_{{\cal J}_N} \frac
{\lambda^N}{{\rm sym({\cal J}_N)}} Z[{\cal J}_N],
\end{equation}
where ${\cal J}_N$ is a Feynman diagram ($2$-complex) with $N$
vertices and $Z[{\cal J}_N]$ is the one given by (\ref{sixteen}).
In fact, the interaction term ${V}[\phi]$ is fixed uniquely
by the $4$-simplex amplitude of the simplicial model while the
kinetic term $I_0[\phi]$ is trivial as we argue below.

Expression (\ref{exp0}) is taken as the discretization (and
manifold) independent definition of the model. Transition
amplitudes between spin network states on boundary graphs
$\gamma_1$ and $\gamma_2$ are shown to be given by the correlation
functions
\begin{equation}\label{exp}
\int {\cal D}[\phi] \underbrace{\phi \cdots
\phi}_{\gamma_1}\underbrace{\phi \cdots \phi}_{\gamma_2}\
e^{-I[\phi]},
\end{equation}
where boundary graphs are determined by the arrangement of fields
in the product. The field can be defined as operators creating
four valent nodes of spin networks  \cite{mik1,a3}. In  \cite{a3}
correlation functions $\left< \phi \cdots \phi \right>$ are
interpreted as a complete family of gauge invariant observables
for quantum gravity. They encode (in principle) the physical
content of the theory and can be used to reconstruct the physical
Hilbert space in a way that mimics Wightman's procedure for
standard QFT  \cite{wi1,wi2}.

How do we construct the GFT out of the spin foam model on a fixed
$2$-complex? In (\ref{exp0}) the combinatoric of diagrams is
completely fixed by the form of the action (\ref{pepito}). To
construct the action of the GFT one starts from the spin foam
model defined on a $2$-complex ${\cal J}_{\Delta}$ dual to a
simplicial complex $\Delta$. The sum over $2$-complexes in
(\ref{exp}) contains only those $2$-complexes that locally look
like the dual of a simplicial decomposition.

Spin foams on such $2$-complexes have edges $e$ with which are
associated a tensor product of $4$ representations
$\rho_{1}\otimes\rho_{2}\otimes\rho_{3}\otimes\rho_{4}$, since they 
bound four colored faces in ${\cal J}_{\Delta}$. If we
want to think of the edge $e$ as associated to the propagator of a
field theory then such a propagator should be a map
\begin{equation}
{\bf{\cal P}}: \rho_{1}\otimes\rho_{2}\otimes\rho_{3}\otimes\rho_{4}
\rightarrow \rho_{1}\otimes\rho_{2}\otimes\rho_{3}\otimes\rho_{4}.
\end{equation}
According to Peter-Weyl theorem, elements of
$\rho_{1}\otimes\cdots\otimes\rho_{4}$ naturally appear in the
mode expansion of a function $\phi(x_1,\cdots,x_4)$ for $x_i \in G
$. Spin labels arise as `momentum' variables in the field theory.
Intertwiners assigned to edges in spin foams impose compatibility
conditions on the representations. In the context of the GFT this
is interpreted as `momentum conservation' which is guaranteed by
the requirement that the field be `translational invariant'.
\begin{equation}\label{recall}
\phi(x_1,x_2,x_3,x_4)=\phi(x_1g,x_2g,x_3g,x_4g) \ \ \ \forall \ \
g\in G.
\end{equation}
One also requires $\phi$ to be invariant under permutations of its
arguments. One can equivalently take the field $\phi$ to be
arbitrary and impose translation invariance by acting with ${\bf P}$
defined as
\begin{equation}
{\bf P }\phi(x_1,x_2,x_3,x_4)=\int dg \ \phi(x_1g,x_2g,x_3g,x_4g).
\end{equation}

All the information of local spin foams is in the vertex
amplitude or fundamental atom ($4$-simplex) amplitude of
Section \ref{sflg}. Therefore, the kinetic term $I_0[\phi]$ is 
simply given by
\begin{equation}\label{gorditas}
I_0[\phi]=\frac{1}{2}\int d^4x \ \left[{\bf P}\phi(x_1,x_2,x_3,x_4)\right]^2,
\end{equation}

Since vertices are $5$-valent in our discretization the
interaction term should contain the product of five field
operators. This is a function of $20$ group elements. If we use
the compact notation $\phi(x_i):=\phi(x_1,\cdots,x_4)$, then the
general `translation invariant' form is
\begin{equation}\label{gordita}
{V}[\phi]=\int d^{20} x \ \ {\cal V}(x_{ij}[x_{ji}]^{-1})\ \
{\bf P}\phi(x_{1i}){\bf P}\phi(x_{2i}){\bf P}\phi(x_{3i}){\bf P}\phi(x_{4i}){\bf P}\phi(x_{5i}),
\end{equation}
where $\cal V$ is a function of $10$ variables evaluated on the
`translation invariant' combinations
$\alpha_{ij}:=x_{ij}[x_{ji}]^{-1}$. For local spin foams the
function ${\cal V}(\alpha_{ij})$ is in one-to-one correspondence with the
fundamental $4$-simplex (atom) amplitude (\ref{atomamp}). 
If we represent the field
$\phi(x_1,x_2,x_3,x_4)$ by $\begin{array}{ccc}
\includegraphics[width=1cm]{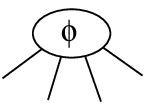}\end{array}$
we can write the action (\ref{pepito}), using (\ref{gorditas})and (\ref{gordita}),  as
\begin{eqnarray}\label{gordi}
{S}[\phi]= \int d^4 x \begin{array}{ccc}
\includegraphics[height=3cm]{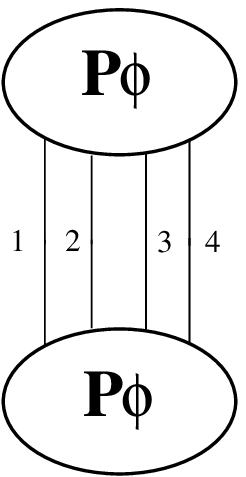}\end{array} + \frac{\lambda}{5!}\int d^{20} x \begin{array}{ccc}
\includegraphics[width=4.5cm]{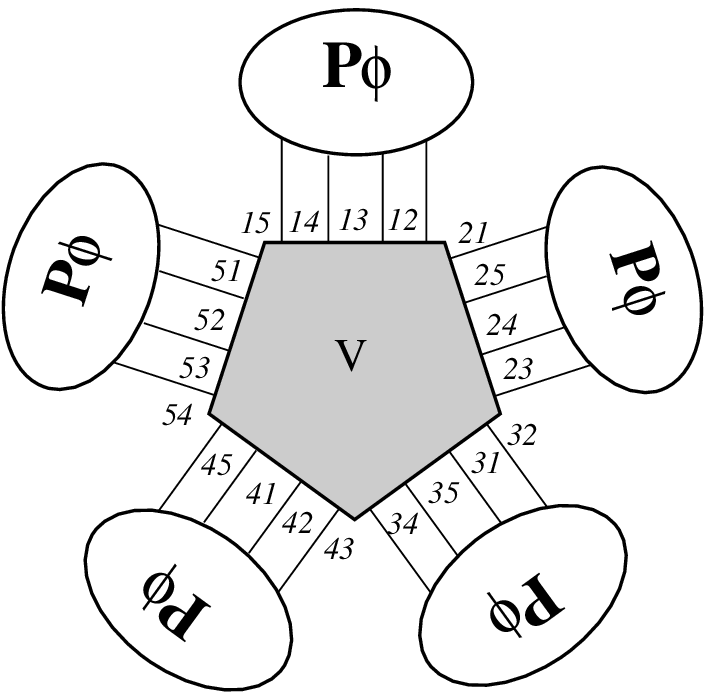}\end{array} ,
\end{eqnarray}
where the straight lines represent the field arguments 
(in the case of the interaction term, the $20$ 
corresponding $x_{ij}$ for $i\not=j=1,\cdots,5$).

The precise way in which
$2$-complexes are generated as Feynman diagrams of the GFT will be
illustrated in the following section. We have shown that any
(local in the sense of Section \ref{sflg}) spin foam model can be given a 
discretization independent formulation in terms of a uniquely determined GFT
theory.

\section{The Barrett-Crane model and its dual GFT-formulation}\label{BCGFT}

In addition to providing a discretization independent formulation
of the Barrett-Crane model, the GFT formulation provides a natural
completion of the definition of the model.  In Section \ref{BCM}
(equation (\ref{BC4})) we noticed that the model on a fixed
discretization is defined up to lower dimensional simplex
amplitudes such as that for faces $f$ (dual to triangles) and
edges $e$ (dual to tetrahedra). The GFT formulation presented here
resolves this ambiguity in a natural way. This normalization of
the Barrett-Crane model was also obtained in  \cite{ori1} using
similar techniques but on a fixed triangulation.

\subsection{The general GFT}

In this section we introduce the general GFT action that can be
specialized to define the various spin foam models described in
the rest of the paper.

\subsubsection{The field theory action and its regularization}\label{yee}

Consider the Lie group $G$ corresponding to either $Spin(4)$ or
$SL(2,\C)$ -- for the GFT dual to Riemannian or Lorentzian
Barrett-Crane model respectively. The field
$\phi(x_1,x_2,x_3,x_4)$ is denoted $\phi(x_i)$ where $i=1\dots 4$,
and $x_i \in G$, symmetric under permutations of its arguments,
i.e.,
$\phi(x_{1},x_{2},x_{3},x_{4})=\phi(x_{\sigma(1)},x_{\sigma(2)},
x_{\sigma(3)},x_{\sigma(4)})$, for $\sigma$ any permutation of
four elements. Define the projectors ${\bf P}$ and ${\bf R}$ as
\begin{equation}
{\bf P}\phi(x_i)\equiv \int \limits_G dg \phi(x_ig) \label{P}
\end{equation}
and
\begin{equation}
{\bf R}\phi(x_i)\equiv \int \limits_{U^4} du_i \phi(x_iu_i), \label{R}
\end{equation}
where $U \subset G$ is a fixed subgroup, and $dg$ and $du$ are the
corresponding invariant measures. The projector ${\bf P}$ imposes the
translation invariance property (\ref{recall}).

Different choices of the subgroup yield different interesting
GFT's. When $U=\{\mathbbm{1}\}$ (${\bf R}=\mathbbm{1}$) the GFT is dual to BF theory and we 
get the model of \cite{oo}.
The GFT is dual to the Riemannian BC model for $U=SU(2) \subset
Spin(4)$. Similarly for the GFT dual to the Lorentzian models the
subgroups $U\subset SL(2,\C)$ are $U=SU(2)$ (leaving invariant
a time-like direction) or $U=SU(1,1)\times \Z_2$ (for an invariant space-like
direction); they result in two different models.

The GFT action is is of the general form (\ref{pepito}) and simply given by
\begin{eqnarray}\label{action}
S[\phi]= \int \limits_{G^4} dx_i \left[{\bf P}\phi(x_i)\right]^2 +
\frac{\lambda}{5!} \int \limits_{G^{10}} dx_{ij}\prod
\limits^{5}_{i=1} {\bf P}{\bf R}{\bf P}\phi(x_{ij}),
\end{eqnarray}
where $i,j=1 \dots 5$, $i\not= j$ and $dx$ is an invariant measure
in $G$ (the normalized Haar measure in $G$ is compact). 

Strictly speaking the operators ${\bf P}$ and ${\bf R}$ are projectors only
when the corresponding groups are compact. Formally we have
\begin{equation}\label{factorg}
{\bf P}^2={{vol}}_G\times {\bf P},\ \ \ {\rm and}\ \ \ {\bf R}^2=vol_U\times {\bf R},
\end{equation}
where $vol_G$, and  $vol_U$ denote the volume
of $G$ and $U$, respectively. These volume factors can be taken to
be one when the $G$ and/or $U$ are compact by using Haar measures.
When $G$ and/or $U$ are non compact the factors are infinite. This
is a rather simple technical problem with which we shall deal
later. Essentially one must drop redundant projectors in the
functional integral.

The partition function can be computed as a perturbative expansion
in Feynman diagrams ${\cal J}_N$, namely
\begin{equation}\label{fevv}
{\cal Z}=\int {\cal D}\phi e^{-S[\phi]}=\sum_{{\cal J}_N}
\frac{\lambda^{N}}{{\rm sym}({\cal J}_N)} A({\cal J}_N),
\end{equation}
where $N$ is the number of vertexes in ${\cal J}_N$ and ${\rm sym}({\cal J}_N)$ is
the standard symmetry factor. 

\begin{figure}[h!]
\centerline{\includegraphics[width=15cm]{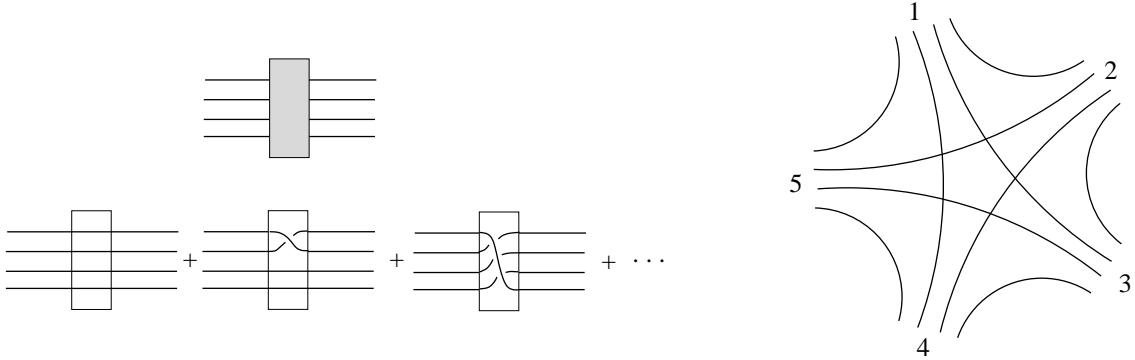}}
\bigskip \caption{The structure of the propagator and the interaction vertex.
Each line represents a delta function as the ones in the integrand of (\ref{kiki})
in the case of the propagator and (\ref{v5duebis})
in the case of the vertex. The shaded box in the propagator represents the sum over
permutations $\sigma$ in (\ref{kiki}). This sum is over 
diagrams including crossings, we have represented three of this terms in the
diagram below the propagator. }
\label{agator}
\end{figure}

In what follows we explain the structure of
the Feynman diagrams of the theory. 
If we use the notation $\phi(x_{1j})=\phi(x_{12},x_{13},x_{14},x_{15})$ we can write
the action as in (\ref{gorditas}) and (\ref{gordita}) 
\begin{eqnarray}\nonumber
    &&S[\phi] = \frac{1}{2} \int
dx_i \, dx^{\prime}_i \ \phi(x_i)\, \sK(x_i,x^{\prime}_i)\,
\phi(x^{\prime}_{i})
\ \\
&&  + \frac{\lambda}{5!} \int dx_{ij}  \ \sV(x_{ij})
~\phi(x_{1j})~\phi(x_{2j})~\phi(x_{3j})~\phi(x_{4j})~\phi(x_{5j}),
\label{actionexpanded}
\end{eqnarray}
where $i \neq j$ and the kinetic $\sK(x_i,x^{\prime}_i)$  and interaction $\sV(x_{ij})$
operators a explicitly given below. The kinetic operator $\sK(x_i,x^{\prime}_i)$  is
given by
\begin{eqnarray}
\label{propo1} &&\nonumber \sK(x_i,x^{\prime}_i)=\sum_{\sigma}\ \int
\limits_{G^2} dg^{\prime} dg^{\prime \prime}\
     \prod
\limits^4_{i=1}
\delta(x_i g^{\prime \prime} g^{\prime -1} x^{\prime -1}_{\sigma(i)})\\
&& \nonumber = \sum_{\sigma}\ \int \limits_{G^2} dg^{\prime} dg\
     \prod
\limits^4_{i=1} \delta(x_i g  x^{\prime -1}_{\sigma(i)})\\ &&
=vol_G\times \sum_{\sigma}\ \int \limits_{G} dg\
     \prod
\limits^4_{i=1} \delta(x_i g  x^{\prime -1}_{\sigma(i)}),
\end{eqnarray}
where the $g^{\prime \prime}$ and $g^{\prime}$ integrations correspond to the
action of the projectors ${\bf P}$ in (\ref{action}). Redefining the
integration variables $g= g^{\prime \prime} g^{\prime -1}$ we obtain the
second line in the previous equation. The $vol_G$ factor comes
from the $g^{\prime}$ integration as in ({\ref{factorg}}). We regularize the kinetic operator by
simply dropping one of the $G$-integrations in the previous
expression, namely
\begin{equation}\label{kiki}
\sK(x_i,x^{\prime}_i)\equiv \sum_{\sigma}\ \int \limits_{G} dg\
     \prod
\limits^4_{i=1} \delta(x_i g  x^{\prime -1}_{\sigma(i)}).
\end{equation}
${\bf P}$ acts by projecting the field $\phi$ into its `translation
invariant' part ${\bf P}\phi(x_i)$. The action (\ref{action}) depends
only on the gauge invariant part of the field, namely
$S[\phi]=S[{\bf P}\phi]$ (recall (\ref{recall})). The inverse of $\sK$
(in the subspace of right invariant fields) corresponds to itself;
therefore the propagator of the theory is simply
\begin{eqnarray}\label{p}
\sP(x_i,x^{\prime}_i)= \sK(x_i,x^{\prime}_i). \label{propo2}
\end{eqnarray}
The propagator is defined by 4 delta functions (plus the
symmetrization and the integration over the group) and it can be
represented as shown on the right diagram of Fig.\ \ref{agator}. 
The potential term (\ref{actionexpanded}) can be written as
\begin{eqnarray}
    \label{v5due}
\sV(x_{ij})&=& \frac{1}{5!}\int dg_i dg^{\prime}_i du_{ij}\ \prod_{i <
j} \delta(x^{-1}_{ji}g^{\prime}_i u_{ij}g^{-1}_ig_ju_{ji}
g^{\prime -1}_jx_{ij}),
\end{eqnarray}
where the $u_{ij}\in U$ correspond to the action of ${\bf R}$ in
(\ref{action}) and $g_i,g^{\prime}_i \in G$ to the action of the
corresponding ${\bf P}$'s, respectively. It is easy to check that in the
evaluation of a closed Feynman diagram the $g^{\prime}_i$'s can be
absorbed by redefinition of the $g_i$'s in the corresponding adjacent
propagators. In this process, the $g^{\prime}_i$'s drop out of the
integrand and each integral over $g^{\prime}_i$ gives a $vol_G$
factor. This means that the second ${\bf P}$ projector in (\ref{action}) is redundant in
the computation of (\ref{fevv}).  The regularization is analogous
to the one implemented in equation (\ref{kiki}): we drop redundant
${\bf P}$'s. The regularized vertex amplitude (for a vertex in the bulk
of a diagram) is then defined as
\begin{equation}
   \sV(x_{ij})= \frac{1}{5!}\int dg_i du_{ij}\
\prod_{i < j} \delta(x^{-1}_{ji}u_{ij}g^{-1}_ig_ju_{ji}x_{ij}).
\label{v5duebis}
\end{equation}
In the case of open diagrams $g^{\prime}$'s remain at external legs.

There is still a redundant $g_i$ integration in (\ref{v5duebis})
which introduces a potentially infinite $vol_G$ factor in the vertex
amplitude. Notice that the (\ref{v5duebis}) depends on the
`translational invariant' combinations $g^{-1}_ig_j$ so that one
of the $g$-integrations is redundant. In other words, one can absorb
one of the $g_i$ by redefining the remaining four using the 
invariance of the measure. In the non compact case this would yield
another $vol_G$ infinite factor.
The regularization now consists
of removing an arbitrary $g_i$ from the expression of the vertex
amplitude. This results in the regularization scheme proposed by
Barrett and Crane in  \cite{BC1}. The regularization presented here
can be applied to any non compact group model on a lattice and can
be regarded as a gauge fixing condition for the internal gauge
invariance (recall Section \ref{anom}). For notational simplicity
we do not implement the regularization explicitly in
(\ref{v5duebis}). The structure of the vertex is represented on
the left diagram in Fig.\ \ref{agator}. Each of the lines represent a delta function
appearing in the integrand of (\ref{v5duebis}) (compare with \ref{gordi}).

The Feynman diagrams of the theory are obtained by connecting the
$5$-valent vertexes with propagators (see Fig. \ref{agator}). At
the open ends of propagators and vertexes there are the four group
variables $x$ corresponding to the arguments of the field (in addition to the 
integration variables $g_i$ and $u_{ij}$ coming from ${\bf P}$ and ${\bf R}$
respectively).  For a
fixed permutation $\sigma$ in each propagator, one can follow the
sequence of delta functions with common arguments across vertexes
and propagators.  On a closed graph, each such sequence must
close.  By associating a surface to each such sequence of
propagators, we construct a $2$-complex $\cal J$  \cite{fre2}.  Thus, by
expanding in Feynman diagrams and in the sum over permutations in
(\ref{propo1}), we obtain a sum over 2-complexes. Each 2-complex
is given by a certain vertex-propagator topology plus a fixed
choice of a permutation on each propagator. 

\subsubsection{Evaluation in configuration space}

Combining (\ref{p}) and (\ref{v5duebis}) and integrating over
internal variables $x_{ij}$ the sequence of delta functions
associated to a face $f\in {\cal J}$ reduces to a single delta function.
Denoting $v_1 \dots v_n$ the ordered set of $n$ vertices bounding
$f$, and $e_{ij}$ the edge connecting $v_i$ with $v_j$, the delta
function corresponding to the face $f$ becomes
\begin{equation}
\delta \left( [u^{\va f}_{v_{1n}}g^{-1}_{v_{1n}} g^{}_{v_{12}}
u^{\va f}_{v_{12}}] g^{}_{e_{12}} [u^{\va f}_{v_{21}}g^{-1}_{v_{21}} \hat
g^{}_{v_{23}} u^{\va f}_{v_{23}}] g^{}_{e_{23}}\cdots g^{}_{e_{(n-1) n}} [\hat
u^{\va f}_{v_{n(n-1)}}g^{-1}_{v_{n(n-1)}} g^{}_{v_{n1}} u^{\va f}_{v_{n1}}] g^{}_{e_{n1}}
\right),\label{cara}
\end{equation}
where  $g_{e_{ij}} \in G$ is the integration variable in (\ref{kiki}) associated to the
edge $e_{ij}$, $g_{v_{ij}}\in G$ denotes the integration variable in (\ref{v5duebis})
associated with the leg $ij$ of the vertex $v_i$ connected to the edge $e_{ij}$, 
and $u^{\va f}_{v_{ij}}\in U$ is the corresponding subgroup integration 
in one of the delta functions in (\ref{v5duebis}).
We use the supra index $f$ to emphasize that the $u$'s appear on single delta functions in 
(\ref{v5duebis}) and therefore do not contribute to more than one face. 
In the previous equation, the product of group
elements between brackets correspond to the vertex contribution to
the face (see (\ref{v5duebis})) while the $g_{e_{ij}}$ comes from
the corresponding propagators (\ref{p}), also we have that
$g_{e_{ij}}=g^{-1}_{e_{ji}}$.  Using (\ref{cara}) the amplitude
$A({\cal J})$ of an arbitrary closed 2-complex becomes a multiple
integral of the form
\begin{eqnarray}\label{config}
&& \!\!\!\!\!\!A({\cal J})= 
\int \prod_{e} dg_e \prod_{v}  dg_v dg_v \\ \nonumber  && \!\!\!\!\!\! \prod_f du^{\va f}_{v}
du^{\va f}_{v}
\delta([u^{\va f}_{v_{1n}}g^{-1}_{v_{1n}} g^{}_{v_{12}}
u^{\va f}_{v_{12}}] g^{}_{e_{12}} [u^{\va f}_{v_{21}}g^{-1}_{v_{21}} \hat
g^{}_{v_{23}} u^{\va f}_{v_{23}}] g^{}_{e_{23}}\cdots g^{}_{e_{(n-1) n}} [\hat
u^{\va f}_{v_{n(n-1)}}g^{-1}_{v_{n(n-1)}} g^{}_{v_{n1}} u^{\va f}_{v_{n1}}] g^{}_{e_{n1}}).
\end{eqnarray}
Notice that if the subgroup $U=\{\mathbbm{1}\}$ the previous expression coincides
with the BF amplitude (\ref{papart}) if we make the change of variables
$g^{\prime}_{{e_{ij}}}=g_{v_{ij}}g_{e_{ij}}g^{-1}_{v_{ji}}$. For $U=\{\mathbbm{1}\}$
we recover Ooguri's GFT \cite{oo}.

\subsubsection{The spin foam representation}

The spin foam representation for the amplitude can be obtained by
expanding the delta functions in terms of irreducible unitary
representations of $G$, namely
\begin{equation}\label{vani}
\delta(g)=\sum_{\rho} \Delta_{\rho} \ {\rm
Tr}\left[{\rho}(g)\right],
\end{equation}
where $\rho$ labels unitary irreducible representations and the
rest of the notation is that of (\ref{deltarep}) when $G$ is
compact. In the non compact case representations are
infinite-dimensional so a formally equivalent expression holds
where $\Delta_{\rho}$ correspond to the so-called Pancharel
measure \cite{a8}.

Using (\ref{vani}) in (\ref{cara}) the face contribution is
\begin{eqnarray}\label{l1}
&& \nonumber \!\!\!\!\!\!\!\!\!\!\!\!\sum_{\rho} \Delta_{\rho} \
{\rm Tr}\left[ {\rho}(u^{\va f}_{v_{1n}})\cdot \rho(g^{-1}_{v_{1n}}
g^{}_{v_{12}}) \cdot \rho(u^{\va f}_{v_{12}})\cdot
{\rho}(g^{}_{e_{12}})\cdot {\rho}(u^{\va f}_{v_{21}})\cdot
\rho(g^{-1}_{v_{21}} g^{}_{v_{23}})\cdot\right. \\ &&
\!\!\!\!\!\!\!\!\!\!\!\!\left.\rho( u^{\va f}_{v_{23}})\cdot
{\rho}(g^{}_{e_{23}})\cdots {\rho}(u^{\va f}_{v_{n(n-1)}})\cdot
\rho(g^{-1}_{v_{n(n-1)}} g^{}_{v_{n1}})\cdot \rho( u^{\va f}_{v_{n1}})\cdot
{\rho}(g^{}_{e_{n1}})\right],
\end{eqnarray}
where we have dropped the integration symbols for simplicity.
The $u$'s appear just once per face so we can perform the
$u$-integrations independently of other faces. Notice that $\int
du \rho(u)$ is the projection onto the invariant subspace under
the action of $U$ in the Hilbert space ${\cal H}_{\rho}$. When
$U=\{\mathbbm{1}\}$ the projector is the identity. In the other cases the
subspace turns out to be 1-dimensional. Consequently the projector
can be written as
\begin{equation} \label{poi}
\int \limits_U du \ \rho(u)= \left|w_{\va \rho^{}} \right> \left<
w_{\va \rho^{}} \right|,
\end{equation}
where  $\left|w_{\va \rho^{}} \right>$ is the corresponding
invariant vector. Equation (\ref{l1}) becomes
\begin{eqnarray}\label{l1:5}\sum_{\rho} \Delta_{\rho} \
\left< w_{\va \rho^{}} \right| \rho(g^{-1}_{v_{1n}} 
g^{}_{v_{12}}) \left. w_{\va \rho^{}} \right> \left< w_{\va
\rho^{}} \right| {\rho}(g^{}_{e_{12}})\left. w_{\va \rho^{}}
\right>\left< w_{\va \rho^{}} \right|
\rho(g^{-1}_{v_{21}} g^{}_{v_{23}})\left. w_{\va \rho^{}} \right> \cdots  \left< w_{\va \rho^{}}
\right| {\rho}(g^{}_{e_{n1}})\left.w_{\va \rho^{}} \right>.
\end{eqnarray}
There are two types of terms in the previous equation: those corresponding to 
edges for which the representation $\rho$ is evaluated at the single $g_{e_{ij}}$
group variable, and those corresponding to vertices for which $\rho$ is evaluated
at the product $g^{-1}_{v_{ij}}g_{v_{ik}}$.
Let us integrate over the edge variables first. The group element 
$g_{e}$ associated to an edge $e\in {\cal J}$ appears
four times as there are four delta functions in the propagator
(\ref{p}). Therefore, integrals over such $g$'s have the general form
\begin{equation}
A_e(\rho^1_e,\dots, \rho^4_e )= \int \limits_{G} \left< w_{\va
\rho^{1}} \right|{\rho^{1}}(g)\left. w_{\va \rho^{1}} \right>
\left< w_{\va \rho^{2}} \right|{\rho^{2}}(g)\left. w_{\va
\rho^{2}} \right> \left< w_{\va \rho^{3}}
\right|{\rho^{3}}(g)\left. w_{\va \rho^{3}} \right> \left< w_{\va
\rho^{4}} \right|{\rho^{4}}(g)\left.w_{\va \rho^{4}} \right>,
\end{equation}
and define the edge amplitude. If we define the {\em kernel}
$K_{\rho}(g)$ as
\begin{equation}\label{kernel}
K_{\rho}(g)\equiv\left< w_{\va \rho^{}} \right|{\rho}(g)\left.
w_{\va \rho^{}} \right>,
\end{equation}
then using the subgroup invariance of the $\left| w_{\va \rho^{}}
\right>$'s the previous equation becomes
\begin{equation}\label{eg}
A_e(\rho_1,\dots, \rho_4 )= vol_U \times \int \limits_{G/U} dy\
K_{\rho_1}(y) K_{\rho_2}(y) K_{\rho_3}(y) K_{\rho_4}(y),
\end{equation}
where the integration is over the homogeneous space $G/U$.
Integration over the group elements associated to vertices are of
the general form
\begin{eqnarray}\label{vg}
&& \nonumber A_v(\rho_{ik} ) =  \int
\limits_{(G/U)^4} \prod \limits_{i=1}^5 dy_i \ \prod \limits_{i<k}
K_{\va \rho_{ik}}(y_i,y_k).
\end{eqnarray}
This corresponds to the vertex amplitude. When $U$ is non compact
the preceding expressions have to be regularized as in Subsection \ref{yee}
by dropping an arbitrary $y$-integration. The
amplitude of a 2-complex ${\cal J}$ is then given by
\begin{equation}\label{genexp}
A({\cal J})=\sum \limits_{{\cal C}_f:\{f\}\rightarrow \{\rho\}}
\prod_{f\in {\cal J}} \Delta_{\rho_f} \prod_{e \in {\cal J}}A_e(\rho_1,\dots,
\rho_4 ) \prod_{v\in {\cal J}}A_v(\rho_1,\dots, \rho_{10} )
\end{equation}
As we shall see in the following the presence of ${\bf R}$ in
(\ref{action}) can be interpreted as imposing the Barrett-Crane
constraints on the GFT dual to BF theory.

\subsubsection{Boundaries}\label{bbbb}

The amplitude of an open diagram, that is, a diagram with a
boundary, is a function of the variables on the boundary, as for
conventional QFT Feynman diagrams. The boundary of a 2-complex is
given by a graph. To start with, the amplitude of the open diagram
is a function of 4 group arguments per each external leg. However,
consider a surface of an open 2-complex and the link $ab$ of the
boundary graph that bounds it.  Let $a$ and $b$ be the nodes on
the boundary graph that bounds $ab$.  The surface determines a
sequence of delta functions that starts with one of the group
elements in $a$, say $g_{a}$, and ends with one of the group
elements in $b$, say $g_{b}$. By integrating internal variables,
all these delta functions can be contracted to a single one of the
form $\delta(g_{a} \cdots
g_{b}^{-1})=\delta(g_{b}^{-1}g_{a}\cdots)$. We can thus define the
group variable $\rho_{ab}=g_{b}^{-1}g_{a}$, naturally associated
to the link $ab$, and conclude that the amplitude of an open
2-complex is a function $A(\rho_{ab})$ of one group element per
each link of its boundary graph. $A(\rho_{ab})$ is gauge invariant
as can be easily checked, and $\rho_{ab}$ represents the 
discrete boundary connection in the language of Section \ref{sflg}.

In ``momentum space", boundary degrees of freedom are encoded in
spin-network states. That is, if $s$ is a spin network given by a
coloring of the boundary graph, then
\begin{equation}\label{spnet}
A(s)=\int d\rho_{ab} \ \bar \psi_{s}(\rho_{ab})\ A(\rho_{ab}),
\end{equation}
where $\psi_{s}(\rho_{ab})$ is the spin network function
 \cite{c4,baez10}. The formula can be inverted
\begin{equation}
A(\rho_{ab})=\sum_s  A(s) \psi_{s}(\rho_{ab}),
\end{equation}
where the sum is performed over all spin network states defined on
the given boundary graph. The argument presented here applies in general for
any GFT of the kind introduced in Section \ref{sec:gft-sf}.
In the case of the GFT's defined here $\rho_{ab}=g_{b}^{-1}g_{a}$ will
appear between $U$-integrations. So strictly speaking, for each $\rho_{ab}=g_{b}^{-1}g_{a}$
there correspond two point $y_a$ and $y_b$ in the homogeneous space $G/U$
and one cannot say that the amplitude depend on the $G$-connection.

\subsection{GFT dual to the Riemannian  Barret-Crane model\label{EU}}

The representation theory of $Spin(4)$ is particularly simple due
to the fact that $Spin(4)=SU(2)\times SU(2)$. Unitary irreducible
representations of $\rho$ of $Spin(4)$ are labeled by two
half-integers $j_{l}$ and $j_r$, and are given by the tensor
product of unitary irreducible representations of SU(2), namely
\begin{equation}
\rho_{j_lj_r}=j_l\otimes j_r.
\end{equation}
In terms of representation matrices we have
\begin{equation}
R^{j_lj_r}(g)_{m m^{\prime}q q^{\prime}}=D^{j_l}_{m
m^{\prime}}(g_r) D^{j_r}_{q q^{\prime}}(g_l),
\end{equation}
where $g=(g_l,g_r)\in Spin(4)$, and $D^{j_l}_{m m^{\prime}}$ are
$SU(2)$-representation matrices. The subgroup $U$ of previous
section is taken as $U=SU(2)\subset Spin(4)$ defined by the
diagonal action on $Spin(4)$ as follows
\begin{equation}\label{chico}
gu \equiv (g_lu,g_ru) \ \ \ {\rm for}\ \ \ u \in U.
\end{equation}
The projector (\ref{poi}) can be explicitly evaluated
\begin{equation} \label{pois}
\int \limits_U du \ \rho_{j_lj_r}(u)= \int du \ D^{j_l}_{m
m^{\prime}}(u) D^{j_r}_{q q^{\prime}}(u)= \frac{\delta_{j_l
j^*_r}}{2j_l+1} \delta_{m q} \delta_{m^{\prime}
q^{\prime}}=\left|w_{\va j_l} \right> \left< w_{\va j_l} \right|,
\end{equation}
where we have used the orthonormality of $SU(2)$ representation
matrices (Footnote \ref{lejos}). The previous equation confirms
that the invariant subspace for each $\rho_{j_lj_r}$ is
1-dimensional as anticipated above. An immediate consequence of
(\ref{pois}) is that the kernel $K_{\rho}$ in (\ref{kernel})
becomes
\begin{equation}\label{matrix}
K_{j}(g)=\left< w_{\va j} \right|{\rho_{j^{\prime}
j^{}}}(g)\left.w_{\va j} \right>=\frac{\delta_{j^{\prime}
j^*}}{{2j+1}} {\rm Tr}\left[D^j(g_lg_r) \right].
\end{equation}
The factor $\delta_{j^{\prime} j}$ restricts the representation to
the simple representations $j_l=j^*_r=j$. 

The projector ${\bf R}$ in
(\ref{action}) imposes the Barrett-Crane simplicity constraints of 
equation (\ref{XX1})!. Simple representations appear in the harmonic analysis of functions on
the 3-sphere, $S_3$. The fact that we encounter them here is not
surprising since $S_3$ is the homogeneous space
$S_3=Spin(4)/SU(2)$ under the $SU(2)$ diagonal insertion
(\ref{chico}). The presence of $R$ in (\ref{action}) projects out
those modes that are not simple or {\em spherical}. Indeed, $\left< w_{\va
j} \right|{\rho_{j^{\prime} j^{*}}}(g)\left.w_{\va j} \right>$ can
be thought of as a function on $S_3$: $\left< w_{\va j}
\right|{\rho_{j^{\prime} j^{*}}}(g)\left.w_{\va j} \right>$
depends on the product $g_lg_r\in SU(2)$ which is isomorphic to
$S_3$. We will see below that in fact ${\bf R}$ imposes all the
constraints of Section \ref{dpc} fixing in addition the value
of the so far undetermined edge amplitude $A_e$.

The invariant measure on $SU(2)$ can be written as a
measure on $S_3$ induced by the isomorphism $SU(2)\rightarrow
S_3$. We can parameterize $h\in SU(2)$ as $h=y_{\va (h)}^{\mu}
\tau_{\mu}$ where $\tau_k=i\sigma_k$ for $k=1,2,3$, $\{
\sigma_{k}\}$ is the set of Pauli matrices, and $\tau_0= 1$. $h\in
SU(2)$ implies $y_{\va (h)}^{\mu}y_{{\va (h)}{\mu}}=1$ (indexes
are lowered and raised with $\delta^{\mu}_{\nu}$), i.e., $y \in
S_3$. In this parameterization the $SU(2)$ Haar measure becomes
\begin{equation}
dh \rightarrow dy= \frac{1}{\pi^2}dy^{\va 4}
\delta(y^{\mu}y_{\mu}-1),
\end{equation}
We can simplify the measure using spherical coordinates for which
an arbitrary point $y\in S^{3}$ is written
\begin{equation}\label{sfefe}
y=\left( {\rm cos}(\psi ),{\rm sin}(\psi ){\rm sin}(\theta ){\rm
cos}(\phi ), {\rm sin}(\psi ){\rm sin}(\theta ){\rm sin}(\phi
),{\rm sin}(\psi ){\rm cos} (\theta )\right) ,
\end{equation}
where $0 \le \psi \le \pi $, $0 \le \theta \le \pi $, and $0\le
\phi \le 2\pi $. The Haar measure becomes
\begin{equation}
dy=\frac{1}{2\pi^2 }\ {\rm sin}^2(\psi )\, {\rm sin}(\theta )\
d\psi\, d\theta \, d\phi . \label{mea}
\end{equation}
Using the known formula for the trace of $SU(2)$ representation
matrices it is easy to check that (\ref{matrix}) becomes
\begin{equation}\label{kerker}
K_{j}(\psi_y)=\frac{\delta_{j^{\prime} j}}{{2j+1}} \frac{{\rm
sin}((2j+1)\psi_y)}{{\rm sin}(\psi_y)},
\end{equation}
where $\psi_h$ is the value of the coordinate $\psi$ in
(\ref{sfefe}) corresponding to $y \in SU(2)$. Notice that we have
rediscovered the kernel (\ref{kernito}) which together
with equation (\ref{intform}) establishes the claimed duality
between the Barrett-Crane model and the GFT of this section.

Now, the general edge and vertex amplitudes ((\ref{eg})
and (\ref{vg})) reduce to multiple integrals over $S_3$. These can
be interpreted as the evaluation of diagrams on the sphere
with propagator given by (\ref{kerker}). This evaluation is referred to as 
{\em relativistic spin network evaluation} in the literature. In the case of the
edge amplitude the evaluation corresponds to a $\theta_4$ graph
with four links labeled by the four spins $j_1,\dots, j_4$. 
In the case of the vertex amplitude the evaluation corresponds to the
$\gamma_5$ spin network of Figure \ref{chunk} for which
the ten links are labeled by the corresponding ten spins 
$j_1,\dots, j_{10}$. There is no intertwiner label at nodes as
the projection ${\bf R}$ fixes the intertwiners to the
unique Barrett-Crane one. 

The amplitude $A_e(j_1,\dots, j_4 )$ can be
computed using (\ref{matrix}) and (\ref{3dp}), it follows that
\begin{equation}\label{eo4}
A_e(j_1,\dots, j_4 )= \frac{{\rm dim}\left[{\rm Inv}({j_1} \otimes 
{j_2} \otimes {j_3} \otimes {j_4} )\right]}
                          {{\rm dim}\left[{j_1} \otimes {j_2} \otimes {j_3}
\otimes {j_4} \right]},
\end{equation}
where ${\rm dim}\left[{\rm Inv}({j_1} \otimes {j_2} \otimes {j_3}
\otimes {j_4} )\right]$ denotes the dimension of the trivial
component of ${\cal H}_{j_1}\otimes {\cal H}_{j_2} \otimes {\cal
H}_{j_3} \otimes {\cal H}_{j_4}$ and ${\rm dim}\left[{j_1} \otimes
{j_2} \otimes {j_3} \otimes {j_4} \right]$ the dimension of the
full space\footnote{There was a mistake in the computation of
the edge amplitude in  \cite{a10} due to the propagation of a typo
in  \cite{fre2}. The erroneous expression of the edge amplitude
contained ${\rm dim}\left[{j_1} \otimes {j_2} \otimes {j_3}
\otimes {j_4} \right]^2$ in the denominator.}. So finally,
\begin{equation}\label{ess1}
A({\cal J})=\sum \limits_{{\cal C}_f:\{f\}\rightarrow \{j\}} \prod_{f\in
{\cal J}} (2j_f+1)^2 \prod_{e \in {\cal J}} A_e(j_1,\dots, j_4 ) \prod_{v\in
{\cal J}}A_v(j_1,\dots, j_{10} ),
\end{equation}
where the vertex amplitude is given by (\ref{vg}) using
(\ref{kerker}). This is the Barrett-Crane model (\ref{BC4}) where
the values of $A_e$ and $A_f$ have been fixed by the GFT
formulation. 

\subsubsection{Finiteness}

In this section we prove that the infinite spin sum in (\ref{ess1}) is convergent 
if the $2$-complex ${\cal J}$ is finite and non-degenerate. In order to do this we 
need to construct bounds for the terms in the sum (edge and vertex amplitudes).
These bounds are provided by the following lemmas.  

\begin{lem} \cite{a7} \label{le}
For any subset of $\kappa$ elements $j_1 \dots j_{\kappa}$ out of
the corresponding four representations appearing in
$A_e(j_1,\dots, j_4 )$, the following bounds hold:
\begin{equation}\nonumber \left| A_e(j_1,\dots, j_4 )
\right| \le \frac {1} {\prod \limits^{\kappa}_{i=1} (2j_i+1)^{\alpha_{\kappa}}}, \ \ \ {where} \ \ \
\alpha_{\kappa}=\left\{ \begin{array}{ccc} 1\ \ { for}\ \ \kappa
\le 3
\\
\frac{3}{4}\ \ { for}\ \  {k=4}
\end{array}\right..
\end{equation}
The inequality for $\kappa \le 3$ is sharp.
\end{lem}

The proof of this lemma is elementary 
and can be found in \cite{a7}.

\begin{lem} \cite{a7}\label{lv}
For any $4$ spins $j_1 \dots j_{4}$ labeling links converging at
the same node in the relativistic spin-network corresponding to
$A_v(j_1,\dots, j_{10} )$ the following bounds hold:
\begin{equation}\nonumber \left| A_v(j_1,\dots, j_{10} )
\right| \le \frac {1} {(2j_1+1)(2j_2+1)(2j_3+1)(2j_4+1)},
\end{equation}
from which follows
\begin{equation}\nonumber \left| A_v(j_1,\dots, j_{10} )
\right| \le \frac {1} {((2j_1+1)\dots(2j_{10}+1))^{2/5}}.
\end{equation}
\end{lem}
\begin{dfi}
A 2-complex ${\cal J}$ is said to be degenerate if it contains some faces
bounded by only one or two edges.
\end{dfi}
\begin{thm} \cite{a7}
The state sum $A({\cal J})$, (\ref{ess1}), converges for any non degenerate
2-complex ${\cal J}$.
\end{thm}
\begin{proof}
The amplitude (\ref{ess1}) can be bounded in the following way:
\begin{equation}\label{lala}
|A({\cal J})| \le \prod_{f \in {\cal J}}  \sum_{j_f} \left( (2j_f+1)
\right)^{2-\frac{3}{4}n_f-\frac{4}{10}n_f}=\prod_{f \in {\cal J}}
\sum_{j_f} \left( (2j_f+1) \right)^{2-\frac{23}{20}n_f},
\end{equation}
where $n_f$ denotes the number of edges of the corresponding face,
and we have used the fact the number of edges equals the number of
vertices in a face of ${\cal J}$. The term $(2j+1)^{2}$ in (\ref{lala})
comes from the face amplitude, $(2j+1)^{-\frac{3}{4} n_f}$ from
Lemma \ref{le} ($\kappa=4$), and $(2j+1)^{-\frac{4}{10} n_f}$ from
Lemma (\ref{lv}) $\kappa=10$. Notice that if the 2-complex
contains faces with more than two edges the previous bound for the
amplitude is finite, since the sum on the RHS of the previous
equation converges for $n_f>2$.
\end{proof}

The fact that the sum over spins converges for a fixed (non degenerate) $2$-complex is
an very encouraging result. It means that for a fixed $2$-complex the sum over
the infinite set of $4$-geometry configurations, represented by the corresponding spin foams,
is well defined. 
Even though the result was obtained in the context of the GFT
formulation, it can be applied to the model defined on a simplicial decomposition of the 
space-time manifold. This is because (\ref{ess1}) refers to the amplitude of a single Feynman diagram.
In this context, the result is very encouraging and appears as a first
necessary step for the construction of the discretization independent generalized projector and physical 
scalar product by studying the refinement limit of Section \ref{dd}.
On the other hand, if the discretization independent formulation is to be obtained directly by means
of the GFT formulation, this result means that the perturbative expansion in $\lambda$
(\ref{exp0}) is finite order by order. The previous statement has not been completely proved
as degenerate $2$-complexes---for which our proof does not hold---will appear in the $\lambda$-expansion.
However, we expect these diagrams to be finite also. A proof of this is lacking at 
the moment but it should be feasible by strengthening the bounds 
we have used in this section.

\subsection{Lorentzian models}\label{LO}

Unitary irreducible representations of $SL(2,\C)$ appearing in the
general expression (\ref{vani}) are the ones in the so-called {\em
principal} series. They are labeled by a natural number $n$ and a
positive real number $\rho$. Unitary irreducible representations
of $SL(2,\C)$ are infinite dimensional. Those in the principal
series are defined by their action on the linear space ${\cal
D}_{n \rho}$ of homogeneous functions of degree
$(\frac{n+i\rho}{2}-1,\frac{-n+i\rho}{2}-1)$ of two complex
variables $z_1$ and $z_2$. Due to the homogeneity properties of
the elements ${\cal D}_{n \rho}$ they can be characterized by
giving their value on the sphere $|z_1|^2+|z_2|^2=1$.
The latter is isomorphic to $SU(2)$. 
The so called canonical basis is defined in terms of functions on $SU(2)$ and is
well suited for studying the following model. The relevant facts
about $SL(2,\C)$ representation theory and the notation used in
this section can be found in the appendix of reference  \cite{a8}
(for all about $SL(2,\C)$ representation theory see
 \cite{gelfand}).

\subsubsection{GFT dual to the Lorentzian Barrett-Crane model}

Equation (\ref{poi}) becomes, in this case,
\begin{equation} \label{poiss}
\int \limits_U du \ \rho_{n \rho}(u)=\int \limits_{SU(2)} du
D^{n\rho}_{\ell m \ell^{\prime} m^{\prime}}(u) =\delta_{n
0}\delta_{\ell 0}\delta_{\ell^{\prime} 0}=\left|w_{\va n \rho}
\right> \left< w_{\va n \rho} \right|,
\end{equation}
in terms of the canonical basis \cite{a8}. The
$SU(2)$ invariant vectors $\left|w_{\va n \rho}\right>$ are given
by $\left<z_1,z_2\right.\left|w_{\va n
\rho}\right>=\delta_{n0}(|z_1|^2+|z_2|^2)^{\frac{i}{2}\rho -1}$
which is indeed a homogeneous function of degree
$(\frac{i\rho}{2}-1,\frac{i\rho}{2}-1)$.

An immediate consequence of (\ref{poiss}) is that the kernel
$K_{\rho}$ in (\ref{kernel}) becomes
\begin{equation}\label{matrixs}
K_{n \rho}(g)=\left< w_{n \rho} \right|{\rho_{n \rho}}(g)\left.
w_{\va n \rho} \right>=\delta_{n0}D^{n\rho}_{0000}(u).
\end{equation}

The kernel $\left< w_{\va 0 \rho} \right|{\rho_{0
\rho}}(g)\left.w_{\va 0 \rho} \right>$ can be thought of as a
function on $H^{+}=SL(2,\C)/SU(2)$ (the upper sheet of $2$-sheeted
hyperboloid in Minkowski space). An arbitrary point $y\in H^{+}$
can be written in hyperbolic coordinates as
\begin{equation}
y=\left( {\rm cosh}(\eta ),{\rm sinh}(\eta ){\rm sin}(\theta ){\rm
cos}(\phi ), {\rm sinh}(\eta ){\rm sin}(\theta ){\rm sin}(\phi
),{\rm sinh}(\eta ){\rm cos} (\theta )\right) ,
\end{equation}
where $0 \le \eta \le \infty $, $0 \le \theta \le \pi $, and $0\le
\phi \le 2\pi $. The invariant measure in these coordinates is
\begin{equation}
dy=\frac{1}{2\pi^2 }\ {\rm sinh}^2(\eta )\, {\rm sin}(\theta )\
d\eta\, d\theta \, d\phi . \label{mea}
\end{equation}
In these coordinates, (\ref{matrixs}) becomes
\begin{equation}\label{kerkerll}
K_{{\rho}}(\eta_y)=\frac{2}{{\rho}} \frac{{\rm sin}(\frac{1}{2}\rho
\eta_y)}{{\rm sinh}(\eta_y)},
\end{equation}
where $\eta_y$ is the value of the coordinate $\eta$ corresponding
to $y\in H^{+}$ (this the generalization of (\ref{kernito})
proposed in  \cite{BC1}). Finally, the amplitude of an arbitrary
diagram (\ref{genexp}) becomes
\begin{equation}\label{ess2}
A(J)=\int \limits_{{\cal C}_f:\{f\}\rightarrow \{{\rho}\}}
\prod_{f\in J} \rho_f^2 d\rho_f \prod_{e \in J} A_e( \{{\rho}_e\}
) \prod_{v\in J} A_v(\{{\rho}_v\}),
\end{equation}
where the formal sum in the general expression (\ref{genexp})
becomes a multiple integral over the coloring ${\cal
C}_f:\{f\}\rightarrow \{{\rho}\}$ of faces, and the weight
$\Delta_{\rho}$ now is given by the Pancherel measure, $\rho^2
d\rho$, of $SL(2,\C)$. The vertex amplitude was proposed in
 \cite{BC1}, the full state sum amplitude including 
the previous normalization was obtained in \cite{a9}.

\subsubsection{Finiteness}

The amplitude of a non degenerate 2-complex turns out to be finite as
in the Euclidean case. We state and prove the main result. Some of
the following lemmas are stated without proof; the reader
interested in the details is referred to the references.
As in the Riemannian case the main idea is to construct appropriate bounds
for the fundamental amplitudes. 

\begin{lem} \cite{a2,a22}\label{four}
For any subset of $\kappa$ elements $\rho_1 \dots \rho_{\kappa}$
out of the corresponding four representations appearing in
$A_e({\rho}_1,\dots, {\rho}_4 )$, the following bounds hold:
\begin{equation}\nonumber \left|A_e({\rho}_1,\dots, {\rho}_4 )
\right| \le \frac {C_{\kappa}} {\left( \prod
\limits^{\kappa}_{i=1} \rho_i \right)^{\alpha_{\kappa}}}, \ \ \
{where} \ \ \ \alpha_{\kappa}=\left\{ \begin{array}{ccc} 1\ \ {
for}\ \ \kappa \le 3
\\
\frac{3}{4}\ \ { for}\ \  {k=4}
\end{array}\right..
\end{equation}
for some positive constant $C_{\kappa}$.
\end{lem}

\begin{lem}\label{bound} \cite{baez3} $A_e({\rho}_1,\dots, {\rho}_4 )$ and
$A_v({\rho}_1,\dots, {\rho}_{10} )$ are bounded by a constant
independent of the $\rho$'s.
\end{lem}

\begin{lem} \cite{a2,a22} \label{ten} For any subset of $\kappa$ elements
$\rho_1 \dots \rho_{\kappa}$ out of the corresponding ten
representations appearing in $A_v({\rho}_1,\dots, {\rho}_{10} )$
the following bounds hold:
\begin{equation}\nonumber \left| A_v({\rho}_1,\dots, {\rho}_{10} ) \right| \le
\frac {K_{\kappa}} {\left( \prod \limits^{\kappa}_{i=1} \rho_i
\right)^{\frac 3 {10}}}
\end{equation}
for some positive constant $K_{\kappa}$.
\end{lem}

\begin{thm} \cite{a2,a22}
The state sum $A({\cal J})$, (\ref{ess2}), converges for any non degenerate
2-complex ${\cal J}$.
\end{thm}
\begin{proof}
We divide each integration region ${\mathbb R}^+$ into the
intervals $[0,1)$, and $[1,\infty)$ so that the multiple integral
decomposes into a finite sum of integrations of the following
types:
\begin{enumerate}
\item[i.] All the integrations are in the range $[0,1)$.  We denote this
term $T(F,0)$, where $F$ is the number of 2-simplexes in the
triangulation.  This term in the sum is finite by Lemma
{\ref{bound}}.
\item[ii.] All the integrations are in the range
$[1,\infty)$.  This term $T(0,F)$ is also finite since, using
Lemmas \ref{four} and {\ref{ten}} for $\kappa=4$, and $\kappa=10$
respectively, we have
\[
T(0,F) \le \prod_{f} \int \limits^{\infty}_{\rho_{f}=1} d \rho_f \
\ \rho_{f}^{2-\frac3 4 {n_e} - \frac3 {10} {n_v}} \le \left(\ \int
\limits^{\infty}_{\rho_{f}=1} d \rho_f \ \
\rho_{f}^{-\frac{46}{40}} \right)^F < \infty.
\]
\item[iii.]  $m$ integrations in $[0,1)$, and $F-m$ in $[1,\infty)$.
In this case $T(m,F-m)$ can be bounded using Lemmas (\ref{four})
and (\ref{ten}) as before.  The idea is to choose the appropriate
subset of representations in the bounds (and the corresponding
values of $\kappa$) so that only the $m-F$ representations
integrated over $[1,\infty)$ appear in the corresponding
denominators.  Since this is clearly possible, the $T(m,F)$ terms
are all finite.
\end{enumerate}
We have bounded ${\cal Z}$ by a finite sum of finite terms which
concludes the proof.
\end{proof}

The `scaling' behavior of amplitudes in the model is very similar to that of 
the Riemannian version of the previous section. If one makes the substitution
$(2j+1)=-i \rho$ and $\psi_y=i \eta_y$ in the Riemannian expression for
the kernel (\ref{kerker}) one obtains the Lorentzian expression 
(\ref{kerkerll}). In addition, with the same substitution the face amplitude
$(2j+1)^2$ in (\ref{ess1}) takes the form $\rho^2$ of the face amplitude 
in (\ref{ess2}). From this viewpoint the finiteness of the Lorentzian
model is intimately related to the finiteness of the Riemannian counterpart.

The non compactness of $SL(2,\C)$ introduces potential
divergences due to the fact that $vol_{SL(2,\C)}=\infty$.
This problem can be dealt with in the GFT formulation by using the
appropriate gauge fixing conditions as explained in Section \ref{BCGFT}.
If we define the model on a simplicial decomposition the same 
gauge fixing can be implemented.

\subsubsection{A new Lorentzian model\label{LOR2}}

In the Euclidean case there was only one way of selecting the
subgroup $U$ of group elements leaving invariant a fixed direction
in Euclidean space-time. In the Lorentzian case there are two
possibilities. The case in which that direction is space-like was
treated in the previous section. When the direction is time-like
the relevant subgroup is $U=SU(1,1)\times \Z_2$.

This case is more complicated due to the non-compactness of $U$.
Consequently, one has to deal with additional infinite volume
factors of the form $U$-volume in (\ref{factorg}). Another
consequence is that the invariant vectors defined in (\ref{poi})
are now distributional and therefore no longer normalizable. All
this makes more difficult the convergence analysis performed in
the previous models and the issue of finiteness remains open.

On the other hand, the model is very attractive as its state sum
representation contains simple representations in both the
continuous and discrete series. As pointed out in  \cite{BC1} and
discussed in the following section one would expect both types of
simple representations to appear in a model of Lorentzian quantum
gravity.

The difficulties introduced by the non-compactness of $U$ make the
calculation of the relevant kernels (\ref{kernel}) more involved.
No explicit formulas are known and they are defined by integral
expressions. We will not derive these expressions here. A complete
derivation can be found in  \cite{a8}. The idea is to use harmonic
analysis on the homogeneous space $SL(2,\C)/SU(1,1)\times \Z_2$
which can be realized as the one-sheeted hyperboloid
$y^{\mu}y^{\nu}\eta_{\mu \nu}=-1$ where $y$ and $-y$ are
identified (imaginary Lobachevskian space, from now on denoted
$H^{-}$). The kernels correspond to eigenfunctions of the massless
wave equation on that space.

The corresponding kernel is given by the following expression:
\begin{eqnarray}\label{carro}
\nonumber K_{\rho,n}(x,y)&=&\int \limits_{C^{+}} d\omega
(\delta_{n,0} \left|y^{\nu} \xi_{\nu} \right|^{i\frac{\rho}2-1}
\left|x^{\nu} \xi_{\nu} \right|^{-i\frac{\rho}2-1} \\ &+&
\delta(\rho)\delta_{n,4k}\frac{32\, \pi e^{-2i k \left[\Theta(x,y)
\right]}}{k} \ \delta(x^{\nu} \xi_{\nu}) \delta(y^{\nu}
\xi_{\nu})),
\end{eqnarray}
where $x,y \in H^{-}$ and $\xi \in C^+$ is a normalized future
pointing null vector in Minkowski space-time. The integration is
performed on the unit sphere defined by these vectors with the
standard invariant measure $d\omega$.

As in the previous cases the expressions for the edge and vertex
amplitudes (equations (\ref{eg}) and (\ref{vg})) reduce to
integrals on the hyperboloid $H^{-}$. The expression for the
amplitude (\ref{genexp}) of an arbitrary diagram is  \cite{a8}
\begin{eqnarray}\label{momodel}
  && \nonumber  A(J)=\sum_{n_f}\int \limits_{\rho_f} d\rho_f \prod_{{f }}\
(\rho_f^2 + n^2_f)  \
    \prod_{{e }} \ A_e( \{ n_e\},\{\rho_{e}\}) \
   \prod_{{ v }} \ A_v( \{ n_v\},\{\rho_{v}\}),
\end{eqnarray}
where now there is a summation over the discrete representations
$n$ and an multiple integral over the continuous representations
$\rho$. For the discrete representations the triangle area
spectrum takes the form (\ref{super}) if we define $k=j+1/2$. The
finiteness properties of this model have not been studied so far.
There is a relative minus sign between the continuous and discrete
eigenvalues of the area squared operators that has been
interpreted as providing a notion of micro causality in  \cite{a8}.
It would be interesting to study this in connection with the
models of Section \ref{fotin}.

\section{Discussion}

We conclude this article with some remarks and a discussion of
recent results and future perspectives in the subject. As we have devoted
the second part of this paper almost exclusively to the Barrett-Crane model
we start by discussing open issues in this context. We go to more general open
questions toward the end of this section. 

\vskip.5cm
\noindent $1$. {\em Normalization}: The Barrett-Crane model is certainly the most well studied 
spin foam model for $4$-dimensional quantum gravity. However, as we pointed out at the end
of Section \ref{BCM}, the definition of the measure is incomplete since
there remains to determine the value of lower dimensional simplex amplitudes such as the
edge (tetrahedron) amplitude $A_e$. Although, the
normalization (\ref{eo4}, \ref{ess1}) of the Barrett-Crane model can be 
fixed in a natural way using the GFT formulation---or using similar (subgroup averaging) 
arguments \cite{ori1,ori3,pfei2}---one would really like to understand
this prescription in relation to the continuum theory. 

In Section \ref{BCM}, we have shown that the Barrett-Crane model can in fact be
interpreted as a `lattice' formulation of the gravity sector of the $SO(4)$
Plebanski theory---this conclusion can be extended to the Lorentzian
theory without further complications. This close relationship between 
the model and a continuum action suggested the
possibility of determining the correct spin foam measure by comparison with the
formal continuum path phase space path integral measure of Plebanski's theory.  
If the value of the edge amplitude (\ref{eo4}) could be found in this way this
would provide further evidence for the GFT-gravity duality and strengthen
the finiteness results. These considerations apply also for other spin foam models
in particular to those defined by constraining BF theory.
A general prescription for the construction of an anomaly free
measure for spin foam models is investigated in  \cite{myo}.  


\vskip.5cm \noindent $2$. {\em Numerical results}:
A very attractive property of the normalization (\ref{ess1}) is however the
finiteness of transition amplitudes on a fixed discretization. 
Having a well defined model---for both Riemannian and Lorentzian gravity---has 
opened for the first time the possibility for numerical explorations. 
An important step toward the numerical evaluation of the BC model 
is the development of efficient algorithms for the computation of the $10j$-symbols  \cite{egan}. 
Numerical calculations by Baez et al. \cite{baez1}
show that the sum over spin foams in our normalization 
of the Barrett-Crane model converges very fast so that
amplitudes are dominated by spin foams where most of the faces are
labeled by zero spin. The leading contribution comes from spin
foams made up from isolated `bubbles of geometry'. 
This problem is however not present in the Lorentzian version (\ref{ess2}) where
simple representations are in the continuum part of the spectrum.
Baez, Christensen, Halford and Tsang propose modifications of the normalization 
which still yield finite amplitudes but avoid this puzzling feature.
However, these modifications consist of ad hoc definitions of the
value of edge and face amplitudes. A clear understanding of this
would have to come with the understanding of the normalization issues discussed above. 

If we take seriously the normalization (\ref{eo4}), the
previous result might not be so worrisome after all. 
According to Section \ref{asy}, the asymptotic expression of the vertex amplitude 
shows that for large spins it is  dominated by degenerate $4$-simplexes.
These configurations are however strongly suppressed by the
form of the edge amplitude (\ref{eo4}) in our normalization.
Assuming that for small spin values the amplitudes behave
appropriately, large spin behavior might not be physically
relevant after all.

As pointed out by Smolin  \cite{lee3}, one could neglect spin zero 
representations according to the argument that they represent
`zero geometry states' (all geometric operators vanish on the spin zero subspace).
If one does so, then transition amplitudes are dominated by the spin one half
representations and the model becomes very similar to a model
of dynamical triangulations---in dynamical triangulations one sums over triangulations
whose $1$-simplexes have the same definite length, while here
the amplitudes are dominated by triangulations where triangles have the same 
fixed area given by (\ref{super}) for $j=1/2$. However one should also point out
an important difference: in our case, configurations are weighted with
complex amplitudes ${\rm exp}(i S)$ which no (a priori) connection to the
Euclidean amplitude  ${\rm exp}(-S)$ of dynamical triangulations. 
In the Lorentzian models the analogy simply disappears as representation 
labels are continuous.

\vskip.5cm
\noindent $3$. {\em The connection with the canonical picture}: 
We began this article motivating the spin foam approach from
the perspective of loop quantum gravity.
Can we establish a rigorous connection between the spin foam
models presented here and the canonical formulation? 
In all the models (for Riemannian gravity) introduced in Section 
\ref{sfm4d}, with the exception of the Barrett-Crane model, this relationship is manifest 
at the kinematical level as the boundary states are given by $SU(2)$-spin networks. 
This kinematic connection exists by construction in these cases.

What happens in the case of the Barrett-Crane model?
The Riemannian BC model is defined in terms of $Spin(4)$, so
naturally one could expect the boundary data to be given by
$Spin(4)$ spin networks (labeled by two half-integers
$(j_{\ell},j_r)$). The simplicity constraints impose
$j_{\ell}=j_r=j$ which could be interpreted as the existence of an
underlying $SU(2)$ connection. However, it can be shown that the 
boundary data of the model cannot be interpreted as given by an $SU(2)$
connection (recall the discussion of Section \ref{bbbb}). Similarly, in the Lorentzian Barrett-Crane model 
boundary states could be naively related to $SL(2,\C)$ spin networks. 
However similar considerations as in the Riemannian sector show that the
simplicity constraints (applied \`a la Barrett-Crane) reduce the
boundary states in a way that cannot be interpreted as a boundary
connection. Alexandrov \cite{alex2,alex3} has been studying
the possibility of defining a $SL(2,\C)$ connection canonical formulation
of general relativity. The second class constraints of the 
theory---related to the simplicity constraints of Plebanski's formulation---are dealt with
using the Dirac bracket in terms of which the original connection
becomes non commutative. In this sense there is no genuine or standard
spin network representation of states. It seems that such approach
could lead to the understanding of the relationship of the Barrett-Crane model
and a canonical formulation of gravity. The are in fact some indications of
this \cite{alex1,liv3}.  

Another difficulty in making contact with LQG is the
fact that operators associated to triangle-areas have continuous
spectrum in the Lorentzian models which is in conflict 
with the canonical result (\ref{aarreeaa}).
From this perspective one is compelled to study the model we have
described in Section \ref{LOR2} for which part of the area
spectrum coincides with (\ref{aarreeaa}). 

In the canonical framework the compact $SU(2)$ formulation is
achieved by means of the introduction of the Imirzi parameter
$\iota$ which defines a one parameter family of unitarily inequivalent
representations of quantum gravity. In particular,
geometric operators are modulated by $\iota$ (see (\ref{aarreeaa})
for example). Can the spin foam approach say something about
$\iota$? There is no conclusive answer to this question so far.
The spin foam quantization of generalizations of Plebanski's
action including the $\iota$ ambiguity  \cite{merced1} are analyzed
in  \cite{liv1,liv2}. 

So far we have only compared spin foams with loop quantum gravity
at the kinematical level. However, one of the main motivations of the approach
was the expected simplification in addressing dynamical questions. 
As we mentioned before, dynamics is not well understood in the canonical formulation. 
This is mostly due to the difficulties 
associated to regularization ambiguities in the construction of the scalar constraint, 
the understanding of its space of solutions and the the definition of
the physical scalar product. As we have seen all these issues can be naturally
addressed in the spin foam formulation. 
In turn, it would be nice to understand whether the approach can help
set a guiding principle that would allow for the selection of a 
preferred regularization of the scalar constraint.
Understanding of this question would strengthen the relationship between
the two formulations. An explicit analysis of the relation between the
covariant (spin foam) and canonical formulation of BF theory is
presented in  \cite{arns}. The reconstruction of the corresponding
scalar constraint operator is studied and carried out
explicitly in simple models. 

\vskip.5cm
\noindent $4$. {\em Alternatives to the Barrett-Crane model}:
The fact that degenerate configurations dominate the asymptotics of the Barrett-Crane
vertex amplitude \cite{baez8} can be interpreted as a serious problem.
A possible solution to the apparent problem  
is proposed in \cite{frei9}, where the evaluation of the
vertex amplitude is modified in a simple way to 
avoid degenerate contributions.  
The potential dominance of degenerate sectors of Plebanski's real formulation
over the gravity sector in a formal path integral has been emphasized 
on general grounds by Reisenberger in  \cite{reis0}. There is so far no 
rigorous result linking the previous two. 
The spin foam quantization of degenerate sectors ($e=0$) of Section \ref{ccc} is an 
attempt to investigate this problem. A puzzling feature of the degenerate model 
is that, even when the vertex and face amplitudes are different, 
it fully contains the Barrett-Crane $4$-simplex configurations.
Is this an indication of the apparent problem raised by Reisenberger?
This motivates the question of whether $e$ in (\ref{cdos}) is well defined in the Barrett-Crane model,
and if so, whether it differs from zero. The vanishing of $e$ would appear as
a serious obstacle to reproducing the gravity sector of Plebanski's theory in the
low energy limit.

\vskip.5cm
\noindent $5$. {\em Coupling with matter}: 
The inclusion of matter into the spin foam formalism is of
clear importance. Here we review recent results obtained in this 
direction. A natural generalization would be gravity 
with cosmological constant. The definition of the analog of 
the Barrett-Crane model with cosmological constant has been 
explored by Roche and Noui in \cite{phil}. The presence of the cosmological `matter'
is modeled using the quantum deformation $U_q(SL(2,\C))$ whose representation
theory is well understood \cite{buffe} and the deformation parameter is determined
by the value of the cosmological constant. In this modification the homogeneous space
$H^+$ is replaced by the  quantum deformation of the hyperboloid defined in terms of
a pile of {\em fuzzy} spheres.   
In a recent paper Oriti and Pfeiffer \cite{ori4} proposed a model
which couples the Riemannian Barrett-Crane model with Yang-Mills
theory. Their construction is likely to be generalizable to other
models. In the context of the GFT formulation of spin foam models Mikovic
proposes a way to include matter by adding certain {\em matter} fields
into the GFT-action. In  \cite{mik2,mik3} he generalizes
the GFT construction described in Section \ref{sec:gft-sf} by
allowing for the inclusion of spinnor fields in finite dimensional
representations of $SO(2)$ and $SO(3)$ representing matter
corresponding to fermions and gauge fields. States in the theory
are given by spin networks with open links; this is consistent
with results obtained in the canonical approach
 \cite{c10bis,c11,baez11,th5}. 

More radical and very appealing possibilities are suggested by Crane. In
 \cite{crane2} he proposes a topological QFT of the type of
BF theory as fundamental theory. The gravitational degrees of
freedom are represented by subset of (constrained) representations
while those of matter are encoded in the remaining ones. The full
spin foam model is topological (no local degrees of freedom). In
order to recover the low energy world (with local excitations) the
author appeals to certain ``symmetry breaking'' of topological
invariance. Crane's other proposal consists of interpreting
topological (conical) singularities naturally arising in the
structure of the Feynman diagrams of the GFT theory as
representing matter degrees of freedom  \cite{crane1}.
These possibilities are very attractive for their purely
geometric character.


\vskip.5cm
\noindent $6$. {\em Discretization dependence}: In Section \ref{sfm3d} we have seen that the discretization
dependence is trivial in three dimensions. In particular this has
been nicely formalized in the definition of continuum spin foams
by Zapata  \cite{za1}. Some definition of the refinement limit---discussed in Section
\ref{dd}---should be investigated for the models which are not trivially
discretization independent. The model defined in Section \ref{ccc}
(corresponding to the degenerate sectors of Plebanski's theory) 
might be an interesting candidate. The model is not topological and
has a clear connection to a continuum action. It is somehow
between the theory we want to define and the simpler theories we
understand well but that do not have local excitations (such as BF
theory and $3$-dimensional gravity). From this viewpoint we
believe that it might be useful to explore its behavior under 
refinement as a `toy model'. The simplicity of the model
might even allow for analytic computations. Incidentally other open issues in the spin foam
approach---such as the problem of the continuum limit, gauge and anomaly questions 
and the construction of the generalized projector into the physical
Hilbert space---should be investigated in this model.

Another proposal for a discretization independent formulation
is given by the GFT formulation.
The GFT formulation is very attractive since it provides a
discretization independent formulation of spin foam models from the
outset. Also it has been very useful for the definition of the
Lorentzian models of Sections {\ref{EU}} and \ref{LO} as a device
for formal manipulations. However, the mathematical consistency of
this definition depends on whether one can make sense of the
expansion in $\lambda$ of equation (\ref{exp0}). Suggestions on
how the series could be summable by complexification of the
coupling $\lambda$ can be found in  \cite{reis1}. A beautiful
example showing that the $\lambda$-series can be summed in certain cases has been
constructed by Freidel and Louapre  \cite{fre10} for certain $3$-dimensional GFT's.

In the standard background dependent realm, the problem of removing the
regulator introduced by the discretization is referred to as the {\em continuum limit}. 
Notice that the nature of the regulator in spin foams is totally different.
The discretization introduces a cut-off in the allowed configurations, but the discrete
and combinatorial nature of the latter is expected to be preserved when the
regulator is removed. When the regulator is removed we expect to obtain a definition
of the generalized projection operator which allows for the computation of
physical transition amplitudes between spin-network states.
On the other hand, physical states are expected to lie outside the
kinematical Hilbert space and so they could be given by elements in the
dual that resemble (in some sense) continuous configurations.
In order to avoid confusion about the matter we have consistently
referred to the notion of {\em removal of the discretization dependence} instead
of that of the {\em continuum limit}.

The approach of dynamical triangulations is a background independent formulation of quantum gravity in which
the removal of the regulator is similar in spirit to both the
problem in the spin foam context and that of (the background dependent) lattice gauge theory.
Here we present an account of the recent results and point out the 
conceptual differences with the spin foam approach. We hope
that, despite the differences, these results might be helpful in 
developing useful techniques in the spin foam approach. 
In dynamical triangulations the nature of the regulator is different from
that of spin foam models. In this approach, (diff-equivalent classes of) smooth metric
configurations are approximated by space-time triangulations where $1$-simplexes
have the same fixed proper length $\ell$. The smaller the length scale the better the
approximation; therefore, the proper length  represents the regulator
in the theory to be removed in a certain $\ell \rightarrow 0$ limit.
The phrase {\em continuum limit} certainly has a clear-cut meaning here. 
The continuum limit in dynamical triangulations has been extensively studied. 
As we previously mentioned, dynamical triangulations is a definition of 
quantum gravity based on the Euclidean path integral, i.e., configurations are weighted with real 
amplitudes ${\rm exp}(-S)$. In order to recover Lorentzian quantum gravity 
one has to define what is meant by a Wick rotation in the background independent
framework. Once this is done, the continuum limit can be studied using
standard techniques of statistical mechanics. Results indicate that 
in Euclidean dynamical triangulations there is no continuum limit \cite{amb1}. 
The path integral is dominated by singular Euclidean geometries.
The situation improves in the Lorentzian models. This is because
of the restrictions imposed by the notion of Wick rotation defined. 
These restrictions are such that in 1+1 gravity the model is exactly soluble                    
and possesses a continuum limit. There is some numerical evidence that could be 
the case in higher dimensions \cite{amb2d,amb3d}.

\vskip.5cm
\noindent $7$. {\em Contact with the low energy world}:
In this article we have concentrated mostly on mathematical and conceptual issues
which are important in the construction of a consistent spin foam model for gravity. 
However, one of the most pressing questions is whether the spin foam
models can reproduce the low energy world of general relativity.
Although this is certainly one of the most important questions, 
there is unfortunately no conclusive evidence of 
this at present and most of the work lies ahead of us. 

Some of the models presented in Section \ref{sfm4d} satisfy some  
notion of `naive' classical-limit in the sense that they are derived from a
continuous action of general relativity. Other models are defined by postulating the fundamental
dynamics and using the kinematical structure discovered by the canonical 
formulation. If any of these models are to be taken as strong candidates for theories of
quantum gravity they must be able to reproduce the physics of gravity 
at low energy and predict the corresponding semi classical corrections.  
In this respect, it is still an unsettled issue whether the differences of the various models
at the fundamental level lead to different theories or rather
are to be expected to disappear in the continuum limit. 
After all, the latter happens in lattice gauge 
theory in the renormalization process. Even when progress has been made in this direction,
there is however no clear-cut formulation of the analog of renormalization theory
for spin foams. The reason for this is the difficulty of applying standard techniques in the 
background independent context. In Section \ref{clew} we mentioned some new ideas and attempts to
tackle this problem. This is an extremely important problem
where new ideas will have to play an important role.

\vskip.5cm
\noindent $8$. {\em The generalized projection vs. the Feynman proper-time propagator}:
Throughout this work we have advocated the viewpoint of the
path integral as a device for the computation of the 
generalized projection operator into the physical 
Hilbert space of quantum gravity. We claimed that this
picture is forced upon us by the general covariance of 
general relativity. Although we have kept the term {\em transition amplitude}
for notational convenience there is no notion of 
{\em time} involved in our definitions.

In other path integral approaches
one aims at the construction of the notion of transition amplitudes between
$3$-geometries at a definite {\em time}. Such transition amplitudes are
governed by the so-called {\em proper time} evolution operator or {\em proper-time} propagator.

This possibility has also been explored in the context of LQG  \cite{reis5}
and is the perspective in which the causal models of
Section \ref{fotin} are defined. There are also modifications
of the Barrett-Crane model which attempt to define such proper time
evolution operator  \cite{liv4,pfei3}. The main difficulty in such approach is 
how to give meaning to the notion of `time' in the 
general covariant context.

\begin{center}
\vskip.5cm
\rule{1cm}{.015cm}\rule{1cm}{.030cm}\rule{2cm}{.05cm}\rule{1cm}{.03cm}\rule{1cm}{.015cm}
\vskip.5cm
\end{center}

The suggestion that quantum gravity should be described in terms
of discrete combinatorial structures can be traced all the way
back to Einstein  \cite{Stachel}. In this review we have described an approach
inspired by the non-perturbative canonical quantization of general relativity
in which these suggestions are concretely realized. 
We have shown that a great deal of progress has been achieved
in understanding the conceptual issues involved. The fundamental motivation was the construction
of a theory that would provide a device to analyze dynamics in quantum gravity.
The conceptual setting is clear: spin foams should provide the definition
of the physical scalar product and hence the physical Hilbert space of 
quantum gravity which encodes all the information about quantum dynamics.
We have seen that these models can be obtained as lattice quantizations of general relativity in 
appropriate variables. The `lattice' action being naively related to the continuum
general relativity action (or some equivalent classical formulation)
in the sense as Wilson's action for lattice gauge theory.
Some of these models have remarkable finiteness properties both in 
the Riemannian and Lorentzian sectors.
We have also shown how the basic structure of spin foams arises from
the canonical formulation of loop quantum gravity. In addition we
described models which are not related to a classical action and 
are constructed from the basic properties of spin networks plus
simple causality requirements. For all these models the
common structure arising is given by spin foams: colored $2$-complexes
where the geometric degrees of freedom are encoded in a fully combinatorial manner.
Spin foam models appear as a beautiful realization of Einstein's idea. 
There are certainly  many difficult open questions and we have tried to 
point out those which we judge the most important ones. We hope that new ideas and 
hard work will continue to contribute to their resolution in the near future.

\section{Acknowledgments}
I thank Abhay Ashtekar, John Baez, Martin Bojowald, Dan Christensen, Rodolfo
Gambini, Amit Ghosh, Jorge Pullin, Michael Reisenberger, Carlo
Rovelli and Lee Smolin for discussions. I thank Martin and Carlo for suggestions that helped
improve the presentation of this work. I am grateful to Florian Girelli, Josh Willis and Jacek Wisniewski
for the careful reading of the manuscript. I specially thank Carlo Rovelli for his strong support. 
This work was supported in part by NSF Grant PHY-0090091, and Eberly Research Funds of Penn State.


\end{document}